\documentclass{article}

\usepackage{epubtk}

\usepackage{latexsym}
\usepackage{epsf}
\usepackage{longtable}

\newcounter{boxcounter}
\newenvironment{mybox}[1]
{\refstepcounter{boxcounter}
 \vspace{1 em}
 \hrule
 \begin{center}
   {\bf Box\protect~$ \mbox{\protect\boldmath $\theboxcounter$} $. #1}
 \end{center}
 \hrule}
{\hrule
 \vspace{1 em}}

\begin{document}

\title{The Confrontation between General Relativity and Experiment}

\author{\epubtkAuthorData{Clifford M. Will}
        {McDonnell Center for the Space Sciences \\
        Department of Physics \\
        Washington University, St. Louis MO 63130 \\
        U.S.A.}
        {}
        {http://wugrav.wustl.edu/People/CLIFF/index.html}
}

\date{}

\maketitle

%%%%%%%%%%%%%%%%%%%%%%%%%%%%%%%%%%%%%%%%%%%%%%%%%%%%%%%%%%%%%%%%%%%%%%%%%%%%%%%%%%%
%%%%%%%%%%%%%%%%%%%%%%%%%%%%%%%%%%%%%%%%%%%%%%%%%%%%%%%%%%%%%%%%%%%%%%%%%%%%%%%%%%%
%%%%%%%%%%%%%%%%%%%%%%%%%%%%%%%%%%%%%%%%%%%%%%%%%%%%%%%%%%%%%%%%%%%%%%%%%%%%%%%%%%%

\begin{abstract}
The status of experimental tests of general relativity and of
theoretical frameworks for analysing them is 
reviewed.
Einstein's equivalence principle (EEP) is well supported by experiments
such as the E\"otv\"os experiment, tests of special relativity, and
the gravitational redshift experiment.  Ongoing tests of EEP and of the
inverse square law are 
searching for new interactions arising from unification or quantum
gravity.  Tests of general
relativity at the post-Newtonian level
have reached high precision, including the light
deflection, the Shapiro time delay, the perihelion advance of
Mercury, and the Nordtvedt effect in lunar motion.
Gravitational-wave damping has been detected in an amount that agrees
with general relativity to better than
half a percent using the Hulse-Taylor binary pulsar,
and other binary pulsar systems have yielded other tests, especially of
strong-field effects.
When direct observation of gravitational radiation from astrophysical
sources begins, new tests of general relativity will be possible.
\end{abstract}

\epubtkKeywords{tests of relativistic gravity,theories of gravity,post-Newtonian
limit,gravitational radiation}

\newpage

%%%%%%%%%%%%%%%%%%%%%%%%%%%%%%%%%%%%%%%%%%%%%%%%%%%%%%%%%%%%%%%%%%%%%%%%%%%%%%%%%%%
%%%%%%%%%%%%%%%%%%%%%%%%%%%%%%%%%%%%%%%%%%%%%%%%%%%%%%%%%%%%%%%%%%%%%%%%%%%%%%%%%%%
%%%%%%%%%%%%%%%%%%%%%%%%%%%%%%%%%%%%%%%%%%%%%%%%%%%%%%%%%%%%%%%%%%%%%%%%%%%%%%%%%%%

\section{Introduction}
\label{S1}

At the time of the birth of general relativity (GR), experimental
confirmation was almost a side issue.  Einstein did calculate observable
effects of general relativity, such as the perihelion advance of Mercury,
which he knew to be an unsolved problem, and the deflection of light,
which was subsequently verified.  But compared to the inner consistency and
elegance of the theory, he regarded such empirical questions as almost
peripheral.  Today, experimental gravitation is a major component of
the field, characterized by continuing efforts to test the theory's
predictions, to search for gravitational imprints of high-energy particle
interactions, and to detect gravitational waves from astronomical sources.

The modern history of experimental relativity can be divided roughly into
four periods: Genesis, Hibernation, a Golden Era, and the Quest for Strong
Gravity.  The Genesis (1887--1919) comprises the period of the two great
experiments which were the foundation of relativistic physics -- the
Michelson-Morley experiment and the E\"otv\"os experiment -- and the two
immediate confirmations of GR -- the deflection of light and the
perihelion advance of Mercury.  Following this was a period of Hibernation
(1920--1960) during which theoretical work temporarily outstripped
technology and experimental possibilities, and, as a consequence, the
field stagnated and was relegated to the backwaters of physics and
astronomy.

But beginning around 1960, astronomical discoveries (quasars, pulsars,
cosmic background radiation) and new experiments pushed GR to the
forefront.  Experimental gravitation experienced a Golden Era (1960--1980)
during which a systematic, world-wide effort took place to understand the
observable predictions of GR, to compare and contrast them with the
predictions of alternative theories of gravity, and to perform new
experiments to test them.  The period began with an experiment to confirm
the gravitational frequency shift of light (1960) and ended with the
reported decrease in the orbital period of the Hulse-Taylor binary pulsar
at a rate consistent with the GR prediction of gravity
wave energy loss (1979).  The results all supported GR, and most
alternative theories of gravity fell by the wayside (for a popular review,
see~\cite{WER}).

Since 1980, the field has entered what might be termed a Quest for Strong
Gravity.  Many of the remaining interesting weak-field predictions of the
theory are extremely small and difficult to check, in some cases requiring
further technological development to bring them into detectable range.
The sense of a systematic assault on the weak-field predictions of GR has
been supplanted to some extent by an opportunistic approach in which novel
and unexpected (and sometimes inexpensive) tests of gravity have arisen
from new theoretical ideas or experimental techniques, often from unlikely
sources.  Examples include the use of laser-cooled atom and ion traps to
perform ultra-precise tests of special relativity;  the proposal of a
``fifth'' force, which led to a host of new tests of the weak equivalence
principle; and recent ideas of large extra dimensions, which have motived
new tests of the inverse square law of gravity at sub-millimeter scales.

Instead, much of the focus has shifted to experiments which can probe the
effects of strong gravitational fields.  The principal figure of merit
that distinguishes strong from weak gravity is the quantity $\epsilon \sim
GM/Rc^2$, where $G$ is the Newtonian gravitational constant, $M$ is the
characteristic mass scale of the phenomenon, $R$ is the characteristic
distance scale, and $c$ is the speed of light.  Near the event horizon of
a non-rotating black hole, or for the expanding observable universe,
$\epsilon \sim 0.5$; for neutron stars, $\epsilon \sim 0.2$.  These are
the regimes of strong gravity.  For the solar system $\epsilon < 10^{-5}$;
this is the regime of weak gravity.  At one extreme are the strong
gravitational fields associated with Planck-scale physics.  Will
unification of the forces, or quantization of gravity at this scale leave
observable effects accessible by experiment?  Dramatically improved tests
of the equivalence principle, of the inverse square law, or of local
Lorentz invariance are being
mounted, to search for or bound the imprinted effects of Planck-scale
phenomena.  At the other extreme are the strong fields associated with
compact objects such as black holes or neutron stars.  Astrophysical
observations and gravitational wave detectors are being planned to explore
and test GR in the strong-field, highly-dynamical regime associated with
the formation and dynamics of these objects.

In this Living Review, we shall survey the theoretical frameworks for
studying experimental gravitation, summarize the current status of
experiments, and attempt to chart the future of the subject. We shall not
provide complete references to early work done in this field but instead
will refer the reader to the appropriate review articles and monographs,
specifically to {\it Theory and Experiment in Gravitational
Physics}~\cite{tegp}, hereafter referred to as TEGP.  Additional recent
reviews in this subject
are~\cite{Will300,sussp,slac,damourreview,DamourPDG,shapiro}. References to
TEGP will be by chapter or section, e.g.\ ``TEGP 8.9~\cite{tegp}''.

\newpage

%%%%%%%%%%%%%%%%%%%%%%%%%%%%%%%%%%%%%%%%%%%%%%%%%%%%%%%%%%%%%%%%%%%%%%%%%%%%%%%%%%%
%%%%%%%%%%%%%%%%%%%%%%%%%%%%%%%%%%%%%%%%%%%%%%%%%%%%%%%%%%%%%%%%%%%%%%%%%%%%%%%%%%%
%%%%%%%%%%%%%%%%%%%%%%%%%%%%%%%%%%%%%%%%%%%%%%%%%%%%%%%%%%%%%%%%%%%%%%%%%%%%%%%%%%%

\section{Tests of the Foundations of Gravitation Theory}
\label{S2}

%%%%%%%%%%%%%%%%%%%%%%%%%%%%%%%%%%%%%%%%%%%%%%%%%%%%%%%%%%%%%%%%%%%%%%%%%%%%%%%%%%%
%%%%%%%%%%%%%%%%%%%%%%%%%%%%%%%%%%%%%%%%%%%%%%%%%%%%%%%%%%%%%%%%%%%%%%%%%%%%%%%%%%%

\subsection{The Einstein equivalence principle}
\label{eep}

The principle of equivalence has historically played an important role in
the development of gravitation theory.  Newton regarded this principle as
such a cornerstone of mechanics that he devoted the opening paragraph of
the {\it Principia} to it.  In 1907, Einstein used the principle as a
basic element in his development of
general relativity.  We now regard the principle of
equivalence as the foundation, not of Newtonian gravity or of GR, but of
the broader idea that spacetime is curved.
Much of
this viewpoint can be traced back to Robert Dicke, who contributed crucial
ideas
about the foundations of gravitation theory between 1960 and 1965.  These
ideas were summarized in his influential Les Houches lectures of 1964
\cite{dicke64},  and resulted in what has come to be called the
Einstein equivalence principle (EEP).

One elementary equivalence principle is the kind Newton had in mind when
he stated that the property of a body called ``mass'' is proportional to
the ``weight'', and is known as the weak equivalence principle (WEP).
An alternative statement of WEP is that the trajectory of a freely
falling ``test'' 
body (one not acted upon by such forces as electromagnetism and
too small to be affected by tidal gravitational forces) is independent of
its internal structure and composition.  In the simplest case of dropping
two different bodies in a gravitational field, WEP states that the bodies
fall with the same acceleration (this is often termed the Universality of
Free Fall, or UFF).

The Einstein equivalence principle (EEP) 
is a more powerful and far-reaching concept; it states that:

\begin{enumerate}
\item
WEP is valid.
\item
The outcome of any local non-gravitational
experiment is independent of the velocity of the freely-falling reference
frame in which it is performed.
\item The outcome of any local
non-gravitational experiment is independent of where and when in the
universe it is performed.
\end{enumerate}

The second piece of EEP is called local Lorentz invariance (LLI), and the
third piece is called local position invariance (LPI).

For example, a measurement of the electric force between two charged
bodies is a local non-gravitational experiment; a measurement of the
gravitational force between two bodies (Cavendish experiment) is not.

The Einstein equivalence principle is the heart and soul of gravitational
theory, for it is possible to argue convincingly that if EEP is valid,
then gravitation must be a ``curved spacetime'' phenomenon, in other
words, the effects of gravity must be equivalent to the effects of living
in a curved spacetime.  As a consequence of this argument, the only
theories of gravity that can fully embody EEP are those that satisfy the
postulates of ``metric theories of gravity'', which are:

\begin{enumerate}
\item
Spacetime is endowed with a symmetric metric.
\item
The trajectories
of freely falling test bodies are geodesics of that metric.
\item
In local
freely falling reference frames, the non-gravitational laws of physics are
those written in the language of special relativity.
\end{enumerate}

The argument that leads to this conclusion simply notes that, if EEP is
valid, then in local freely falling frames, the laws governing experiments
must be independent of the velocity of the frame (local Lorentz
invariance), with constant values for the various atomic constants (in
order to be independent of location).  The only laws we know of that
fulfill this are those that are compatible with special relativity, such
as Maxwell's equations of electromagnetism.  Furthermore, in local freely
falling frames, test bodies appear to be unaccelerated, in other words
they move on straight lines; but such ``locally straight'' lines simply
correspond to ``geodesics'' in a curved spacetime (TEGP 2.3~\cite{tegp}).

General relativity is a metric theory of gravity, but then so are
many others, including the Brans-Dicke theory and its generalizations.  
Theories in which varying non-gravitational constants are
associated with dynamical fields that couple to matter directly are
not metric theories.  Neither, in
this narrow sense, is
superstring theory (see Sec.~\ref{newinteractions}), which, while
based fundamentally on a spacetime metric, introduces additional
fields (dilatons, moduli)
that can couple to material stress-energy in a
way that can lead to violations, say, of WEP.  It is important to point out,
however, that there is some ambiguity in whether one treats such fields as
EEP-violating gravitational fields, or simply as additional matter fields, like
those that carry electromagnetism or the weak interactions.
Still,
the notion of curved spacetime is a very general and fundamental
one, and therefore it is important to test the various aspects of
the Einstein equivalence principle thoroughly.
We first survey the experimental tests, and describe some of the theoretical
formalisms that have been developed to interpret them.  
For other reviews of
EEP and its experimental and theoretical significance,
see~\cite{hauganlammer2,lammer03}.

%%%%%%%%%%%%%%%%%%%%%%%%%%%%%%%%%%%%%%%%%%%%%%%%%%%%%%%%%%%

\subsubsection{Tests of the weak equivalence principle}
\label{wep}

A direct test of WEP is the comparison of the acceleration of two
laborat\-ory-sized bodies of different composition in an external
gravitational field.
If the principle were violated, then the
accelerations of different bodies would differ.  The simplest
way to quantify such possible violations of WEP in a form
suitable for comparison with experiment is to suppose that for
a body with inertial mass $m_{\rm I}$, the passive gravitational
mass $m_{\rm P}$ is no longer equal to $m_{\rm I}$, so that in a
gravitational field $g$, the acceleration is given
by $m_{\rm I} a= m_{\rm P} g$.  Now the inertial mass of a typical
laboratory body is made up of several types of mass-energy:  rest
energy, electromagnetic energy, weak-interaction energy, and so
on.  If one of these forms of energy contributes to $m_{\rm P}$
differently than it does to $m_{\rm I}$, a violation of WEP would
result.  One could then write
\begin{equation}\label{E1}
    m_{\rm P} = m_{\rm I} + \sum_A \eta^A E^A /c^2,
\end{equation}
where $E^A$ is the internal energy of the body generated by
interaction $A$, and $\eta^A$ is a dimensionless parameter that
measures the strength of the violation of WEP induced by that
interaction, and $c$ is the speed of light.  A measurement or limit
on the fractional difference in acceleration between two bodies
then yields a quantity called the ``E\"otv\"os ratio'' given by
\begin{equation}\label{E2}
 \eta \equiv {{2 | a_1  -  a_2 |} \over {| a_1 +  a_2 |}}
     = \sum_A \eta^A
 \left(   {{E_1^A} \over {m_1 c^2}}
      -  {{E_2^A} \over {m_2 c^2} }
\right),
\end{equation}
where we drop the subscript ``I'' from the inertial masses.
Thus, experimental limits on $\eta$ place limits on the
WEP-violation parameters $\eta^A$.

Many high-precision E\"otv\"os-type experiments have been performed, from
the pendulum experiments of Newton, Bessel and Potter, to the classic
torsion-balance measurements of E\"otv\"os~\cite{eotvos},
Dicke~\cite{dicke1}, Braginsky~\cite{braginsky} and their collaborators.
In the modern torsion-balance experiments, two objects of different
composition are connected by a rod or placed on a tray and suspended in a
horizontal orientation by a fine wire.  If the gravitational acceleration
of the bodies differs, and this difference has a component perpendular to
the suspension wire, there will be a torque induced on the 
wire, related to the angle between the wire and the direction of the
gravitational acceleration {\boldmath $g$}.  If the entire apparatus is
rotated about some direction with angular velocity $\omega$, the torque
will be modulated with period $2 \pi / \omega$. In the experiments of
E\"otv\"os and his collaborators, the wire and {\boldmath $g$} were not
quite parallel because of the centripetal acceleration on the apparatus
due to the Earth's rotation; the apparatus was rotated about the direction
of the wire.  In the Dicke and Braginsky experiments, {\boldmath $g$} was
that of the Sun, and the rotation of the Earth provided the modulation of
the torque at a period of 24~hr (TEGP 2.4~(a)~\cite{tegp}).  Beginning in
the late 1980s, numerous experiments were carried out primarily to search
for a ``fifth force'' (see Sec.~\ref{fifthforce}), but their null
results also constituted tests of WEP.  In the ``free-fall Galileo
experiment'' performed at the University of Colorado, the relative
free-fall acceleration of two bodies made of uranium and copper was
measured using a laser interferometric technique. The ``E\"ot-Wash''
experiments carried out at the University of Washington used a
sophisticated torsion balance tray to compare the accelerations of various
materials toward local topography on Earth, movable laboratory masses, the
Sun and the galaxy~\cite{Su94,baessler99}, and have reached levels of
$3 \times 10^{-13}$~\cite{adelberger01}.  
The resulting upper limits on $\eta$ are summarized
in Figure~\ref{wepfig} (TEGP 14.1~\cite{tegp}; for a bibliography of
experiments up to 1991, see~\cite{fischbach92}). 

\epubtkImage{}{
\begin{figure}[hptb]
  \def\epsfsize#1#2{0.5#1}
  \centerline{\epsfbox{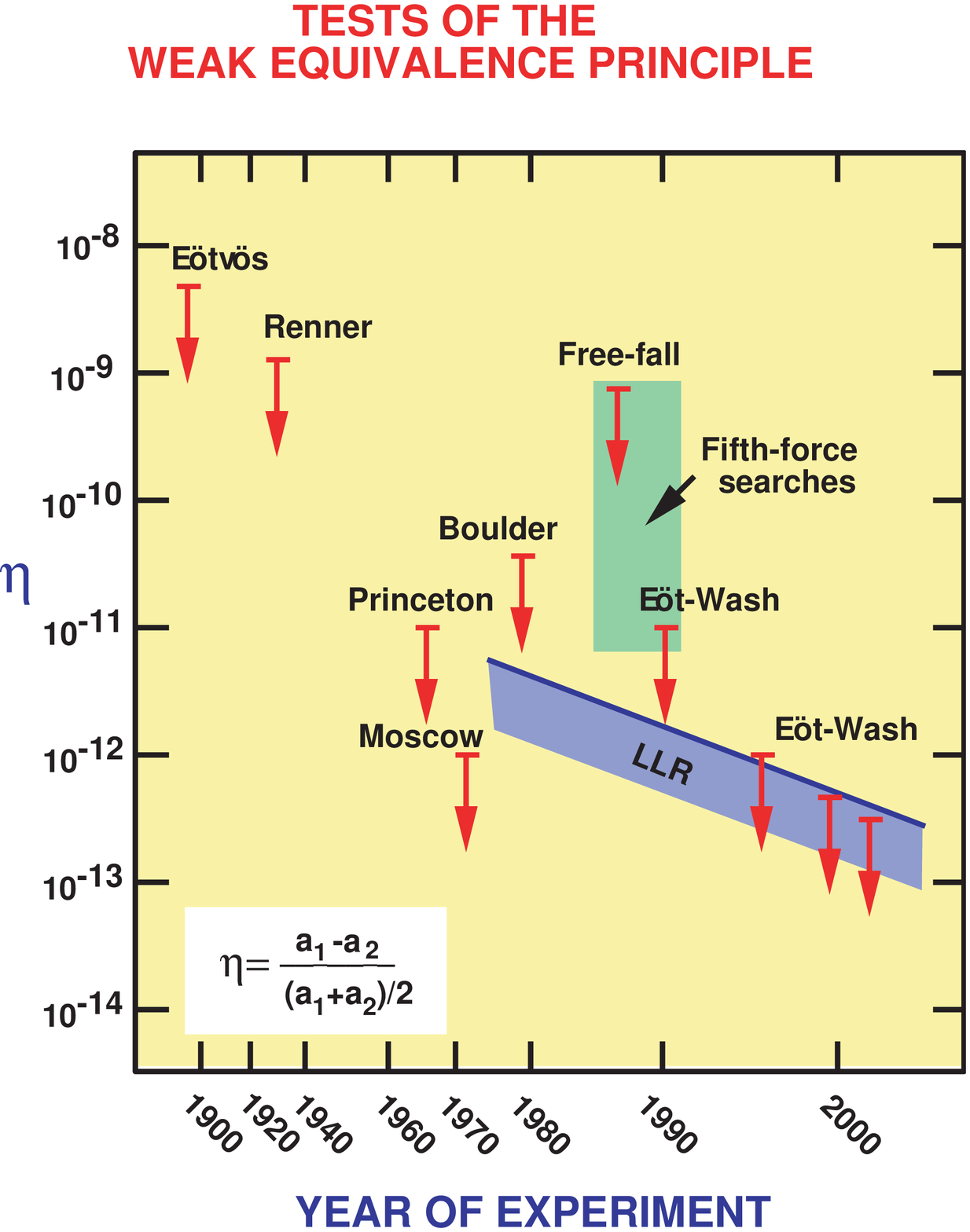}}
  \caption{\it Selected tests of the weak equivalence principle,
    showing bounds on $\eta$, which measures fractional difference in
    acceleration of different materials or bodies.  The free-fall and
    E$\mbox{\it \"o}$t-Wash experiments were originally performed to search for a
    fifth force (green region, representing many experiments). 
    The blue band shows evolving bounds on $\eta$ for
    gravitating bodies from lunar laser ranging (LLR).}
  \label{wepfig}
\end{figure}}

A number of projects are in the development or planning stage to push
the bounds on $\eta$ even lower.  The project MICROSCOPE
(MICRO-Satellite \`a Train\'ee Compens\'ee pour l'Observation du
Principe d'\'Equivalence) is
designed to test WEP to $10^{-15}$.  It is being developed
by the French space agency CNES for a possible launch in March, 2008,
for a one-year mission~\cite{microscope}.
The drag-compensated 
satellite will be in a Sun-synchronous polar orbit at 700 km
altitude, with a payload consisting of two differential
accelerometers, one with elements made of the same material
(platinum), and another with elements made of different materials
(platinum and titanium).  

Another, known as
Satellite Test of the Equivalence Principle (STEP)~\cite{STEP},
is under consideration as a possible
joint effort of NASA  and the European Space Agency (ESA), with the
goal of a $10^{-18}$ test.  STEP would improve upon MICROSCOPE by
using cryogenic techniques to reduce thermal noise, among other
effects.  At present, STEP (along with a number of variants, called MiniSTEP
and QuickSTEP) has not been approved by any agency beyond the level of
basic design studies or supporting research and development.
An alternative concept for a space test of WEP is
Galileo-Galilei~\cite{gg}, which uses a rapidly rotating differential
accelerometer as its basic element.  Its goal is a bound on $\eta$ at
the $10^{-13}$ level on the ground and $10^{-17}$ in space.  

%%%%%%%%%%%%%%%%%%%%%%%%%%%%%%%%%%%%%%%%%%%

\subsubsection{Tests of local Lorentz invariance}
\label{lli}

Although special relativity itself never benefited from the kind of
``crucial'' experiments, such as the perihelion advance of Mercury and
the deflection of light, that contributed so much to the initial acceptance
of
GR and to the fame of Einstein, the steady
accumulation of experimental support, together with the successful
merger of special relativity
with quantum mechanics, led to its being accepted
by mainstream physicists by the late 1920s, ultimately to become part of
the standard toolkit of every working physicist.
This accumulation
included

\begin{itemize}

\item
the classic Michelson-Morley experiment and its
descendents\cite{mm,shankland,jaseja,brillethall},
\item
the Ives-Stillwell, Rossi-Hall and
other tests of time-dilation\cite{ives,rossi,farley},
\item
tests of the independence of the speed of
light of the velocity of the source, using both binary X-ray stellar
sources and high-energy pions\cite{brecher,alvager},
\item
tests of the isotropy of the speed of light\cite{Champeney,riis,krisher90}

\end{itemize}

In addition to these direct experiments, there was the Dirac
equation of quantum mechanics and its prediction of anti-particles and
spin; later would come the stunningly
successful
relativistic theory of quantum electrodynamics.

In 2005, on the 100th anniversary of the introduction of special relativity,
one might ask ``what is there to test?''.  Special relativity has
been so thoroughly
integrated into the fabric of modern physics that its validity is
rarely
challenged, except by cranks and crackpots.
It is ironic then, that during the past several years, a vigorous
theoretical and experimental effort has been launched, on an
international
scale, to find violations of special relativity.
The motivation for this effort is not a desire
to repudiate Einstein, but to look for
evidence of new physics ``beyond'' Einstein, such as apparent
violations
of Lorentz invariance that might result from certain models of quantum
gravity.
Quantum gravity asserts that there is a fundamental length scale
given by the Planck length, $L_p = (\hbar G/c^3)^{1/2} = 1.6 \times
10^{-33}
\, {\rm cm}$, but since length is not an invariant quantity
(Lorentz-FitzGerald contraction), then there could be a violation of
Lorentz
invariance at some level in quantum gravity.   In brane world
scenarios, while
physics may be locally Lorentz invariant in the higher dimensional
world,
the confinement of the interactions of normal physics to our
four-dimensional ``brane'' could induce apparent Lorentz violating
effects.
And in models such as string theory, the presence of additional
scalar,
vector and tensor long-range fields that couple to matter of the
standard
model could induce effective violations of Lorentz symmetry.
These and other ideas have motivated
a serious
reconsideration of how to test Lorentz invariance with better
precision and
in new ways.

A simple and
useful way of interpreting some of these modern experiments, called the
$c^2$-formalism,  is to suppose that the
electromagnetic interactions suffer a slight violation of Lorentz
invariance, through a change in the speed of electromagnetic radiation $c$
relative to the limiting speed of material test particles ($c_0$, made 
to take the value unity via a choice of units), in other words, $c \ne 1$ (see
Sec.~\ref{c2formalism}).  Such a violation necessarily selects a preferred
universal rest frame, presumably that of the cosmic background radiation,
through which we are moving at about 370~km/s~\cite{lineweaver96}.  Such a
Lorentz-non-invariant electromagnetic interaction would cause shifts in
the energy levels of atoms and nuclei that depend on the orientation of
the quantization axis of the state relative to our universal velocity
vector, and on the quantum numbers of the state.  The presence or absence
of such energy shifts can be examined by measuring the energy of one such
state relative to another state that is either unaffected or is affected
differently by the supposed violation.  One way is to look for a shifting
of the energy levels of states that are ordinarily equally spaced, such as
the Zeeman-split
$2J+1$ ground states of a nucleus of total spin $J$ in a magnetic field;
another is to compare the levels of a complex nucleus
with the atomic hyperfine levels of a hydrogen maser clock.  
The magnitude of these ``clock
anisotropies'' would be proportional to
$\delta \equiv | c^{-2}-1|$.

The earliest clock anisotropy experiments were the
Hughes-Drever experiments, performed in the period
1959--60 independently by Hughes and collaborators at Yale University, and
by Drever at Glasgow University, although their original motivation was 
somewhat different~\cite{hughes,drever}.  
The Hughes-Drever 
experiments yielded extremely accurate results, quoted as limits
on the parameter $\delta \equiv c^{-2}-1$ in Figure~\ref{llifig}.  
Dramatic improvements were made in the 1980s using
laser-cooled trapped atoms and ions\cite{prestage85,lamoreaux86,chupp}.
This technique made
it possible
to reduce the broading of resonance lines caused by collisions,
leading to improved bounds on $\delta$ shown in Figure \ref{llifig}
(experiments
labelled NIST, U. Washington and Harvard, respectively).

Also
included for comparison is the corresponding limit obtained from
Michelson-Morley type experiments (for a review, see~\cite{hauganwill}).
In those experiments, when viewed from the preferred frame, the speed of
light down the two arms of the moving interferometer is $c$, while it can be
shown using the electrodynamics of the $c^2$ formalism,
that the compensating
Lorentz-FitzGerald contraction of the parallel arm is governed by the speed 
$c_0=1$.  Thus the Michelson-Morley
experiment and its descendants also measure the coefficient
$c^{-2}-1$.  One of these is the Brillet-Hall experiment~\cite{brillethall},
which used a Fabry-Perot laser interferometer. 
In a  recent series of experiments,
the frequencies of electromagnetic cavity oscillators in various
orientations
were compared with each other or
with atomic clocks as a function of
the orientation of the 
laboratory~\cite{wolf03,lipa03,muller03,antonini05,stanwix05}.  
These
placed bounds on $c^{-2}-1$ at the level of better than a part in $10^9$.
Haugan and L\"ammerzahl~\cite{hauganlammer1} have considered
the bounds that Michelson-Morley type experiments could place on a modified
electrodynamics involving a ``vector-valued'' effective photon mass.

\epubtkImage{}{
\begin{figure}[hptb]
  \def\epsfsize#1#2{0.5#1}
  \centerline{\epsfbox{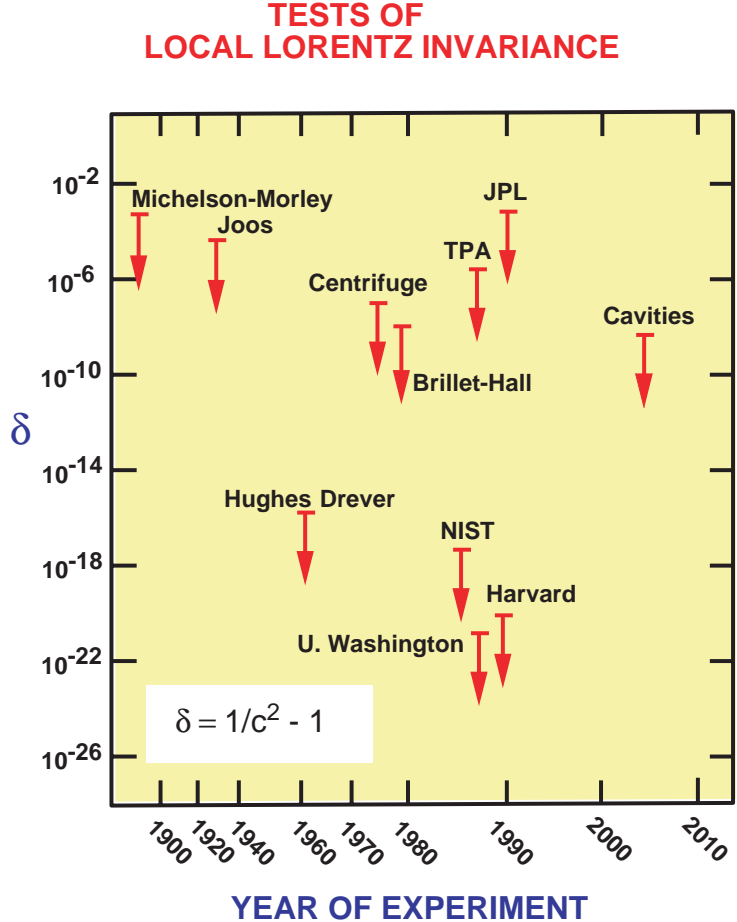}}
  \caption{\it Selected tests of local Lorentz invariance showing the
    bounds on the parameter $\delta$, which measures the degree of
    violation of Lorentz invariance in electromagnetism.  The
    Michelson-Morley, Joos, Brillet-Hall and cavity experiments test the
    isotropy of the round-trip speed of light.  
    The centrifuge, two-photon absorption (TPA) and
    JPL experiments test the isotropy of light speed using one-way
    propagation.  The most precise experiments test isotropy of
    atomic energy levels.  The limits assume a speed of Earth of
    370\protect~km/s relative to the mean rest frame of the universe.}
  \label{llifig}
\end{figure}}

The $c^2$ framework focusses exclusively on classical electrodynamics.
It has recently been extended to the entire standard model of particle physics
by Kosteleck\'y and
colleagues~\cite{colladay97,colladay98,kosteleckymewes02}.  The ``Standard
Model Extension'' (SME) has a large number of Lorentz-violating
parameters, opening up many new opportunities for experimental tests (Sec.
\ref{SME}).
A variety of clock anisotropy experiments have been carried out to
bound the electromagnetic parameters of the SME 
framework~\cite{kosteleckylane99}.  
For example,
the cavity experiments described above~\cite{wolf03,lipa03,muller03}
placed bounds on the coefficients of the
tensors $\tilde{\kappa}_{e-}$ and $\tilde{\kappa}_{o+}$ 
(see Sec.~ \ref{SME} for definitions) at the levels of
$10^{-14}$ and $10^{-10}$, respectively.
Direct comparisons between atomic clocks based on different nuclear species
place bounds on SME parameters in the neutron and proton sectors, depending
on the nature of the transitions involved.  The bounds achieved range from
$10^{-27}$ to $10^{-32} \, {\rm GeV}$.  

Astrophysical observations have also been used to bound Lorentz violations.
For example, if photons satisfy the Lorentz violating dispersion relation
\begin{equation}
E^2 =  p^2 c^2 + E_{Pl} f^{(1)} |p|c + f^{(2)} p^2c^2 +
\frac{f^{(3)}}{E_{Pl}} |p|^3c^3 + \dots \,,
\label{dispersion}
\end{equation}
where $E_{Pl}= (\hbar c^5/G)^{1/2}$ is the Planck energy, 
then the speed of light 
$v_\gamma=\partial E /\partial p$ would
be given, to linear order in the $f^{(n)}$ by
\begin{equation}
\frac{v_\gamma}{c} \approx 1 + \sum_{n \ge 1} 
\frac{(n-1)f_\gamma^{(n)} E^{n-2}}{2E_{Pl}^{n-2}} \,.
\end{equation}
Such a Lorentz-violating dispersion relation could be a relic of
quantum gravity, for instance.
By bounding the difference in arrival time of high-energy photons from a
burst source at large distances, one could bound contributions to the
dispersion for $n >2$.  One limit, $|f^{(3)}| < 128$ comes 
from observations of  1 and 2 TeV gamma rays from the blazar Markarian 421
\cite{biller}.  Another limit comes from birefringence in photon
propagation: in many Lorentz violating models, different photon polarizations  
may propagate with different speeds, causing the plane of polarization
of a wave to rotate.  If the frequency dependence of this rotation has
a dispersion relation similar to Eq. (\ref{dispersion}), then by
studying ``polarization diffusion'' of light from a polarized source
in a given bandwidth, one can effectively place a bound 
$|f^{(3)}| < 10^{-4}$ \cite{gleiser}.
Other testable effects of Lorentz invariance violation include threshold
effects in particle reactions, 
gravitational Cerenkov radiation, and neutrino oscillations.

Mattingly \cite{mattingly} gives 
a thorough and up-to-date review of both the theoretical
frameworks and the experimental results for tests of LLI.

%%%%%%%%%%%%%%%%%%%%%%%%%%%%%%%%%%%%%%%%%%%%%%%%%%%%%%%%%

\subsubsection{Tests of local position invariance}
\label{lpi}

The principle of local position invariance, the third part of
EEP, can be tested by the gravitational redshift
experiment,
 the first experimental test of gravitation proposed by Einstein. Despite
the fact that Einstein regarded this as a crucial test of GR, we now
realize that it does not distinguish between GR and any other metric
theory of gravity, but is only a test of EEP.  A typical gravitational
redshift experiment measures the frequency or wavelength shift $Z \equiv
\Delta \nu / \nu = - \Delta \lambda / \lambda$ between two identical
frequency standards (clocks) placed at rest at different heights in a
static gravitational field.  If the frequency of a given type of atomic
clock is the same when measured in a local, momentarily comoving freely
falling frame (Lorentz frame), independent of the location or velocity of
that frame, then the comparison of frequencies of two clocks at rest at
different locations boils down to a comparison of the velocities of two
local Lorentz frames, one at rest with respect to one clock at the moment
of emission of its signal, the other at rest with respect to the other
clock at the moment of reception of the signal.  The frequency shift is
then a consequence of the first-order Doppler shift between the frames.
The structure of the clock plays no role whatsoever.  The result is a
shift

\begin{equation} \label{E3}
Z = \Delta U/ c^2 \,,
\end{equation}
where $\Delta U$ is the difference in the Newtonian gravitational
potential between the receiver and the emitter.  If LPI is not
valid, then it turns out that the shift can be written
\begin{equation} \label{E4}
Z = (1+ \alpha ) \Delta U /c^2 \,,
\end{equation}
where the parameter $\alpha$ may depend upon the nature of the
clock whose shift is being measured (see TEGP 2.4~(c)~\cite{tegp} for
details).

The first successful, high-precision redshift measurement was the
series of Pound-Rebka-Snider experiments of 1960--1965 that
measured the frequency shift of gamma-ray photons from $^{57}$Fe
as they ascended or descended the Jefferson Physical Laboratory
tower at Harvard University.  The high accuracy
achieved -- one percent -- was obtained by making
use of the M\"ossbauer effect to produce a narrow resonance line
whose shift could be accurately determined.  Other experiments
since 1960 measured the shift of spectral lines in the Sun's
gravitational field and the change in rate of atomic clocks
transported aloft on aircraft, rockets and satellites.  Figure~\ref{lpifig}
summarizes the important redshift experiments that have been
performed since 1960 (TEGP 2.4~(c)~\cite{tegp}).

\epubtkImage{}{
\begin{figure}[hptb]
  \def\epsfsize#1#2{0.5#1}
  \centerline{\epsfbox{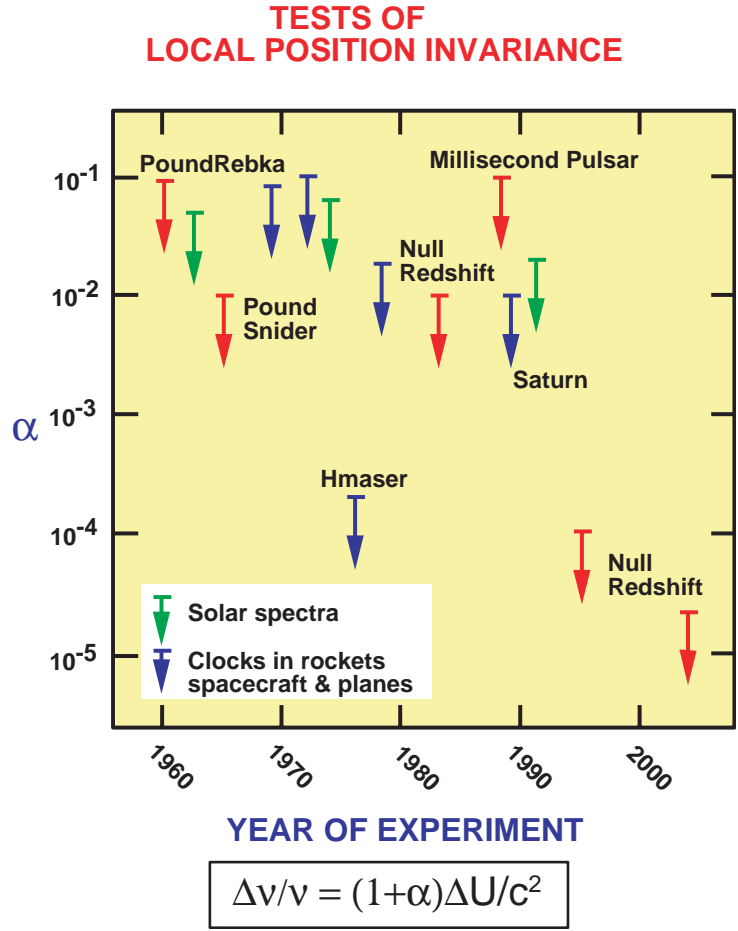}}
  \caption{\it Selected tests of local position invariance via
    gravitational redshift experiments, showing bounds on $\alpha$,
    which measures degree of deviation of redshift from the formula
    $\Delta \nu / \nu = \Delta U/c^2$.  In null redshift experiments, the
    bound is on the difference in $\alpha$ between different kinds of
    clocks.}
  \label{lpifig}
\end{figure}}

After almost 50 years of inconclusive or contradictory measurements, the
gravitational redshift of solar spectral lines was finally  measured
reliably.  During the early years of GR, the failure to
measure this effect in solar lines
was siezed upon by some as reason to doubt the theory.
Unfortunately, the measurement is not simple.
Solar spectral lines are subject to the ``limb effect'', a variation
of spectral line wavelengths between the center of the solar disk and
its edge or ``limb''; this effect is actually a Doppler shift caused by complex
convective and turbulent motions in the photosphere and lower
chromosphere. and is expected to be minimized by observing at the
solar limb, where the motions are predominantly transverse.  The
secret is to use strong, symmetrical lines, leading to unambiguous
wavelength measurements.  Successful measurements were finally made
in 1962 and 1972 (TEGP 2.4~(c)~\cite{tegp}).  In 1991, LoPresto et
al.~\cite{lopresto91} 
measured the solar shift in agreement with LPI to about 2 percent by 
observing the oxygen triplet lines both in
absorption in the limb and in emission just off the limb.

The most precise standard redshift
test to date was the Vessot-Levine rocket
experiment that took place in June 1976~\cite{vessot}.  A
hydrogen-maser clock was flown on a rocket to an altitude of
about 10,000~km and its frequency compared to a similar clock on
the ground.  The experiment took advantage of the masers'
frequency stability by monitoring the frequency shift as a
function of altitude.  A sophisticated data acquisition scheme
accurately eliminated all effects of the first-order Doppler
shift due to the rocket's motion, while tracking data were used
to determine the payload's location and the velocity (to evaluate
the potential difference $\Delta U$, and the special relativistic
time dilation).  Analysis of the data yielded a limit
$| \alpha | < 2 \times 10^{-4}$.

A ``null'' redshift experiment performed in 1978 tested whether the {\it
relative} rates of two different clocks depended upon position.  Two
hydrogen maser clocks and an ensemble of three superconducting-cavity
stabilized oscillator (SCSO) clocks were compared over a 10-day period.
During the period of the experiment, the solar potential $U/c^2$ changed
sinusoidally with a 24-hour period by $3 \times10^{-13}$ because of the
Earth's rotation, and changed linearly at $3 \times 10^{-12}$ per day
because the Earth is 90 degrees from perihelion in April.  However,
analysis of the data revealed no variations of either type within
experimental errors, leading to a limit on the LPI violation parameter $|
\alpha^{\rm H} - \alpha^{\rm SCSO} | < 2 \times 10^{-2}$~\cite{turneaure}.
This bound has been improved using more stable frequency
standards, such as atomic fountain clocks~\cite{Godone,prestage95,bauch02}.  The
current bound, from comparing a Cesium atomic fountain with a Hydrogen maser
for a year, is $|\alpha^{\rm H} - \alpha^{\rm Cs} | < 2.1 \times
10^{-5}$~\cite{bauch02}.  

The varying gravitational redshift of
Earth-bound clocks relative to the highly stable millisecond pulsar PSR
1937+21, caused by the Earth's motion in the solar gravitational field
around the Earth-Moon center of mass (amplitude 4000~km), was 
measured to about 10 percent~\cite{Taylor87}.
Two measurements of the  redshift using stable oscillator clocks
on spacecraft were made at the one percent level: one
used the Voyager spacecraft in Saturn's gravitational 
field~\cite{krisher90a}, while another  
used the Galileo spacecraft in the Sun's field~\cite{krisher93}.

The gravitational redshift could be improved to the $10^{-10}$
level
using an array of laser cooled atomic clocks on board
a spacecraft which would travel to
within four solar radii of the Sun~\cite{maleki01}.

Modern advances in navigation using Earth-orbiting atomic clocks and
accurate time-transfer must routinely take gravitational redshift and
time-dilation effects into account. For example, the Global Positioning
System (GPS) provides absolute positional
accuracies of around 15 m (even better in
its military mode), and 50 nanoseconds
in time transfer accuracy, anywhere on Earth.
Yet the difference in rate between
satellite and ground clocks as a result of 
relativistic effects is a whopping 39~{\it microseconds} per day
($46~\mu{\rm s}$ from the gravitational redshift, and $-7~\mu {\rm s}$
from time dilation). If these effects were not accurately accounted for,
GPS would fail to function at its stated accuracy. This represents a
welcome practical application of GR! (For the role of GR in GPS,
see~\cite{ashby1,ashby2}; for a popular essay, see~\cite{physicscentral}.)

Local position invariance also refers to position in time.  If LPI is
satisfied, the fundamental constants of non-gravitational physics
should be constants in time.  Table~\ref{varyconstants} shows current bounds on
cosmological variations in selected dimensionless constants.  For
discussion and references to early work, see TEGP 2.4~(c)~\cite{tegp} or
\cite{dyson72}.  For a comprehensive recent review 
both of experiments and of 
theoretical ideas that underly proposals for
varying constants, 
see~\cite{uzan03}.

Experimental bounds on varying constants come in two types: bounds on
the present rate of variation, and bounds on the difference between
today's value and a value in the distant past.  The main example of
the former type is the clock comparison test, in which highly stable
atomic clocks of different fundamental type are intercompared over
periods ranging from months to years (variants of the null redshift
experiment).  If the frequencies of the
clocks depend differently on the electromagnetic
fine structure constant $\alpha_{\rm EM}$, the
electron-proton mass ratio $m_e/m_p$, or the gyromagnetic ratio of the proton
$g_p$, for example,
then a limit on a drift of the fractional frequency
difference translates into a limit on a drift of the constant(s).  The
dependence of the frequencies on the constants may be quite complex,
depending on the atomic species involved.  The most recent experiments
have exploited the techniques of laser cooling and trapping, and of
atom fountains, in order to achieve extreme clock stability, and
compared the Rubidium-87 hyperfine transition~\cite{salomon03}, 
the Mercury-199 ion electric quadrupole transition~\cite{bize03}, 
the atomic Hydrogen $1S-2S$ transition~\cite{fischer04}, 
or an optical transition in Ytterbium-171~\cite{peik04},
against the ground-state hyperfine transition in Cesium-133.  These
experiments show that, today,
$\dot \alpha_{\rm EM}/\alpha_{\rm EM} < 3 \times 10^{-15} \, {\rm
yr}^{-1}$.  

The second type of bound involves measuring the relics of or signal from
a process that occurred in the distant
past and comparing the inferred value of the constant with the value
measured in the laboratory today.  
One sub-type uses astronomical
measurements of spectral lines at large redshift, while the other uses
fossils of nuclear processes on Earth to infer values of constants
early in geological history.  

\begin{table}[t]
\begin{center}
\begin{tabular}{p{3.0 cm}lll}
\hline\hline
		&Limit on $ \dot{k} / k$ &&\\ 
Constant $k$    & $({\rm yr}^{-1})$  &Redshift & Method \\
\hline\hline
Fine structure 
&$< 30 \times 10^{-16}$ & $0$ &
Clock comparisons~\cite{salomon03,bize03,fischer04,peik04}\\
constant &$< 0.5 \times 10^{-16}$ & $0.15$ &
  Oklo Natural Reactor~\cite{damourdyson,fujii04,petrov05} \\
$\alpha_{\rm EM}=e^2/\hbar c$&$< 3.4 \times 10^{-16}$ & $0.45$ &
    $^{187}{\rm Re}$ decay in meteorites~\cite{olive04}\\
&$(6.4 \pm 1.4) \times 10^{-16}$ & $0.2 - 3.7$ &
      Spectra in distant quasars~\cite{webb99,murphy01}\\
&$< 1.2 \times 10^{-16}$ & $0.4 - 2.3$ &
	Spectra in distant quasars~\cite{petitjean1,petitjean2} \\
\hline
Weak interaction  &$< 1 \times 10^{-11}$& $0.15$ &
Oklo Natural Reactor~\cite{damourdyson} \\
constant & $< 5 \times 10^{-12}$& $10^9$ &
Big Bang nucleosynthesis~\cite{malaney,reeves} \\
$\alpha_{\rm W}=G_{\rm f} m_{\rm p}^2 c/\hbar^3$ && \\
\hline
e-p mass ratio & $ < 3 \times 10^{-15}$ & $2.6 - 3.0$ &
Spectra in distant quasars~\cite{ivanchik05} \\
\hline\hline
\end{tabular}
\caption{\it Bounds on cosmological variation of fundamental constants
of non-gravitational physics.  For an in-depth review, see~\cite{uzan03}.}
\label{varyconstants}
\end{center}
\end{table}

Earlier comparisons of
spectral lines of different atoms or transitions in distant galaxies
and quasars produced bounds $\alpha_{\rm EM}$ or $g_p(m_e/m_p)$ on the order of
a part in 10 per Hubble time~\cite{wolfe76}.  Dramatic improvements in
the precision of astronomical and laboratory spectroscopy, in the
ability to model the complex astronomical environments where emission
and absorption lines are produced, and in the ability to reach large
redshift have made it possible to improve the bounds significantly.
In fact, in 1999, Webb {\em et al.}~\cite{webb99,murphy01} announced that
measurements of 
absorption lines in Mg, Al, Si, Cr, Fe, Ni and Zn in quasars in the
redshift range $0.5 < Z < 3.5$ indicated a smaller value of $\alpha_{\rm
EM}$
in earlier epochs, namely 
$\Delta \alpha_{\rm EM}/\alpha_{\rm EM} = (-0.72 \pm 0.18) \times 10^{-5}$, 
corresponding to $\dot \alpha_{\rm EM}/\alpha_{\rm EM} = (6.4 \pm 1.4)
\times 10^{-16}  \, {\rm yr}^{-1}$ (assuming a linear drift with time).
Measurements by other groups have so far 
failed to confirm 
this non-zero  effect~\cite{petitjean1,petitjean2,quast04};
a recent analysis of Mg absorption systems in quasars at $0.4 < Z < 2.3$
gave 
$\dot \alpha_{\rm EM}/\alpha_{\rm EM} = 
(-0.6 \pm 0.6) \times 10^{-16} \, {\rm yr}^{-1}$~\cite{petitjean1}.

Another important set of bounds arises from studies of the ``Oklo''
phenomenon, a group of natural, sustained $^{235}{\rm U}$  fission reactors that
occurred in the Oklo region of Gabon, Africa, around 1.8 billion years ago.
Measurements of ore samples yielded an abnormally low value for the ratio
of two isotopes of Samarium, $^{149}{\rm Sm}/^{147}{\rm Sm}$.  Neither of
these isotopes is a fission product, but $^{149}{\rm Sm}$ can be depleted by
a flux of neutrons.  Estimates of the neutron fluence (integrated dose)
during the reactors' ``on'' phase, combined with the measured abundance
anomaly, yield a value for the neutron cross-section for  $^{149}{\rm Sm}$
1.8 billion years ago that agrees with the modern value.  However, the
capture cross-section is extremely sensitive to the energy of a low-lying
level ($E \sim 0.1 \, {\rm eV}$), so that a variation in the energy of this
level of only 20 milli-eV over a billion years would change the capture
cross-section from its present value by more than the observed amount.  This
was first analysed in 1976 by Shlyakter~\cite{shlyakter}.  
Recent reanalyses of the
Oklo data~\cite{damourdyson,fujii04,petrov05} lead to a bound on $\dot
\alpha_{\rm EM}$ at the level of around $5 \times 10^{-17} \, {\rm yr}^{-1}$.

In a similar manner, recent 
re-analyses of decay rates of $^{187}{\rm Re}$ in ancient
meteorites (4.5 billion years old) gave the 
bound $\dot \alpha_{\rm EM}/\alpha_{\rm EM}
< 3.4 \times 10^{-16} \, {\rm yr}^{-1}$~\cite{olive04}.

%%%%%%%%%%%%%%%%%%%%%%%%%%%%%%%%%%%%%%%%%%%%%%%%%%%%%%%%%%%%%%%%%%%%%%%%%%%%%%%%%%%
%%%%%%%%%%%%%%%%%%%%%%%%%%%%%%%%%%%%%%%%%%%%%%%%%%%%%%%%%%%%%%%%%%%%%%%%%%%%%%%%%%%

\subsection{Theoretical Frameworks for Analyzing EEP}
\label{EEPframeworks}

%%%%%%%%%%%%%%%%%%%%%%%%%%%%%%%%%%%%%%%%%%%%%%%%%%%%%%%%%%%%%%%%%%%%%%%%%%%%%%%%%%%

\subsubsection{Schiff's conjecture}
\label{Schiff}

Because the three parts of the Einstein equivalence principle
discussed above are so very different in their empirical
consequences, it is tempting to regard them as independent
theoretical principles.  On the other hand, any complete and
self-consistent gravitation theory must possess sufficient
mathematical machinery to make predictions for the outcomes of
experiments that test each principle, and because there are
limits to the number of ways that gravitation can be meshed with
the special relativistic laws of physics, one might not be
surprised if there were theoretical connections between the three
sub-principles.  For instance, the same mathematical formalism
that produces equations describing the free fall of a hydrogen
atom must also produce equations that determine the energy levels
of hydrogen in a gravitational field, and thereby the ticking
rate of a hydrogen maser clock.  Hence a violation of EEP in the
fundamental machinery of a theory that manifests itself as a
violation of WEP might also be expected to show up as a violation
of local position invariance.  Around 1960, Schiff conjectured
that this kind of connection was a necessary feature of any
self-consistent theory of gravity.  More precisely,
Schiff's conjecture states that {\it any complete, self-consistent
theory of gravity that embodies WEP necessarily embodies EEP}.
In other words, the validity of WEP alone guarantees the validity
of local Lorentz and position invariance, and thereby of EEP.

If Schiff's conjecture is correct, then E\"otv\"os experiments may
be seen as the direct empirical foundation for EEP, hence for the
interpretation of gravity as a curved-spacetime phenomenon.  Of course,
a rigorous proof of such a conjecture is impossible (indeed, some
special counter-examples are known~\cite{Ohanian74,Ni77,Coley82}),
yet a number
of powerful ``plausibility'' arguments can be formulated.

The most general and elegant of these arguments is based upon the
assumption of energy conservation.  This assumption allows one to
perform very simple cyclic gedanken experiments in which the
energy at the end of the cycle must equal that at the beginning
of the cycle.  This approach was pioneered by Dicke, Nordtvedt and
Haugan (see, e.g.~\cite{haugan}).
A system in a quantum state $A$ decays to state $B$,
emitting a quantum of frequency $\nu$.  The quantum falls a height
$H$ in an external gravitational field and is shifted to frequency
$\nu'$, while the system in state $B$ falls with acceleration
$g_B$.  At the bottom, state $A$ is rebuilt out of state $B$, the
quantum of frequency $ \nu'$, and the kinetic energy $m_B g_B H$
that state $B$ has gained during its fall.  The energy left over
must be exactly enough, $m_A g_A H$, to raise state $A$ to its original
location.  (Here an assumption of local Lorentz invariance
permits the inertial masses $m_A$  and $m_B$ to be identified with
the total energies of the bodies.)  If $g_A$ and $g_B$ depend on
that portion of the internal energy of the states that was
involved in the quantum transition from $A$ to $B$ according to
\begin{equation} \label{E5}
   g_A = g(1 + \alpha E_A / m_A c^2 ), \qquad
   g_B = g(1 + \alpha E_B / m_B c^2 ), \qquad
   E_A - E_B \equiv h \nu
\end{equation}
(violation of WEP), then by conservation of energy, there must be
a corresponding violation of LPI in the frequency shift of the
form (to lowest order in $h \nu /mc^2$)
\begin{equation}\label{E6}
   Z = ( \nu' - \nu )/ \nu'
     = (1+\alpha ) gH/c^2 = (1+\alpha ) \Delta U/c^2.
\end{equation}
Haugan generalized this approach to include violations of
LLI~\cite{haugan}, (TEGP 2.5 \cite{tegp}).

\newpage

\begin{mybox}{The $ \mbox{\protect\boldmath $ TH\epsilon\mu $} $ formalism}
  \begin{enumerate}
  \item{\bf Coordinate system and conventions:}
    $x^0 =t=$ time coordinate associated with the static nature of the
    static spherically symmetric (SSS) gravitational field;
    ${\bf x} =(x,y,z) =$ isotropic quasi-Cartesian spatial coordinates; spatial
    vector and gradient operations as in Cartesian space.
  \item{\bf Matter and field variables:}
    \begin{itemize}
    \item
      $m_{0a} =$ rest mass of particle $a$.
    \item
      $e_a =$ charge of particle $a$.
    \item
      $x_a^\mu (t) =$ world line of particle $a$.
    \item
      $v_a^\mu = dx_a^\mu/dt =$ coordinate velocity of particle $a$.
    \item
      $A_\mu =$ electromagnetic vector potential;
      ${\bf E}= \nabla A_0-\partial {\bf A}/\partial t,\,{\bf B}=\nabla \times {\bf A}$
    \end{itemize}
  \item{\bf Gravitational potential:}  $U( {\bf x} )$
  \item{\bf Arbitrary functions:}
    $T(U)$, $H(U)$, $\epsilon (U)$, $\mu (U)$;
    EEP is satisfied if $\epsilon = \mu = (H/T)^{1/2}$ for all $U$.
  \item{\bf Action:}
    {\setlength{\arraycolsep}{0.14 em}
      \begin{eqnarray*}
        I &=& - \sum_a m_{0a} \int
        (T-Hv_a^2)^{1/2} dt
        + \sum_a e_a \int
        A_\mu (x_a^\nu ) v_a^\mu dt  \\
        && \quad + (8 \pi)^{-1} \int
        (\epsilon E^2-\mu^{-1} B^2) d^4 x.
      \end{eqnarray*}}%
  \item{\bf Non-Metric parameters:}
    {\setlength{\arraycolsep}{0.14 em}
      \begin{eqnarray*}
        \Gamma_0 &=& -c_0^2(\partial /\partial U) \ln [\epsilon (T/H)^{1/2}]_0,\\
        \Lambda_0 &=& -c_0^2(\partial /\partial U) \ln [\mu (T/H)^{1/2}]_0,\\
        \Upsilon_0 &=& 1- (TH^{-1} \epsilon\mu)_0,
      \end{eqnarray*}}%
    where $c_0 = (T_0 /H_0 )^{1/2}$ and subscript
    ``0'' refers to a chosen point in space.
    If EEP is satisfied, $\Gamma_0 \equiv \Lambda_0 \equiv \Upsilon_0 \equiv 0$.
  \end{enumerate}
  \label{box1}
\end{mybox}

\newpage

%%%%%%%%%%%%%%%%%%%%%%%%%%%%%%%%%%%%%%%%%%%%%%%%%%%%%%%%%%%%%%%%%%%%%%%%%%%%%%%%%%%

%\subsubsection{The $ \mbox{\protect\boldmath $TH\epsilon\mu$} $ formalism}
\subsubsection{The $TH\epsilon\mu$ formalism}
\label{theuformalism}

The first successful attempt to prove Schiff's conjecture more
formally was made by Lightman and Lee~\cite{lightmanlee}.  They developed a
framework called the $TH\epsilon\mu$ formalism that encompasses
all metric theories of gravity and many non-metric theories
(Box~\ref{box1}).  It restricts attention to the behavior of charged
particles (electromagnetic interactions only) in an external
static spherically symmetric (SSS) gravitational field, described
by a potential $U$.  It characterizes the motion of the charged
particles in the external potential by two arbitrary functions
$T(U)$ and $H(U)$, and characterizes the response of
electromagnetic fields to the external potential (gravitationally
modified Maxwell equations) by two functions $\epsilon (U)$ and
$\mu (U)$.  The forms of $T$, $H$, $\epsilon$ and $\mu$ vary from theory
to theory, but every metric theory satisfies
\begin{equation}\label{E7}
    \epsilon = \mu = (H/T)^{1/2},
\end{equation}
for all $U$.  This consequence follows from the action of
electrodynamics with a ``minimal'' or metric coupling:
{\setlength{\arraycolsep}{0.14 em}
\begin{eqnarray}
  I & = & - \sum_a m_{0a} \int
        (-g_{\mu\nu} v_a^\mu v_a^\nu )^{1/2} dt
    + \sum_a e_a \int
        A_\mu (x_a^\nu ) v_a^\mu dt \nonumber\\
  & & - {1 \over {16 \pi}} \int \sqrt{-g}
          g^{\mu \alpha} g^{\nu \beta}
          F_{\mu \nu} F_{\alpha \beta}
          d^4 x, \label{E8}
\end{eqnarray}}%
where the variables are defined in Box~\ref{box1}, and where
$F_{\mu\nu} \equiv A_{\nu, \mu} - A_{\mu, \nu}$.  By identifying
$g_{00} = T$ and $g_{ij} = H \delta_{ij}$ in a SSS field,
$F_{i0} = E_i$ and $F_{ij} = \epsilon_{ijk} B_k$,
one obtains Eq.~(\ref{E7}).
Conversely, every theory within this class that satisfies Eq.~(\ref{E7})
can have its electrodynamic equations cast into ``metric'' form.
In a given non-metric theory, the functions $T$, $H$, $\epsilon$ and
$\mu$ will depend in general on the full gravitational environment,
including the potential of the Earth, Sun and Galaxy, as well as on
cosmological boundary conditions.  Which of these factors has the most
influence on a given experiment will depend on the nature of the
experiment.

Lightman and Lee then calculated explicitly the rate of
fall of a ``test'' body made up of interacting charged particles,
and found that the rate was independent of the internal
electromagnetic structure of the body (WEP) if and only if Eq.~(\ref{E7})
was satisfied.  In other words, WEP $\Rightarrow$ EEP and Schiff's
conjecture was verified, at least within the restrictions built
into the formalism.

Certain combinations of the functions $T$, $H$, $\epsilon$ and $\mu$
reflect different aspects of EEP.  For instance, position or
$U$-dependence of either of the combinations $\epsilon (T/H)^{1/2}$
and $\mu (T/H)^{1/2}$ signals violations of LPI, the first
combination playing the role of the locally measured electric
charge or fine structure constant.  The ``non-metric parameters''
$\Gamma_0$ and $\Lambda_0$ (Box~\ref{box1}) are measures of such
violations of EEP.  Similarly, if the parameter
$\Upsilon_0 \equiv 1-(TH^{-1} \epsilon \mu )_0$ is non-zero anywhere,
then violations of LLI will occur.  This parameter is related to the
difference between the speed of light $c$, and the
limiting speed of material test particles $c_0$, given by
\begin{equation}\label{E9}
    c = ( \epsilon_0 \mu_0 )^{- 1/2},\qquad
    c_0 = ( T_0 / H_0 )^{1/2}.
\end{equation}
In many applications,
by suitable definition of units, $c_0$ can be set equal to unity.
If EEP is valid, $\Gamma_0 \equiv \Lambda_0 \equiv \Upsilon_0 = 0$
everywhere.

The rate of fall of a composite spherical test body of
electromagnetically interacting particles then has the form
\begin{equation}
    {\bf a} = (m_{\rm P} /m) \nabla U,\label{E10}
\end{equation}
{\setlength{\arraycolsep}{0.14 em}
\begin{eqnarray}
          m_{\rm P} /m
 &=&1 + (E_{\rm B}^{\rm ES} /Mc_0^2 )
          \left[ 2 \Gamma_0 - {8 \over 3} \Upsilon_0 \right]
      + (E_{\rm B}^{\rm MS} /Mc_0^2 )
      \left[ 2 \Lambda_0 - {4 \over 3} \Upsilon_0 \right]
      \nonumber \\
      && + \dots,\label{E11}
\end{eqnarray}}%
where $E_{\rm B}^{\rm ES}$ and $E_{\rm B}^{\rm MS}$ are the electrostatic
and magnetostatic binding energies of the body, given by
\begin{equation}
          E_{\rm B}^{\rm ES}
 =-{1 \over 4} T_0^{1/2} H_0^{-1} \epsilon_0^{-1}
\left< \sum_{ab} \frac{e_a e_b}{r_{ab}} \right>, \label{E12}
\end{equation}
\begin{equation}
       E_{\rm B}^{\rm MS} =
  - {1 \over 8} T_0^{1/2} H_0^{-1} \mu_0
  \left< \sum_{ab} \frac{e_a e_b}{r_{ab}}
       [ {\bf v}_a \cdot {\bf v}_b
       + ( {\bf v}_a \cdot {\bf n}_{ab} )
           ( {\bf v}_b \cdot {\bf n}_{ab} )]
\right>, \label{E13}
\end{equation}
where $r_{ab} = | {\bf x}_a  -  {\bf x}_b |$,
${\bf n}_{ab}  =  ( {\bf x}_a  -  {\bf x}_b )/r_{ab}$, and the
angle brackets denote an expectation value of the enclosed
operator for the system's internal state.  E\"otv\"os experiments
place limits on the WEP-violating terms in Eq.~(\ref{E11}), and
ultimately place limits on the non-metric parameters
$| \Gamma_0 | < 2 \times 10^{-10}$ and
$| \Lambda_0 |  <  3  \times  10^{-6}$.
(We set
$\Upsilon_0 = 0$ because of very tight constraints on it from
tests of LLI; see Figure \ref{llifig}, where $\delta = -\Upsilon$.)  
These limits are
sufficiently tight to rule out a number of non-metric theories of
gravity thought previously to be viable (TEGP 2.6~(f)~\cite{tegp}).

The $TH \epsilon\mu$ formalism also yields a
gravitationally modified Dirac equation that can be used to
determine the gravitational redshift experienced by a variety of
atomic clocks.  For the redshift parameter $\alpha$
(Eq.~(\ref{E4})), the results are (TEGP 2.6~(c)~\cite{tegp}):
\begin{equation}\label{E14}
\alpha = \!\!\left\{\!\!\! \begin{array}{ll}
  -3\Gamma_0 + \Lambda_0  & \mbox{hydrogen hyperfine transition,
    H-Maser clock} \\
-{1 \over 2} (3\Gamma_0 + \Lambda_0)  & \mbox{electromagnetic mode in
  cavity, SCSO clock,} \\
-2\Gamma_0  & \mbox{phonon mode in solid, principal transition in hydrogen.}
\end{array} \right.
\end{equation}

The redshift is the standard one $( \alpha = 0)$, independently
of the nature of the clock if and only if
$\Gamma_0 \equiv \Lambda_0 \equiv 0$.  Thus the Vessot-Levine
rocket redshift experiment sets a limit on the parameter
combination $| 3 \Gamma_0 - \Lambda_0 |$ (Figure~\ref{lpifig}); the
null-redshift experiment comparing hydrogen-maser and SCSO clocks
sets a limit on
$| \alpha_{\rm H} - \alpha_{\rm SCSO} | = {3 \over 2} | \Gamma_0 - \Lambda_0 |$.
Alvarez and Mann~\cite{AlvarezMann96a,AlvarezMann96b,AlvarezMann97a,
AlvarezMann97b,AlvarezMann97c} extended the $TH\epsilon\mu$ formalism to
permit analysis of such effects as the Lamb shift, anomalous magnetic moments
and non-baryonic effects, and placed interesting bounds on EEP violations.

%%%%%%%%%%%%%%%%%%%%%%%%%%%%%%%%%%%%%%%%%%%%%%%%%%%%%%%%%%%%%%%%%%%%%%%%%%%%%%%%%%%

%\subsubsection{The $ \mbox{\protect\boldmath ${c^2}$} $ formalism}
\subsubsection{The ${c^2}$ formalism}
\label{c2formalism}

The $TH \epsilon \mu$ formalism can also be applied to tests of
local Lorentz invariance, but in this context it can be
simplified.  Since most such tests do not concern themselves with
the spatial variation of the functions $T$, $H$, $\epsilon$, and $\mu$,
but rather with observations made in moving frames, we can treat
them as spatial constants.  Then by rescaling the time and space
coordinates, the charges and the electromagnetic fields, we can
put the action in Box~\ref{box1} into the form (TEGP 2.6~(a)~\cite{tegp})
\begin{equation}\label{E15}
  I = - \sum_a m_{0a} \int
        (1-v_a^2)^{1/2} dt
    + \sum_a e_a \int
        A_\mu (x_a^\nu ) v_a^\mu dt
  + (8 \pi)^{-1} \int
        (E^2-c^2B^2) d^4 x,
\end{equation}
where $c^2 \equiv H_0 /T_0 \epsilon_0 \mu_0=(1-\Upsilon_0)^{-1}$.
This amounts to using units in which the limiting speed $c_0$
of massive test particles is unity, and the speed of light is $c$.
If $c \ne 1$, LLI is violated; furthermore, the form of the
action above must be assumed to be valid only in some preferred
universal rest frame.  The natural candidate for such a frame is
the rest frame of the microwave background.

The electrodynamical equations which follow from Eq.~(\ref{E15})
yield the behavior of rods and clocks, just as in the full $TH
\epsilon \mu$
formalism.  For example, the length of a rod 
which moves with velocity $ \bf V$ relative to the rest frame
in a direction parallel to its length
will be observed by a rest observer to be contracted relative to
an identical rod perpendicular to the motion by a factor
$1 - V^2 /2 +  O(V^4 )$.  Notice that $c$ does not appear in this
expression, because only electrostatic interactions are involved, and $c$
appears only in the magnetic sector of the action (\ref{E15}).  
The energy and momentum of an electromagnetically
bound body 
moving with velocity $ \bf V$ relative to the rest frame
are given by
{\setlength{\arraycolsep}{0.14 em}
\begin{equation}\label{E16}
  \begin{array}{rcl}
  E &=& M_{\rm R} + {1 \over 2}  M_{\rm R} V^2
  + {1 \over 2} \delta M_{\rm I}^{ij} V^i V^j
  + O(MV^4) \,, \vspace{0.5 em} \\
  P^i &=& M_{\rm R} V^i +  \delta M_{\rm I}^{ij} V^j + O(MV^3) \,,
\end{array}
\end{equation}}
where $M_{\rm R} = M_0 - E_{\rm B}^{\rm ES}$, $M_0$ is the
sum of the particle rest masses, $E_{\rm B}^{\rm ES}$ is the
electrostatic binding energy of the system (Eq.~(\ref{E12}) with
$T_0^{1/2}H_0 \epsilon_0^{-1}=1$), and
\begin{equation}\label{E17}
  \delta M_{\rm I}^{ij}  =  -  2
\left( {1 \over {c^2}} -  1 \right)
\left[ {4 \over 3} E_{\rm B}^{\rm ES} \delta^{ij}
    +   \tilde E_{\rm B}^{{\rm ES}\,ij} \right],
\end{equation}
where
\begin{equation} \label{E18}
       \tilde E_{\rm B}^{{\rm ES}\,ij}
 = -{1 \over 4} \left< \sum_{ab} \frac{e_a e_b}{r_{ab}} \left(
       n_{ab}^i n_{ab}^j
   - {1 \over 3} \delta^{ij} \right) \right>.
\end{equation}
Note that $(c^{-2} - 1)$ corresponds to the parameter $\delta$
plotted in Figure~\ref{llifig}.

The electrodynamics given by Eq.~(\ref{E15}) can also be
quantized, so that we may treat the interaction of photons with
atoms via perturbation theory.  The energy of a photon is $\hbar$
times its frequency $\omega$, while its momentum is $\hbar \omega /c$.
Using this approach, one finds that the difference in round trip
travel times of light along the two arms of the interferometer in the
Michelson-Morley experiment is given by
$L_0 (v^2 /c)(c^{-2}  - 1)$.
The experimental null result then leads to the bound on
$(c^{-2} - 1)$ shown on Figure~\ref{llifig}.  Similarly the anisotropy in
energy levels is clearly illustrated by the tensorial terms
in Eqs.~(\ref{E16}) and (\ref{E18}); by
evaluating $\tilde E_{\rm B}^{{\rm ES}\,ij}$ for each nucleus in the
various Hughes-Drever-type experiments and comparing with the
experimental limits on energy differences, one obtains the extremely
tight bounds also shown on Figure~\ref{llifig}.

The behavior of moving
atomic clocks can also be analysed in detail, and bounds
on $(c^{-2} - 1)$ can be placed using results from tests of
time dilation and of the propagation of light.  In some cases, it is
advantageous to combine the $c^2$ framework with a ``kinematical''
viewpoint that treats a general class of boost transformations between
moving frames.  Such kinematical approaches have been discussed by
Robertson, Mansouri and Sexl, and Will (see~\cite{Will92b}).

For example,
in the ``JPL'' experiment, in which
the phases of two hydrogen masers connected by a fiberoptic link were
compared as a function of the Earth's orientation,
the predicted phase difference as a function of
direction is, to first order in $\bf V$, the velocity of the Earth
through the cosmic background,
\begin{equation}\label{E19}
    \Delta \phi / \tilde \phi \approx - {4 \over 3} (1-c^2)
    ( {\bf V} \cdot {\bf n} ~-~ {\bf V} \cdot {\bf n}_0 ),
\end{equation}
where $\tilde \phi  = 2 \pi \nu L$, $\nu$ is the maser frequency,
$L=21$ km is the baseline, and where ${\bf n}$ and ${\bf n}_0$ are
unit vectors along the direction of propagation of the light at
a given time and at the initial time of the experiment, respectively.  The
observed limit on a diurnal variation in the relative phase
resulted in the bound
$| c^{-2}-1 | < 3 \times 10^{-4}$.
Tighter bounds were obtained from a ``two-photon absorption'' (TPA)
experiment, and a 1960s series of
``M\"ossbauer-rotor'' experiments, which tested the isotropy of
time dilation between a gamma ray emitter on the rim of a rotating
disk and an absorber placed at the center~\cite{Will92b}.

\subsubsection{The Standard Model Extension (SME)}
\label{SME}

Kosteleck\'y and collaborators developed a useful and elegant framework for
discussing violations of Lorentz symmetry in the context of the standard
model of particle physics \cite{colladay97,colladay98,kosteleckymewes02}.
Called the Standard Model Extension (SME), it takes the standard
SU(3) $\times$ SU(2) $\times$ U(1) field theory of particle physics, and
modifies the terms in the action by inserting a variety of tensorial
quantities in the quark, lepton, Higgs, and gauge boson sectors 
that could explicitly violate LLI.  SME extends the earlier classical
$TH\epsilon\mu$ and  $c^2$ frameworks, and the $\chi-g$ framework of
Ni~\cite{Ni77}
to quantum field theory and
particle physics.
The modified terms split naturally 
into those that are
odd under CPT (i.e. that violate CPT) and terms that are even under CPT.
The result is a rich and complex framework, with many parameters to be
analysed and tested by experiment.  
Such details are  beyond the scope of this review; for a review of SME and
other frameworks, the reader is referred to the Living Review by Mattingly
\cite{mattingly}.

Here we confine our attention to the electromagnetic sector, in order to
link the SME with the $c^2$ framework discussed above.  In the SME, the
Lagrangian for a scalar particle $\phi$ with charge $e$ interacting 
with electrodynamics takes the form
\begin{eqnarray}
{\cal L} &=& [\eta^{\mu\nu} + (k_\phi)^{\mu\nu}] (D_\mu \phi)^\dag D_\nu
\phi - m^2 \phi^\dag \phi 
\nonumber \\
&& - \frac{1}{4} [ \eta^{\mu \alpha} \eta^{\nu \beta} + 
(k_F)^{\mu \nu \alpha\beta} ]
F_{\mu \nu} F_{\alpha \beta} \,,
\label{ESMaction}
\end{eqnarray}
where $D_\mu \phi = \partial_\mu \phi + ieA_\mu \phi$, and where 
$(k_\phi)^{\mu\nu}$ is a real symmetric trace-free tensor, and 
$(k_F)^{\mu \nu\alpha \beta}$ is a tensor with the symmetries of the
Riemann tensor, and with vanishing double trace.  It has 19 independent
components.  There could also be a CPT-odd term in $\cal L$ of the form 
$(k_A)^\mu \epsilon_{\mu\nu\alpha\beta} A^\nu F^{\alpha \beta}$, but because
of a
variety of pre-existing theoretical and experimental constraints,
it is generally set to zero.

The  tensor $(k_F)^{\mu \alpha\nu \beta}$ can be decomposed into
``electric'', ``magnetic'' and ``odd-parity'' components, by defining
\begin{eqnarray}
(\kappa_{DE})^{jk} &=& -2 (k_F)^{0j0k} \,,
\nonumber \\
(\kappa_{HB})^{jk} &=& \frac{1}{2} \epsilon^{jpq}\epsilon^{krs}(k_F)^{pqrs} \,,
\nonumber \\
(\kappa_{DB})^{kj} &=& -(k_{HE})^{jk} = \epsilon^{jpq} (k_F)^{0kpq} \,.
\end{eqnarray}
In many applications it is useful to use the further decomposition
\begin{eqnarray}
\tilde{\kappa}_{tr} &=& \frac{1}{3} (\kappa_{DE})^{jj} \,,
\nonumber \\
(\tilde{\kappa}_{e+})^{jk} &=& \frac{1}{2} (\kappa_{DE}+\kappa_{HB})^{jk}
\,,
\nonumber \\
(\tilde{\kappa}_{e-})^{jk} &=& \frac{1}{2} (\kappa_{DE}-\kappa_{HB})^{jk}
-\frac{1}{3} \delta^{jk} (\kappa_{DE})^{ii} \,,
\nonumber \\
(\tilde{\kappa}_{o+})^{jk} &=& \frac{1}{2} (\kappa_{DB}+\kappa_{HE})^{jk}
\,,
\nonumber \\
(\tilde{\kappa}_{o-})^{jk} &=& \frac{1}{2} (\kappa_{DB}-\kappa_{HE})^{jk}
\,.
\label{kappatensors}
\end{eqnarray}
The first expression is a single number, the next three are symmetric
trace-free matrices, and the final is an antisymmetric matrix, accounting
thereby for the 19 components of the original tensor $(k_F)^{\mu \alpha\nu
\beta}$.  

In the rest frame of the universe, these tensors have some form that is
established by the global nature of the solutions of the overarching theory
being used.  In a frame that is moving relative to the universe, the tensors
will have components that depend on the velocity of the frame, and on the
orientation of the frame relative to that velocity.

In the case where the theory is rotationally symmetric in the preferred
frame, the tensors $(k_\phi)^{\mu\nu}$ and 
$(k_F)^{\mu \nu \alpha\beta}$ can be expressed in the form
\begin{eqnarray}
(k_\phi)^{\mu\nu} &=& \tilde{\kappa}_\phi (u^\mu u^\nu + \frac{1}{4}
\eta^{\mu\nu}  )\,,
\nonumber \\
(k_F)^{\mu \nu\alpha \beta} &=& \tilde{\kappa}_{tr} (4u^{[\mu}\eta^{\nu
][\alpha}u^{\beta]} - \eta^{\mu [\alpha}\eta^{\beta ]\nu} ) \,,
\end{eqnarray}
where $[\,]$ around indices denote antisymmetrization, and where $u^\mu$ is
the four-velocity of an observer at rest in the preferred frame.  
With this assumption, all the tensorial quantities in Eq.
(\ref{kappatensors}) vanish in the preferred frame, 
and, after suitable rescalings of coordinates and fields, 
the action (\ref{ESMaction}) can be put into
the form of the $c^2$ framework, with 
\begin{equation}
c = \left (\frac{1-\frac{3}{4} \tilde{\kappa}_\phi}{1+\frac{1}{4}
\tilde{\kappa}_\phi} \right )^{1/2} \left ( \frac{1-\tilde{\kappa}_{tr}}{1+\tilde{\kappa}_{tr}}
\right )^{1/2} \,.
\end{equation}

%%%%%%%%%%%%%%%%%%%%%%%%%%%%%%%%%%%%%%%%%%%%%%%%%%%%%%%%%%%%%%%%%%%%%%%%%%%%%%%%%%%
%%%%%%%%%%%%%%%%%%%%%%%%%%%%%%%%%%%%%%%%%%%%%%%%%%%%%%%%%%%%%%%%%%%%%%%%%%%%%%%%%%%

\subsection{EEP, particle physics, and the search for new interactions}
\label{newinteractions}

Thus far, we have discussed EEP as a principle that strictly divides the
world into metric and non-metric theories, and have implied
that a failure of EEP might invalidate metric theories (and thus general
relativity).  
On the other hand, there is mounting 
theoretical evidence to suggest that EEP is {\it
likely} to be violated at some level, whether by quantum gravity
effects, by effects arising from string theory, or by hitherto
undetected interactions.  
Roughly speaking, in addition to the pure Einsteinian
gravitational interaction, which respects EEP, theories such as string
theory predict
other interactions which do not.  In string theory, for example, the
existence of such EEP-violating fields is assured, but the theory is not
yet mature enough to enable a robust calculation of their strength relative to
gravity, or a determination of whether they are long range, like gravity, or
short range, like the nuclear and weak interactions,
and thus too short range to be detectable.

In one simple example~\cite{dick98}, one can write the Lagrangian for the low-energy
limit of a string-inspired theory in the so-called ``Einstein frame'', in which
the gravitational Lagrangian is purely general relativistic:
{\setlength{\arraycolsep}{0.14 em}
\begin{eqnarray}
\tilde {\cal L} &=& \sqrt{- \tilde g} \biggl(
{\tilde g}^{\mu\nu} \biggl[ {1 \over {2\kappa}} {\tilde
R}_{\mu\nu} - {1 \over 2} \tilde G (\varphi) \partial_\mu \varphi
\partial_\nu \varphi  \biggr]
-  U(\varphi) {\tilde g}^{\mu\nu}{\tilde g}^{\alpha\beta}
F_{\mu\alpha} F_{\nu\beta} \nonumber \\
&&+ \overline{\tilde{\psi}} \biggl [ i {\tilde e}^\mu_a \gamma^a
(\partial_\mu + {\tilde \Omega}_\mu +qA_\mu ) - \tilde M (\varphi) \biggr
]
\tilde \psi \biggr),
\label{stringlagrangian1}
\end{eqnarray}}
where ${\tilde g}_{\mu\nu}$ is the non-physical metric,
${\tilde R}_{\mu\nu}$ is the Ricci tensor derived from
it,
$\varphi$ is a
dilaton field, and $\tilde G$, $U$ and $\tilde M$ are functions of
$\varphi$.  The Lagrangian includes that for the electromagnetic field
$F_{\mu\nu}$, and that for
particles, written in terms of Dirac spinors $\tilde \psi$.  This is
not a metric representation because of the coupling of $\varphi$ to
matter via $\tilde M (\varphi)$ and $U(\varphi)$.
A conformal transformation ${\tilde g}_{\mu\nu} =
F(\varphi) g_{\mu\nu}$, $\tilde \psi = F(\varphi)^{-3/4} \psi$,
puts the Lagrangian in the form (``Jordan'' frame)
{\setlength{\arraycolsep}{0.14 em}
\begin{eqnarray}
{\cal L} &=& \sqrt{- g} \biggl( {g}^{\mu\nu}
\biggl[ {1 \over {
2\kappa}} F(\varphi) {R}_{\mu\nu}
- {1 \over 2} F(\varphi)
\tilde G (\varphi) \partial_\mu \varphi
\partial_\nu \varphi
+{3 \over {4\kappa F(\varphi)}} \partial_\mu F
\partial_\nu F \biggr] \nonumber \\
&&- U(\varphi) {g}^{\mu\nu}{g}^{\alpha\beta}
F_{\mu\alpha} F_{\nu\beta}
\nonumber \\
&&
+ \overline{\psi} \biggl [ i {e}^\mu_a
\gamma^a
(\partial_\mu + {\Omega}_\mu +qA_\mu ) - \tilde M (\varphi)
F^{1/2} \biggr ]
\psi \biggr).
\label{stringlagrangian2}
\end{eqnarray}}
One may choose $F(\varphi)= {\rm const.}/\tilde M (\varphi)^2$
so that the particle Lagrangian takes the
metric form (no explicit
coupling to $\varphi$), but the electromagnetic Lagrangian
will still couple non-metrically to $U(\varphi)$.  The gravitational
Lagrangian here takes the form of a scalar-tensor theory (Sec.~\ref{scalartensor}).  But the non-metric electromagnetic term will, in
general, produce violations of EEP.  For examples of specific models,
see~\cite{TaylorVeneziano,DamourPolyakov}.  Another class of non-metric
theories are included in the ``varying speed of light (VSL)'' theories; for
a detailed review, see~\cite{magueijo03}

On the other hand, whether one views such effects as a violation of EEP or
as effects arising from additional ``matter'' fields whose interactions,
like those of the electromagnetic field, do not fully embody EEP, is to some
degree a matter of semantics.  Unlike the fields of the standard model of
electromagnetic, weak and strong interactions, which couple to properties
other than mass-energy and are either short range or are strongly screened,
the fields inspired by string theory {\em could} be long range (if they
remain massless by virtue of a symmetry, or at best, acquire a very small
mass), and {\em can} couple to mass-energy, and thus can mimic gravitational
fields.  Still, there appears to be no way to make this precise.

As a result, EEP and related tests are now viewed as ways to discover or place
constraints on new physical interactions, or as a branch of
``non-accelerator particle physics'', searching for the possible imprints
of high-energy particle effects in the low-energy realm of gravity.
Whether current or proposed experiments
can actually probe these phenomena meaningfully is an open
question at the moment, largely because of a dearth of firm
theoretical predictions.  

\subsubsection{The ``fifth'' force}
\label{fifthforce}

On the phenomenological side, the idea of
using EEP tests in this way may have originated in the middle 1980s,
with the search for a ``fifth'' force.
In 1986, as a result
of a detailed reanalysis of E\"otv\"os' original data,
Fischbach {\em et al.}~\cite{fischbach5} suggested the existence of a fifth
force of nature, with a strength of about a percent that of
gravity, but with a range (as defined by the range $\lambda$ of a Yukawa
potential, $e^{-r/\lambda} /r$) of a few hundred meters.
This proposal dovetailed with earlier hints of a deviation from the
inverse-square law of Newtonian gravitation derived from measurements of
the gravity profile down deep mines in Australia,
and with emerging
ideas from particle physics suggesting the possible
presence of very low-mass particles with gravitational-strength couplings.
During the next four years
numerous experiments looked for evidence of the fifth force by searching
for composition-dependent differences in acceleration, with variants of
the E\"otv\"os experiment or with free-fall Galileo-type experiments.
Although two early experiments reported positive evidence, the others
all yielded null results.  Over the range between one and $10^4$ meters,
the null experiments produced upper limits on the strength of a postulated
fifth force between $10^{-3}$ and $10^{-6}$ of the strength of gravity.
Interpreted as tests of WEP (corresponding to the limit of
infinite-range forces), the results of two representative experiments from
this period, the free-fall Galileo experiment
and the early E\"ot-Wash experiment, are shown in
Figure~\ref{wepfig}.  At the same time, tests of the inverse-square
law of gravity were carried out by comparing variations in gravity
measurements up tall towers or down mines or boreholes with gravity
variations predicted using the inverse square law together with Earth
models and surface gravity data mathematically ``continued'' up
the tower or down the hole.  Despite early reports of anomalies,
independent tower, borehole and seawater measurements 
ultimately showed no evidence of a
deviation.  Analyses of orbital data from planetary range
measurements, lunar laser ranging, and laser tracking of the LAGEOS
satellite verified the inverse-square law to parts in $10^8$ over
scales of $10^3$ to $10^5$ km, and to parts in $10^9$ over planetary
scales of several astronomical units~\cite{talmadge}.
A consensus emerged that there was no
credible experimental evidence for a fifth force of nature, of a type
and range
proposed by Fischbach {\em et al.}
For reviews and bibliographies of this episode,
see~\cite{fischbach92,FischbachTalmadge,
FischbachTalmadge2,Adelberger91,WillSky}.

\subsubsection{Short-range modifications of Newtonian gravity}
\label{shortrange}

Although the idea of an intermediate-range violation of Newton's
gravitational law was dropped,  new ideas emerged to suggest the possibility
that the inverse-square law could be violated at very short ranges, below
the centimeter range of existing laboratory verifications of the $1/r^2$
behavior.  One set of ideas~\cite{add98,antoniadis98,randall1,randall2} posited 
that some of the extra spatial dimensions that come with string theory
could extend over macroscopic scales, rather than being rolled
up at the Planck scale of $10^{-33} \, {\rm cm}$, which was then
the conventional
viewpoint.  On laboratory distances large
compared to the relevant scale 
of the extra dimension, gravity would fall off as the
inverse square, whereas on short scales, gravity would fall off as
$1/R^{2+n}$,  where $n$ is the number of large extra dimensions.  Many
models favored $n=1$ or $n=2$.
Other possibilities for effective modifications of gravity at short range
involved the exchange of light scalar particles.

Following these proposals,
many of the high-precision, low-noise methods that were
developed for tests of WEP were adapted to carry out
laboratory tests of the inverse square law of
Newtonian gravitation at millimeter scales and below.  
The challenge of these experiments has been to
distinguish gravitation-like interactions from electromagnetic and
quantum
mechanical (Casimir) effects.  No deviations from
the inverse square law have been found to date at distances between $10
\,\mu
{\rm m}$ and $10 \, {\rm mm}$\cite{long99,hoyle01,hoyle04,kapitulnik,long03}.
For a comprehensive review of both the theory and the experiments,
see~\cite{adelberger03}.

%\newpage

%%%%%%%%%%%%%%%%%%%%%%%%%%%%%%%%%%%%%%%%%%%%%%%%%%%%%%%%%%%%%%%%%%%%%%%%%%%%%%%%%%%
%%%%%%%%%%%%%%%%%%%%%%%%%%%%%%%%%%%%%%%%%%%%%%%%%%%%%%%%%%%%%%%%%%%%%%%%%%%%%%%%%%%
%%%%%%%%%%%%%%%%%%%%%%%%%%%%%%%%%%%%%%%%%%%%%%%%%%%%%%%%%%%%%%%%%%%%%%%%%%%%%%%%%%%

\section{Tests of Post-Newtonian Gravity}
\label{S3}

%%%%%%%%%%%%%%%%%%%%%%%%%%%%%%%%%%%%%%%%%%%%%%%%%%%%%%%%%%%%%%%%%%%%%%%%%%%%%%%%%%%
%%%%%%%%%%%%%%%%%%%%%%%%%%%%%%%%%%%%%%%%%%%%%%%%%%%%%%%%%%%%%%%%%%%%%%%%%%%%%%%%%%%

\subsection{Metric theories of gravity and the strong equivalence principle}
\label{metrictheories}

%%%%%%%%%%%%%%%%%%%%%%%%%%%%%%%%%%%%%%%%%%%%%%%%%%%%%%%%%%%%%%%%%%%%%%%%%%%%%%%%%%%

\subsubsection{Universal coupling and the metric postulates}
\label{universal}

The empirical evidence supporting the Einstein
equivalence principle, discussed in the previous section,
supports the conclusion that the only theories of
gravity that have a hope of being viable are metric
theories, or possibly theories that are metric apart from very weak
or short-range non-metric couplings (as in string theory).  Therefore for
the remainder of this review, we shall turn our attention
exclusively to metric theories of gravity, which assume that
(i)~there exists a
symmetric metric, (ii)~test bodies follow geodesics of the
metric, and (iii)~in local Lorentz frames, the non-gravitational
laws of physics are those of special relativity.

The property that all non-gravitational fields should couple in
the same manner to a single gravitational field is sometimes
called ``universal coupling''.  Because of it, one can discuss the
metric as a property of spacetime itself rather than as a field
over spacetime.  This is because its properties may be measured
and studied using a variety of different experimental devices,
composed of different non-gravitational fields and particles,
and, because of universal coupling, the results will be
independent of the device.  Thus, for instance, the proper time
between two events is a characteristic of spacetime and of the
location of the events, not of the clocks used to measure it.

Consequently, if EEP is valid, the non-gravitational laws of
physics may be formulated by taking their special relativistic
forms in terms of the Min\-kowski metric {\boldmath $\eta$} and simply
``going over'' to new forms in terms of the curved spacetime
metric {\boldmath $g$}, using the mathematics of differential geometry.
The details of this ``going over'' can be found in standard
textbooks (\cite{MTW,Weinberg}, TEGP 3.2.~\cite{tegp}).

%%%%%%%%%%%%%%%%%%%%%%%%%%%%%%%%%%%%%%%%%%%%%%%%%%%%%%%%%%%%%%%%%%%%%%%%%%%%%%%%%%%

\subsubsection{The strong equivalence principle}
\label{sep}

In any metric theory of gravity, matter and non-gravitational
fields respond only to the spacetime metric {\boldmath $g$}.  In
principle, however, there could exist other gravitational fields
besides the metric, such as scalar fields, vector fields, and so
on.  If, by our strict definition of metric theory,
matter does not couple to these fields, what can their role
in gravitation theory be?  Their role must be that of mediating
the manner in which matter and non-gravitational fields generate
gravitational fields and produce the metric; once determined,
however, the metric alone acts back on the matter in the manner
prescribed by EEP.

What distinguishes one metric theory from another, therefore, is
the number and kind of gravitational fields it contains in
addition to the metric, and the equations that determine the
structure and evolution of these fields.  From this viewpoint,
one can divide all metric theories of gravity into two fundamental
classes:  ``purely dynamical'' and ``prior-geometric''.

By ``purely dynamical metric theory''
 we mean any metric theory
whose gravitational fields have their structure and evolution
determined by coupled partial differential field equations.  In
other words, the behavior of each field is influenced to some
extent by a coupling to at least one of the other fields in the
theory.  By ``prior geometric''
 theory, we mean any metric theory
that contains ``absolute elements'', fields or equations whose
structure and evolution are given {\it a priori}, and are
independent of the structure and evolution of the other fields of
the theory.  These ``absolute elements'' typically include flat
background metrics {\boldmath $\eta$} or cosmic time coordinates
$t$.

General relativity is a purely dynamical theory since it contains
only one gravitational field, the metric itself, and its
structure and evolution are governed by partial differential
equations (Einstein's equations).  Brans-Dicke theory and its
generalizations are purely
dynamical theories; the field equation for the metric involves the
scalar field (as well as the matter as source), and that for the
scalar field involves the metric.  Rosen's bimetric theory is a
prior-geometric theory:  It has a non-dynamical, Riemann-flat
background metric
{\boldmath $\eta$}, and the field equations for the physical metric
{\boldmath $g$} involve {\boldmath $\eta$}.

By discussing metric theories of gravity from this broad point of
view, it is possible to draw some general conclusions about the
nature of gravity in different metric theories, conclusions that
are reminiscent of the Einstein equivalence principle, but that
are subsumed under the name ``strong equivalence principle''.

Consider a local, freely falling frame in any metric theory of
gravity.  Let this frame be small enough that inhomogeneities in
the external gravitational fields can be neglected throughout its
volume.  On the other hand, let the frame be large enough to
encompass a system of gravitating matter and its associated
gravitational fields.  The system could be a star, a black hole,
the solar system or a Cavendish experiment.  Call this frame a
``quasi-local Lorentz frame''.  To determine the behavior
of the system we must calculate the metric.  The computation
proceeds in two stages.  First we determine the external
behavior of the metric and gravitational fields, thereby
establishing boundary values for the fields generated by the
local system, at a boundary of the quasi-local frame ``far'' from
the local system.  Second, we solve for the fields generated by
the local system.  But because the metric is coupled directly or
indirectly to the other fields of the theory, its structure and
evolution will be influenced by those fields, and in particular
by the boundary values taken on by those fields far from the
local system.  This will be true even if we work in a coordinate
system in which the asymptotic form of $g_{\mu\nu}$ in the
boundary region between the local system and the external world
is that of the Minkowski metric.  Thus the gravitational
environment in which the local gravitating system resides can
influence the metric generated by the local system via the
boundary values of the auxiliary fields.  Consequently, the
results of local gravitational experiments may depend on the
location and velocity of the frame relative to the external
environment.  Of course, local {\it non}-gravitational
experiments are unaffected since the gravitational fields they
generate are assumed to be negligible, and since those
experiments couple only to the metric, whose form can always be
made locally Minkowskian at a given spacetime event.
Local gravitational experiments might
include Cavendish experiments, measurement of the acceleration of
massive self-gravitating bodies,
studies of the structure of stars and planets, or
analyses of the periods of ``gravitational clocks''.  We
can now make several statements about different kinds of metric
theories.

(i)   A theory which contains only the metric {\boldmath $g$} yields
local gravitational physics which is independent of the location
and velocity of the local system.  This follows from the fact
that the only field coupling the local system to the environment
is {\boldmath $g$}, and it is always possible to find a coordinate
system in which {\boldmath $g$} takes the Minkowski form at the boundary
between the local system and the external environment (neglecting
inhomogeneities in the external gravitational field).  Thus the
asymptotic values of $g_{\mu\nu}$ are constants independent of
location, and are asymptotically Lorentz invariant, thus
independent of velocity.  General relativity is an example of
such a theory.

(ii)  A theory which contains the metric {\boldmath $g$} and dynamical
scalar fields $\varphi_A$ yields local gravitational physics
which may depend on the location of the frame but which is
independent of the velocity of the frame.  This follows from the
asymptotic Lorentz invariance of the Minkowski metric and of the
scalar fields, but now the asymptotic values of the scalar fields
may depend on the location of the frame.  An example is
Brans-Dicke theory, where the asymptotic scalar field determines
the effective value of the gravitational constant, which can thus vary as
$\varphi$ varies.  On the other hand, a form of velocity dependence in
local physics can enter indirectly if the asymptotic values of the
scalar field vary with time cosmologically.  Then the {\it rate} of
variation of the gravitational constant could
depend on the velocity of the frame.

(iii) A theory which contains the metric {\boldmath $g$} and additional
dynamical vector or tensor fields or prior-geometric fields
yields local gravitational physics which may have both location
and velocity-dependent effects.

These ideas can be summarized in
the strong equivalence principle (SEP), which states that:
\begin{enumerate}
\item
WEP is valid for self-gravitating bodies as well as for test bodies.
\item
The outcome of any local test experiment is
independent of the velocity of the (freely falling) apparatus.
\item
The outcome of any local test experiment is
independent of where and when in the universe it is performed.
\end{enumerate}
The distinction between SEP and EEP is the inclusion of bodies
with self-gravitational interactions (planets, stars) and of
experiments involving gravitational forces (Cavendish
experiments, gravimeter measurements).  Note that SEP contains
EEP as the special case in which local gravitational forces are
ignored.

The above discussion of the coupling of auxiliary fields to local
gravitating systems indicates that if SEP is strictly valid, there must be
one and only one gravitational field in the universe, the metric
{\boldmath $g$}.  These arguments are only suggestive however, and no
rigorous proof of this statement is available at present.
Empirically it has been found that almost every metric theory other than
GR introduces auxiliary gravitational fields,
either dynamical or prior geometric, and thus predicts violations
of SEP at some level (here we ignore quantum-theory inspired
modifications to GR involving ``$R^2$'' terms).  The one exception is
Nordstr\"om's 1913 conformally-flat scalar theory~\cite{nordstrom13}, which
can be written purely in terms of the metric; the theory satisfies SEP, but
unfortunately violates experiment by predicting no deflection of light.
General relativity seems to be the only viable
metric theory that embodies SEP completely.  In
Sec.~\ref{septests}, we shall discuss experimental evidence for the
validity of SEP.

%%%%%%%%%%%%%%%%%%%%%%%%%%%%%%%%%%%%%%%%%%%%%%%%%%%%%%%%%%%%%%%%%%%%%%%%%%%%%%%%%%%
%%%%%%%%%%%%%%%%%%%%%%%%%%%%%%%%%%%%%%%%%%%%%%%%%%%%%%%%%%%%%%%%%%%%%%%%%%%%%%%%%%%

\subsection{The parametrized post-Newtonian formalism}
\label{ppn}

Despite the possible existence of long-range gravitational fields
in addition to the metric in various metric theories of gravity,
the postulates of those theories demand that matter and
non-gravitational fields be completely oblivious to them.  The
only gravitational field that enters the equations of motion is
the metric {\boldmath $g$}.  The role of the other fields that a theory may
contain can only be that of helping to generate the spacetime
curvature associated with the metric.  Matter may create these
fields, and they plus the matter may generate the metric, but they
cannot act back directly on the matter.  Matter responds only to
the metric.

Thus the metric and the equations of motion for matter become the
primary entities for calculating observable effects,
and all that distinguishes one
metric theory from another is the particular way in which matter
and possibly other gravitational fields generate the metric.

The comparison of metric theories of gravity with each other and
with experiment becomes particularly simple when one takes the
slow-motion, weak-field limit.  This approximation, known as the
post-Newtonian limit, is sufficiently accurate to encompass most
solar-system tests that can be performed in the foreseeable
future.  It turns out that, in this limit, the spacetime metric
{\boldmath $g$} predicted by nearly every metric theory of gravity has the
same structure.  It can be written as an expansion about the
Minkowski metric ($ \eta_{\mu\nu} = {\rm diag}(-1,1,1,1)$) in terms
of dimensionless gravitational potentials of varying degrees of
smallness.
These potentials are constructed from the matter
variables (Box~\ref{box2}) in imitation of the Newtonian gravitational
potential
\begin{equation}\label{E20}
    U ( {\bf x}, t) \equiv \int \rho ( {\bf x}', t)
   | {\bf x} - {\bf x}' |^{-1}
    d^3 x'.
\end{equation}
The ``order of smallness'' is determined according to the rules
$U \sim~v^2 \sim \Pi \sim p/ \rho \sim \epsilon$,
$v^i \sim | d/dt | / | d/dx | \sim \epsilon^{1/2}$, and so on (we use units
in which $G=c=1$; see Box~\ref{box2}).

\begin{table}[t]
  \begin{center}
    \begin{tabular}{cp{2.8 cm}ccc}
      \hline\hline
      \multicolumn{1}{p{1 cm}}{Para\-meter} &
      What it measures relative to GR &
      \multicolumn{1}{p{1.85 cm}}{Value in GR} &
      \multicolumn{1}{p{2.15 cm}}{Value in semi\-conservative theories} &
      \multicolumn{1}{p{2.15 cm}}{Value in fully\-conservative theories} \\
      \hline \hline
      $\gamma$ & How much space-curvature produced by unit rest mass?
      & 1 & $\gamma$ & $\gamma$ \\
      \hline
      $\beta$ & How much ``nonlinearity'' in the superposition law for
      gravity? & 1 & $\beta$ & $\beta$ \\
      \hline
      $\xi$ & Preferred-location effects? & 0 & $\xi$ & $\xi$ \\
      \hline
      $\alpha_1$ & Preferred-frame & 0 & $\alpha_1$ & 0 \\
      $\alpha_2$ & effects? & 0 & $\alpha_2$ & 0 \\
      $\alpha_3$ & & 0 & 0 & 0 \\
      \hline
      $\alpha_3$ & Violation of con- & 0 & 0 & 0 \\
      $\zeta_1$ & servation of total & 0 & 0 & 0 \\
      $\zeta_2$ & momentum?& 0 & 0 & 0 \\
      $\zeta_3$ & & 0 & 0 & 0 \\
      $\zeta_4$ & & 0 & 0 & 0 \\
      \hline \hline
    \end{tabular}
    \caption{\it The PPN Parameters and their significance (note that
      $\alpha_3$ has been shown twice to indicate that it is a measure
      of two effects).}
    \label{ppnmeaning}
  \end{center}
\end{table}

A consistent post-Newtonian limit requires determination of $g_{00}$
correct through $O(\epsilon^2)$,
$g_{0i}$ through $O(\epsilon^{3/2})$ and $g_{ij}$
through $O(\epsilon)$ (for details see TEGP 4.1~\cite{tegp}).  The only way that
one metric theory differs from another is in the numerical values
of the coefficients that appear in front of the metric
potentials.  The parametrized post-Newtonian (PPN) formalism
inserts parameters in place of these coefficients, parameters
whose values depend on the theory under  study.  In the current
version of the PPN  formalism, summarized in Box~\ref{box2}, ten
parameters are used, chosen in such a manner that they measure or
indicate general properties of metric theories of gravity
(Table~\ref{ppnmeaning}).  Under reasonable assumptions about the
kinds of potentials that can be present at post-Newtonian order
(basically only Poisson-like potentials), one finds that ten PPN
parameters exhaust the possibilities.

The parameters $\gamma$ and $\beta$ are the usual
Eddington-Robertson-Schiff parameters used to describe the
``classical'' tests of GR, and are in some sense the most important; they
are the only non-zero parameters in GR and scalar-tensor gravity.
The parameter $\xi$ is non-zero in
any theory of gravity that predicts preferred-location effects
such as a galaxy-induced anisotropy in the local gravitational
constant $G_{\rm L}$ (also called ``Whitehead'' effects);
$\alpha_1$, $\alpha_2$, $\alpha_3$ measure whether or
not the theory predicts post-Newtonian preferred-frame
effects; $\alpha_3$, $\zeta_1$, $\zeta_2$,
$\zeta_3$, $\zeta_4$ measure whether or not the theory
predicts violations of global conservation laws for total
momentum.  
In Table~\ref{ppnmeaning} we show the values these
parameters take (i)~in GR, (ii)~in any theory
of gravity that possesses conservation laws for total momentum,
called ``semi-conservative'' (any theory that is based on an
invariant action principle is semi-conservative), and
(iii)~in any theory that in addition possesses six global
conservation laws for angular momentum, called ``fully
conservative'' (such theories automatically predict no
post-Newtonian preferred-frame effects).  Semi-conservative
theories have five free PPN  parameters ($\gamma$, $\beta$, $\xi$,
$\alpha_1$, $\alpha_2$) while fully conservative theories
have three ($\gamma$, $\beta$, $\xi$).

The PPN  formalism was pioneered by
Kenneth Nordtvedt~\cite{nordtvedt2}, who studied the post-Newtonian
metric of a system of gravitating point masses, extending earlier
work by Eddington, Robertson and Schiff (TEGP 4.2~\cite{tegp}).  
Will~\cite{Will71a} generalized the framework to perfect fluids.  A
general and unified version of the PPN  formalism was developed by
Will and Nordtvedt.  The canonical version, with
conventions altered to be more in accord with standard textbooks
such as~\cite{MTW}, is discussed in detail in TEGP 4~\cite{tegp}.  
Other versions
of the PPN  formalism have been developed to deal with point
masses with charge, fluid with anisotropic stresses,
bodies with strong internal gravity, and
post-post-Newtonian effects (TEGP 4.2, 14.2~\cite{tegp}).

\newpage

\begin{mybox}{The Parametrized Post-Newtonian Formalism}
  \begin{enumerate}
  \item{\bf Coordinate system:}
    The framework uses a nearly
    globally Lorentz coordinate system in which the coordinates are
    $(t, x^1, x^2, x^3)$.  Three-dimensional, Euclidean vector
    notation is used throughout.  All coordinate arbitrariness
    (``gauge freedom'') has been removed by specialization of the
    coordinates to the standard PPN  gauge (TEGP 4.2~\cite{tegp}).
    Units are chosen so that $G = c = 1$, where $G$ is the physically
    measured Newtonian constant far from the solar system.
  \item{\bf Matter variables:}
    \begin{itemize}
    \item $\rho$:
      density of rest mass as measured in a local freely falling
      frame momentarily comoving with the gravitating matter;
    \item $v^i=(dx^i /dt)$: coordinate velocity of the matter;
    \item $w^i$:
      coordinate velocity of the PPN coordinate system relative
      to the mean rest-frame of the universe;
    \item $p$:
      pressure as measured in a local freely falling frame
      momentarily comoving with the matter;
    \item $\Pi$:
      internal energy per unit rest mass.  It includes all forms
      of non-rest-mass, non-gravitational energy, e.g.\ energy
      of compression and thermal energy.
    \end{itemize}
  \item{\bf PPN parameters:}
    $\gamma, \beta, \xi, \alpha_1, \alpha_2, \alpha_3, \zeta_1,\zeta_2,\zeta_3, \zeta_4. $
  \item{\bf Metric:}
    {\setlength{\arraycolsep}{0.14 em}
      \begin{eqnarray*}
        g_{00} &=&
        -1 + 2U - 2 \beta U^2 - 2 \xi \Phi_W
        + (2 \gamma +2+ \alpha_3 + \zeta_1 - 2 \xi ) \Phi_1 \\
        &&+ 2(3 \gamma - 2 \beta + 1 + \zeta_2 + \xi ) \Phi_2
        + 2(1 + \zeta_3 ) \Phi_3
        + 2(3 \gamma + 3 \zeta_4 - 2 \xi ) \Phi_4 \\
        &&- ( \zeta_1 - 2 \xi ) {\cal A}
        - ( \alpha_1 - \alpha_2 - \alpha_3 ) w^2 U
        - \alpha_2 w^i w^j U_{ij}
        + (2 \alpha_3 - \alpha_1 ) w^i V_i \\
        && +  O(\epsilon^3), \\
        g_{0i}
        &=& - {1 \over 2 }
        (4 \gamma + 3 + \alpha_1 - \alpha_2
        + \zeta_1 - 2 \xi ) V_i
        - {1 \over 2}
        (1 + \alpha_2 - \zeta_1 + 2 \xi )W_i \\
        &&- {1 \over 2} ( \alpha_1 - 2 \alpha_2 ) w^i U
        - \alpha_2 w^j U_{ij} + O(\epsilon^{5/2}), \\
        g_{ij}
        &=& (1 + 2 \gamma U ) \delta_{ij} + O(\epsilon^2).
      \end{eqnarray*}}%
  \item{\bf Metric potentials:}
    {\setlength{\arraycolsep}{0.14 em}
      \begin{eqnarray*}
        U&=&\int {{\rho' } \over {| {\bf x}-{\bf x}' |}}
        d^3x', \\
        U_{ij}&=&
        \int {{\rho' (x-x')_i (x-x')_j}
          \over {| {\bf x}-{\bf x}' |^3}} d^3x', \\
        \Phi_W &=&
        \int {{\rho' \rho'' ({\bf x}-{\bf x}')}
          \over {| {\bf x}-{\bf x}' |^3}} \cdot \left(
          {{{\bf x}' -{\bf x}'' }
            \over {| {\bf x}-{\bf x}'' |}}-
          {{{\bf x}-{\bf x}''}
            \over {| {\bf x}'-{\bf x}'' |}}
        \right) d^3x' d^3x'', \\
        {\cal A} &=& \int {{\rho' [{\bf v}'
            \cdot ({\bf x}-{\bf x}')]^2 }
          \over {| {\bf x}-{\bf x}' |^3}}
        d^3x', \\
        \Phi_1 &=&
        \int {{\rho' v'^2}
          \over {| {\bf x}-{\bf x}' |}} d^3x', \\
        \Phi_2 &=&
        \int {{\rho' U'} \over {| {\bf x}-{\bf x}' |}}
        d^3x', \\
        \Phi_3 &=&
        \int {{\rho' \Pi'} \over {| {\bf x}-{\bf x}' |}}
        d^3x', \\
        \Phi_4&=&
        \int {{p' } \over {| {\bf x}-{\bf x}' |}} d^3x', \\
        V_i &=&
        \int {{\rho' v_i'} \over {| {\bf x}-{\bf x}' |}} d^3x', \\
        W_i&=&
        \int {{\rho' [{\bf v}' \cdot
            ({\bf x}-{\bf x}')](x-x')_i}
          \over {| {\bf x}-{\bf x}' |^3}} d^3x'.
      \end{eqnarray*}}%
  \item{\bf Stress--energy tensor} (perfect fluid){\bf:}
    {\setlength{\arraycolsep}{0.14 em}
      \begin{eqnarray*}
        T^{00} &=& \rho (1+ \Pi + v^2 +2U), \\
        T^{0i} &=& \rho v^i (1+ \Pi + v^2 +2U + p/\rho), \\
        T^{ij} &=& \rho v^iv^j (1+ \Pi + v^2 +2U + p/\rho) +
        p\delta^{ij}(1-2\gamma U).
      \end{eqnarray*}}%
  \item{\bf Equations of motion:}
    \begin{itemize}
    \item Stressed matter: \quad
      ${T^{\mu\nu}}_{;\nu}= 0$,
    \item Test bodies: \quad
      $d^2 x^\mu /d \lambda^2 + {\Gamma^\mu}_{\nu \lambda}
      (dx^\nu /d \lambda )( dx^\lambda / d \lambda ) = 0$,
    \item Maxwell's equations: \quad
      ${F^{\mu \nu}}_{; \nu} = 4 \pi J^\mu$, \qquad
      $F_{\mu \nu} = A_{\nu ; \mu} - A_{\mu ; \nu }$.
    \end{itemize}
  \end{enumerate}
  \label{box2}
\end{mybox}

\newpage

%%%%%%%%%%%%%%%%%%%%%%%%%%%%%%%%%%%%%%%%%%%%%%%%%%%%%%%%%%%%%%%%%%%%%%%%%%%%%%%%%%%
%%%%%%%%%%%%%%%%%%%%%%%%%%%%%%%%%%%%%%%%%%%%%%%%%%%%%%%%%%%%%%%%%%%%%%%%%%%%%%%%%%%

\subsection{Competing theories of gravity}
\label{theories}

One of the important applications of the PPN formalism is the
comparison and classification of alternative metric theories of
gravity.  The population of viable theories has fluctuated over
the years as new effects and tests have been discovered, largely
through the use of the PPN  framework, which eliminated many
theories thought previously to be viable.  The theory population
has also fluctuated as new, potentially viable theories have been
invented.

In this review, we shall focus on GR, the
general class of scalar-tensor modifications of it, of which the
Jordan-Fierz-Brans-Dicke theory (Brans-Dicke, for short)
is the classic example, and vector-tensor theories.  
The reasons are several-fold:

\begin{itemize}
\item 
A full compendium of alternative theories circa 1981 
is given in TEGP 5~\cite{tegp}.
\item 
Many alternative metric theories developed during the 1970s and
1980s could be viewed as ``straw-man'' theories, invented to prove
that such theories exist or to illustrate particular properties.  Few
of these could be regarded as well-motivated theories from the point
of view, say, of field theory or particle physics.  
\item 
A number of theories fall into the class of ``prior-geometric''
theories, with absolute elements such as a flat background metric in
addition to the physical metric.  Most of these theories predict
``preferred-frame'' effects, that have been tightly constrained by
observations (see Sec.~\ref{preferred}).
An example is Rosen's bimetric theory.
\item 
A large number of alternative theories of gravity predict
gravitational wave emission substantially different from that of general
relativity,
in strong disagreement with observations of the binary pulsar (see
Sec.~\ref{S5}).
\item 
Scalar-tensor modifications of GR have become very
popular in unification schemes
such as string theory, and in cosmological model building.
Because the scalar fields could be massive, the potentials in the
post-Newtonian limit could be modified by Yukawa-like terms.
\item
Vector-tensor theories have attracted recent attention, 
in the spirit of the SME (Sec.~\ref{SME}), 
as models for 
violations of Lorentz invariance in the gravitational sector.
\end{itemize}

%%%%%%%%%%%%%%%%%%%%%%%%%%%%%%%%%%%%%%%%%%%%%%%%%%%%%%%%%%%%%%%%%%%%%%%%%%%%%%%%%%%

\subsubsection{General relativity}
\label{generalrelativity}

The metric {\boldmath $g$} is the sole
dynamical field, and the theory contains no arbitrary functions or
parameters, apart from the value of the Newtonian coupling constant
$G$, which is measurable in laboratory experiments.  Throughout this
article, we ignore the cosmological constant $\Lambda_{\rm C}$.  
We do this despite recent evidence, from supernova data, of an accelerating
universe, which would indicate either a non-zero
cosmological constant or a dynamical
``dark energy'' contributing about 70 percent of the critical density.
Although
$\Lambda_{\rm C}$ has significance for quantum field theory, quantum
gravity, and cosmology, on the scale of the solar-system or of stellar
systems its effects are negligible, for the values of $\Lambda_{\rm C}$
inferred from supernova observations.  

The field equations of GR are derivable from an invariant action
principle $\delta I=0$, where
\begin{equation}\label{E21}
I=(16\pi G)^{-1} \int R (-g)^{1/2} d^4x + I_{\rm m}(\psi_{\rm m}, g_{\mu\nu}),
\end{equation}
where $R$ is the Ricci scalar, and $I_{\rm m}$ is the matter action, which
depends on matter fields $\psi_{\rm m}$ universally coupled to the metric
{\boldmath $g$}.  By varying the action with respect to $g_{\mu\nu}$, we
obtain the field equations
\begin{equation}\label{E22}
G_{\mu\nu} \equiv R_{\mu\nu} - {1 \over 2} g_{\mu\nu} R = 8\pi G T_{\mu\nu},
\end{equation}
where $T_{\mu\nu}$ is the matter energy-momentum tensor.  General
covariance of the matter action implies the equations of motion
${T^{\mu\nu}}_{;\nu}=0$; varying $I_{\rm m}$ with respect to
$\psi_{\rm m}$
yields the matter field equations of the Standard Model.  
By virtue of the {\it absence} of
prior-geometric elements, the equations of motion are also a
consequence of the field equations via the Bianchi identities
${G^{\mu\nu}}_{;\nu}=0$.

The general procedure for deriving the post-Newtonian limit of 
metric theories is spelled
out in TEGP 5.1~\cite{tegp}, and is described in detail for GR in TEGP 5.2~\cite{tegp}.  The PPN
parameter values are listed in Table~\ref{ppnvalues}.

\begin{table}[t]
  \begin{center}
    \begin{tabular}{@{}l@{}ccccccc@{}}
      \hline \hline
      & Arbitrary & Cosmic & \multicolumn{5}{c}{PPN Parameters} \\
      & Functions & Matching & \multicolumn{5}{c}{\hrulefill} \\
      Theory & or Constants & Parameters & $\gamma$ & $\beta$ &
      $\xi$ & $\alpha_1$ & $\alpha_2$ \\
      \hline \hline
       General Relativity
      & none & none & 1 & 1 & 0 & 0 & 0 \\
      \hline
      Scalar-Tensor \\
      \quad Brans-Dicke & $\omega_{\rm BD}$ & $\phi_0$ &
      $ \displaystyle \frac{1+\omega_{\rm BD}}{2+\omega_{\rm BD}}$ & 1 & 0 & 0 & 0 \\
      \quad General & $A(\varphi)$, $V(\varphi)$ & $\varphi_0$ &
      $ \rule{0 cm}{0.6cm} \displaystyle \frac{1+\omega}{2+\omega}$ &
      $1+\Lambda$ & 0 & 0 & 0 \\
      \hline
      Vector-Tensor\\
      \quad Unconstrained & $\omega$, $c_1$, $c_2$, $c_3$, $c_4$ & $u$ &
      $\gamma^\prime$&$\beta^\prime$&0&$\alpha_1^\prime$&$\alpha_2^\prime$\\
      \quad Einstein-{\AE}ther & $c_1$, $c_2$, $c_3$, $c_4$& none&
      1&1&0&$\alpha_1^\prime$&$\alpha_2^\prime$\\
      \hline
        Rosen's Bimetric
      & none & $c_0$, $c_1$ & 1 & 1 & 0 & 0 &
      $ \rule{0 cm}{0.5cm} \displaystyle \frac{c_0}{c_1}-1$ \\
      \hline \hline
    \end{tabular}
    \caption{\it Metric theories and their PPN parameter values
      ($\alpha_3 = \zeta_i=0$ for all cases).  The parameters
      $\gamma^\prime$, $\beta^\prime$ 
      $\alpha_1^\prime$ and $\alpha_2^\prime$
      denote complicated functions of $u$ and of the
      arbitrary constants. Here $\Lambda$ is not the cosmological
      constant $\Lambda_{\rm C}$, but is defined by Eq. (\ref{E27}).}
    \label{ppnvalues}
  \end{center}
\end{table}

%%%%%%%%%%%%%%%%%%%%%%%%%%%%%%%%%%%%%%%%%%%%%%%%%%%%%%%%%%%%%%%%%%%%%%%%%%%%%%%%%%%

\subsubsection{Scalar-tensor theories}
\label{scalartensor}

These theories contain the metric {\boldmath $g$}, a
scalar field $\varphi$, a potential  function $V(\varphi)$, and a
coupling function $A(\varphi)$ (generalizations to more than one scalar
field have also been carried out~\cite{DamourEspo92}).
For some purposes, the action is conveniently written in a non-metric
representation, sometimes denoted the ``Einstein frame'', in which the
gravitational action looks exactly like that of GR:
\begin{equation}\label{E23}
\tilde I=(16\pi G)^{-1} \int [\tilde R -2\tilde g^{\mu\nu} \partial_\mu \varphi
\partial_\nu
\varphi -V(\varphi)] (-\tilde g)^{1/2} d^4x + I_{\rm m}(\psi_{\rm m}, A^2(\varphi)
\tilde g_{\mu\nu}),
\end{equation}
where $\tilde R \equiv \tilde g^{\mu\nu} \tilde R_{\mu\nu}$ is the
Ricci scalar of the
``Einstein'' metric $\tilde g_{\mu\nu}$.  (Apart from the scalar potential term
$V(\varphi)$, this corresponds to Eq.~(\ref{stringlagrangian1})
with $\tilde G(\varphi) \equiv (4\pi G)^{-1}$, $U(\varphi) \equiv 1$,
and $\tilde M(\varphi) \propto A(\varphi)$.)  This representation is a
``non-metric'' one because the matter fields $\psi_{\rm m}$ couple to a
combination of $\varphi$ and $\tilde g_{\mu\nu}$.
Despite appearances, however,
it is a metric theory, because it can be put
into a metric representation by identifying the ``physical metric''
\begin{equation}\label{E24}
g_{\mu\nu} \equiv A^2(\varphi) \tilde g_{\mu\nu}.
\end{equation}
The action can then be rewritten in the metric form
\begin{equation}\label{E25}
I=(16\pi G)^{-1} \int [\phi R - \phi^{-1} \omega(\phi) g^{\mu\nu}\partial_\mu
\phi \partial_\nu \phi - \phi^2 V] (-g)^{1/2} d^4x + I_{\rm
  m}(\psi_{\rm m},
g_{\mu\nu}),
\end{equation}
where
{\setlength{\arraycolsep}{0.14 em}
\begin{eqnarray}
\phi &\equiv& A(\varphi)^{-2}, \nonumber \\
3+2\omega(\phi) &\equiv& \alpha(\varphi)^{-2}, \label{E26} \\
 \alpha(\varphi) &\equiv& d (\ln A(\varphi))/d\varphi. \nonumber
\end{eqnarray}}
The Einstein frame is useful for discussing general characteristics of
such theories, and for some cosmological applications, while the metric
representation is most useful for calculating observable effects.
The field equations, post-Newtonian limit and PPN  parameters are
discussed in TEGP 5.3~\cite{tegp}, and the values of the PPN  parameters are
listed in Table~\ref{ppnvalues}.

The
parameters that enter the post-Newtonian limit are
\begin{equation}\label{E27}
\omega \equiv \omega(\phi_0),
\qquad
\Lambda \equiv [(d\omega/d\phi)(3+2\omega)^{-2}(4+2\omega)^{-1}]_{\phi_0},
\end{equation}
where $\phi_0$ is the value of $\phi$ today far from the
system being studied, as determined by appropriate cosmological boundary
conditions.
In Brans-Dicke theory ($\omega(\phi) \equiv \omega_{\rm BD}=$ constant),
the larger the value of
$\omega_{\rm BD}$, the smaller the effects of the scalar field, and in the
limit $\omega_{\rm BD} \to \infty$ ($\alpha_0 \to 0$),
the theory becomes indistinguishable from
GR in all its predictions.  In more general
theories, the function $\omega ( \phi )$ could have the
property that, at the present epoch, and in weak-field situations,
the value of the scalar field $\phi_0$ is such that
$\omega$ is very large and $\Lambda$ is very small (theory almost
identical to GR today), but that for past or
future values of $\phi$, or in strong-field regions such as the
interiors of neutron stars, $\omega$ and $\Lambda$ could take on values
that would lead to significant differences from GR.  It is useful to point
out that all versions of scalar-tensor gravity predict that $\gamma \le 1$
(Table \ref{ppnvalues}).

Damour and Esposito-Far\`ese~\cite{DamourEspo92} have adopted an alternative
parametrization of scalar-tensor theories, in which  one
expands $\ln A(\varphi)$
about a cosmological background field value $\varphi_0$:
\begin{equation}
\ln A(\varphi) = \alpha_0 (\varphi -\varphi_0) + {1 \over 2} \beta_0
(\varphi -\varphi_0)^2 + \dots.
\label{alphaexpand}
\end{equation}
A precisely linear coupling function produces Brans-Dicke theory, with
$\alpha_0^2 = 1/(2 \omega_{\rm BD} +3)$, or  
$1/(2+\omega_{\rm BD})=2\alpha_0^2/(1+\alpha_0^2)$.
The function $\ln A(\varphi)$ acts
as a potential for the scalar field $\varphi$ within matter, 
and, if $\beta_0
>0$, then during cosmological evolution,
the scalar field naturally evolves toward the minimum of the
potential, i.e.\ toward $\alpha_0 \approx 0$, 
$\omega \to \infty$, or toward
a theory close to, though not precisely GR~\cite{DamourNord93a,DamourNord93b}.
Estimates of the
expected relic deviations from GR today in such theories depend on the
cosmological model, but range from $10^{-5}$ to a few times $10^{-7}$
for $|\gamma-1|$.

Negative
values of $\beta_0$ correspond to a ``locally unstable'' 
scalar potential (the overall theory is still stable in the sense of having
no tachyons or ghosts).
In this case, objects such as neutron stars can experience a
``spontaneous scalarization'', whereby the interior values of $\varphi$
can take on values very different from the exterior values, through
non-linear interactions between strong gravity and the scalar field,
dramatically affecting the stars' internal structure and leading to
strong violations of SEP.  On the other hand, in the case 
$\beta_0 <0$, one must confront that fact that, 
with an unstable $\varphi$ potential,
cosmological evolution would presumably drive the system away from the
peak where $\alpha_0 \approx 0$,
toward parameter values that could be excluded
by solar system experiments.  

Scalar fields coupled to gravity or matter are also
ubiquitous in particle-physics-inspired models of unification, such as
string theory~\cite{TaylorVeneziano,maeda88,DamourPolyakov,
damourpiazza02a,damourpiazza02b}.
In some models, the coupling to matter may lead to
violations of EEP, which could be  tested or bounded
by the experiments described in Sec.~\ref{eep}.  In
many models the scalar field could be massive; if the Compton wavelength is
of macroscopic scale, its effects are those of a ``fifth force''.
Only if the theory can be cast as a metric theory with a
scalar field of infinite range or of range long compared to the scale
of the system in question (solar system) can the PPN  framework be
strictly
applied.  If the mass of the scalar field is sufficiently large that its
range is microscopic, then, on solar-system scales, the scalar field is
suppressed, and the theory is essentially equivalent to general
relativity.  

\subsubsection{Vector-Tensor theories}
\label{vectortensor}

These theories contain the metric {\boldmath $g$} and a dynamical, 
typically timelike, four-vector
field $u^\mu$.   In some models, the four-vector is unconstrained, while
in others, called Einstein-{\AE}ther theories it is constrained to be timelike
with unit norm.  The most general action for such theories
that is quadratic in derivatives of 
the vector is given by
\begin{equation}
I = (16\pi G)^{-1} \int \left [(1+ \omega u_\mu u^\mu )R
   - K^{\mu\nu}_{\alpha\beta} \nabla_\mu u^\alpha \nabla_\nu
   u^\beta + \lambda (u_\mu u^\mu +1) \right ] (-g)^{1/2} d^4x +
   I_{\rm m}(\psi_{\rm m}, g_{\mu\nu}),
     \,,
   \label{aetheraction}
\end{equation}
where
\begin{equation}
 K^{\mu\nu}_{\alpha\beta} = c_1 g^{\mu\nu} g_{\alpha\beta}
	+ c_2 \delta^\mu_\alpha \delta^\nu_\beta
	+ c_3 \delta^\mu_\beta\delta^\nu_\alpha 
	- c_4 u^\mu u^\nu g_{\alpha\beta} \,.
	\label{ktensor}
\end{equation}
The coefficients $c_i$ are arbitrary.  In the unconstrained theories,
$\lambda \equiv 0$ and  $\omega$ is arbitrary.  In the constrained theories,
$\lambda$ is a Lagrange multiplier, and by virtue of the constraint
$u_\mu u^\mu = -1$, the factor $\omega u_\mu u^\mu$
in front of the Ricci scalar can be
absorbed into a rescaling of $G$; equivalently, in the constrained theories,
we can set $\omega=0$.  Note that the possible term $u^\mu u^\nu
R_{\mu\nu}$ can be shown under integration by parts to be equivalent to a
linear combination of the terms involving $c_2$ and $c_3$.  

Unconstrained theories were studied during the 1970s as ``straw-man''
alternatives to GR.  In addition to having up to four
arbitrary parameters, they also left the magnitude of the vector field 
arbitrary, since it satisfies a linear homogenous vacuum field equation
of the form ${\cal L} u^\mu =0$ ($c_4=0$ in all such cases studied). 
Indeed, this latter fact was one of most serious defects of these theories.
Each theory studied corresponds to a special case of the
action (\ref{aetheraction}), all with $\lambda \equiv 0$:

\begin{itemize}
\item
{\em General vector-tensor theory ($\omega$, $\tau$, $\epsilon$, $\eta$)}, TEGP 5.4~\cite{tegp}:
The gravitational Lagrangian for this class of theories had the form
$R + \omega u_\mu u^\mu R + \eta u^\mu u^\nu R_{\mu\nu}
-\epsilon F_{\mu\nu}F^{\mu\nu} + \tau \nabla_\mu u_\nu \nabla^\mu u^\nu$,
where $F_{\mu\nu} = \nabla_\mu u_\nu -\nabla_\nu u_\mu$, corresponding to
the values
$c_1 = 2\epsilon - \tau$, $c_2 = -\eta$, $c_1+c_2+c_3= -\tau$, $c_4=0$.
In these theories $\gamma$, $\beta$, $\alpha_1$ and
$\alpha_2$ are complicated functions of the parameters and of 
$u^2 = -u^\mu u_\mu$, while the rest vanish.

\item
{\em Will-Nordtvedt theory}~\cite{willnordtvedt72}: This is the special case
$c_1
= -1$, $c_2 = c_3 = c_4 =0$.  In this theory, the PPN parameters are given
by $\gamma = \beta = 1$, $\alpha_2 = u^2/(1+u^2/2)$, and zero for the rest.

\item
{\em Hellings-Nordtvedt theory ($\omega$)}~\cite{hellings73}: This is 
the special case $c_1 =2$, $c_2 =2\omega$,
$c_1 + c_2 +c_3 = 0 = c_4$.  Here $\gamma$, $\beta$, $\alpha_1$ and
$\alpha_2$ are complicated functions of the parameters and of $u^2$, while
the rest vanish.

\end{itemize}

The Einstein-{\AE}ther theories were motivated in part by a desire to explore
possibilities for violations of Lorentz invariance in gravity, in parallel
with similar studies in matter interactions, such as the SME.  The general
class of theories was analysed by Jacobson and
collaborators~\cite{jacobson01,mattingly02,jacobson04,eling04,foster05}, 
motivated
in part by~\cite{kosteleckysamuel}.  

Analysing the post-Newtonian limit, they were able
to infer values of the PPN parameters $\gamma$ and $\beta$ 
as follows~\cite{foster05}:

\begin{equation}
\gamma = 1 \,, \quad \beta =1 \,,
\nonumber
\end{equation}
\begin{equation}
\xi = \alpha_3 = \zeta_1 = \zeta_2 = \zeta_3 = \zeta_4 \,,
\nonumber
\end{equation}
\begin{equation}
\alpha_1 = -\frac{8(c_3^2+c_1c_4)}{2c_1-c_1^2+c_3^2} \,,
\nonumber
\end{equation}
\begin{eqnarray}
\alpha_2 &=& \frac{(2c_{13}-c_{14})^2}{c_{123}(2-c_{14})} 
\nonumber \\
&& -
\frac{12c_3c_{13}+2c_2c_{14}(1-2c_{14})+(c_1^2-c_3^2)(4-6c_{13}+7c_{14})}
{(2-c_{14})(2c_1-c_1^2+c_3^2)} \,,
\end{eqnarray}
where $c_{123}=c_1+c_2+c_3$,  $c_{13}=c_1-c_3$, $c_{14}=c_1-c_4$, subject to
the constraints $c_{123} \ne 0$, $c_{14} \ne 2$, $2c_1-c_1^2+c_3^2 \ne 0$.

By requiring that gravitational-wave modes have real (as
opposed to imaginary) frequencies, one can impose the bounds
$c_1/(c_1+c_4) \ge 0$ and $(c_1+c_2+c_3)/(c_1+c_4) \ge 0$.  
Considerations of positivity of energy impose the constraints $c_1 > 0$,
$c_1+c_4>0$ and $c_1+c_2+c_3 >0$.  

%%%%%%%%%%%%%%%%%%%%%%%%%%%%%%%%%%%%%%%%%%%%%%%%%%%%%%%%%%%%%%%%%%%%%%%%%%%%%%%%%%
%%%%%%%%%%%%%%%%%%%%%%%%%%%%%%%%%%%%%%%%%%%%%%%%%%%%%%%%%%%%%%%%%%%%%%%%%%%%%%%%%%%

\subsection{Tests of the parameter $\gamma$}
\label{gamma}

With the PPN formalism in hand, we are now ready to confront
gravitation theories with the results of solar-system
experiments.  In this section we focus on tests of the parameter
$\gamma$, consisting of the deflection of light and the time delay
of light.

%%%%%%%%%%%%%%%%%%%%%%%%%%%%%%%%%%%%%%%%%%%%%%%%%%%%%%%%%%%%%%%%%%%%%%%%%%%%%%%%%%%

\subsubsection{The deflection of light}
\label{deflection}

A light ray
(or photon) which passes the Sun at a distance $d$ is deflected by
an angle
\begin{equation}\label{E28}
    \delta \theta = {1 \over 2}(1+ \gamma ) (4m_\odot/d)
    [(1+\cos\Phi )/2]
\end{equation}
(TEGP 7.1~\cite{tegp}), where $m_\odot$ is the mass of the Sun
and $\Phi$ is the angle between the Earth-Sun line and the
incoming direction of the photon (Figure~\ref{deflectiongeom}).
For a grazing ray,
$d \approx d_\odot$, $\Phi \approx 0$, and
\begin{equation}\label{E29}
     \delta \theta ~\approx~{1 \over 2} (1+ \gamma ) 1.''7505,
\end{equation}
independent of the frequency of light.  Another, more useful
expression gives the change in the relative angular separation
between an observed source of light and a nearby reference source
as both rays pass near the Sun:
\begin{equation}\label{E30}
          \delta \theta = {1 \over 2}(1+ \gamma )
\left[ - {{4m_\odot} \over d} \cos\chi
      + {{4m_\odot} \over d_{\rm r}}
\left( {{1+ \cos\Phi_{\rm r}} \over 2} \right) \right],
\end{equation}
where $d$ and $d_{\rm r}$ are the distances of closest approach of
the source and reference rays respectively, $\Phi_{\rm r}$ is the
angular separation between the Sun and the reference source, and
$\chi$ is the angle between the Sun-source and the Sun-reference
directions, projected on the plane of the sky (Figure~\ref{deflectiongeom}).
Thus, for example, the relative angular separation between the
two sources may vary if the line of sight of one of them passes
near the Sun ($d \sim R_\odot$, $d_{\rm r} \gg d$,
$\chi$ varying with time).

\epubtkImage{}{
\begin{figure}[hptb]
  \def\epsfsize#1#2{0.7#1}
  \centerline{\epsfbox{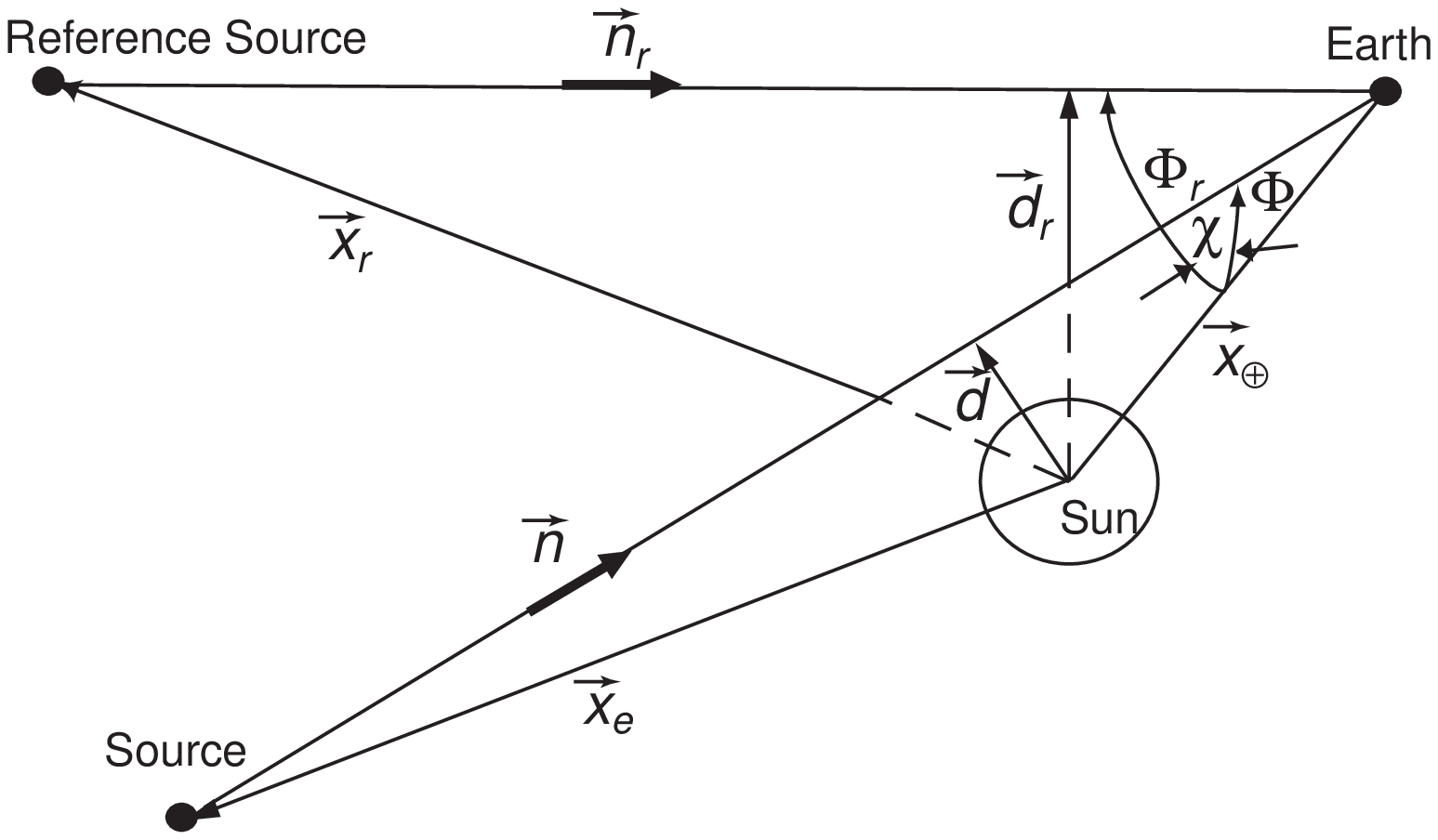}}
  \caption{\it Geometry of light deflection measurements.}
  \label{deflectiongeom}
\end{figure}}

It is interesting to note that the classic derivations of the
deflection of light that use only 
the corpuscular theory of light (Cavendish 1784, von Soldner
1803~\cite{willcavendish}), or
the principle of equivalence (Einstein 1911),
yield only the ``1/2'' part of the coefficient in front of the
expression in Eq.~(\ref{E28}).  But the result of these calculations
is the deflection of light relative to local straight lines, as
established for example by rigid rods; however, because of space
curvature around the Sun, determined by the PPN  parameter
$\gamma$, local straight lines are bent relative to asymptotic
straight lines far from the Sun by just enough to yield the
remaining factor ``$\gamma /2$''.  The first factor ``1/2''
holds in any metric theory, the second ``$\gamma /2$'' varies
from theory to theory.  Thus, calculations that purport to derive
the full deflection using the equivalence principle alone are
incorrect.

The prediction of the full bending of light by the Sun was one of
the great successes of Einstein's GR.
Eddington's confirmation of the bending of optical starlight
observed during a solar eclipse in the first days following World
War I helped make Einstein famous.  However, the experiments of
Eddington and his co-workers had only 30~percent accuracy, and
succeeding experiments were not much better:  The results were
scattered between one half and twice the Einstein value
(Figure~\ref{gammavalues}),
and the accuracies were low.

\epubtkImage{}{
\begin{figure}[hptb]
  \def\epsfsize#1#2{0.6#1}
  \centerline{\epsfbox{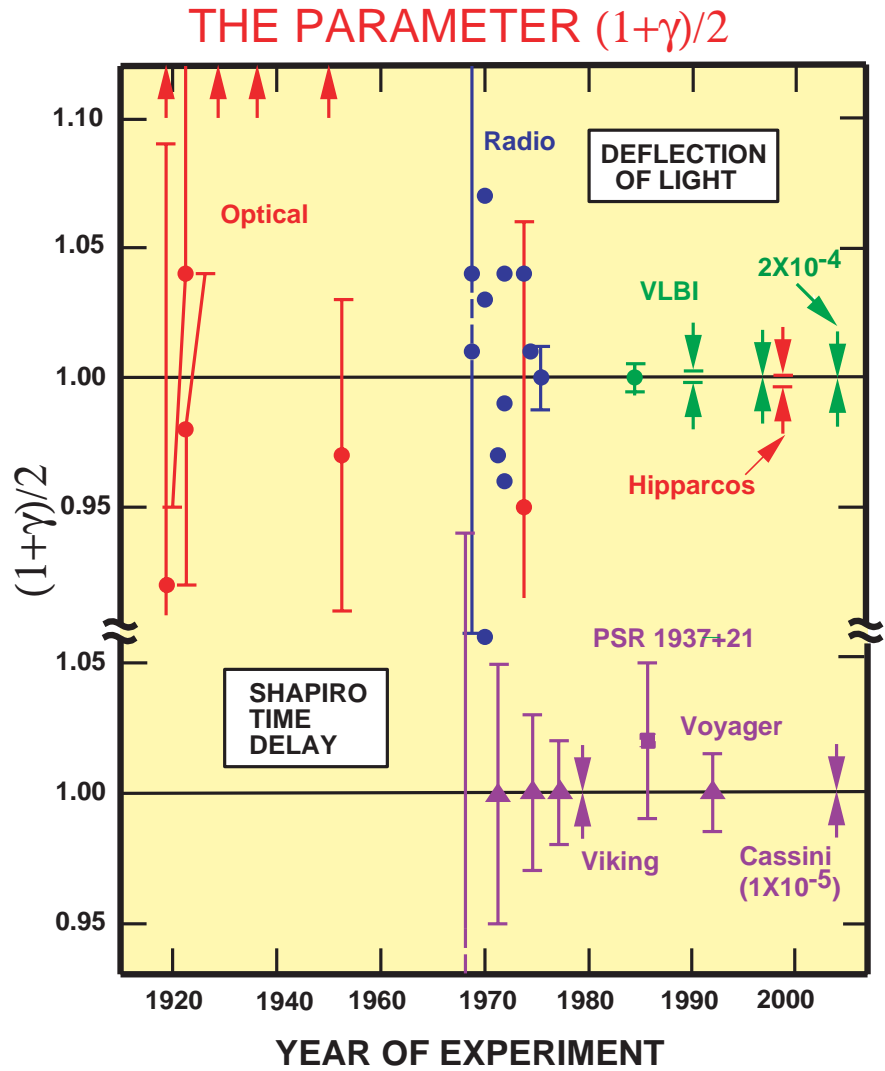}}
  \caption{\it Measurements of the coefficient $(1 + \gamma )/2$ from
    light deflection and time delay measurements.  Its GR
    value is unity.  The arrows at the top denote anomalously large
    values from early eclipse expeditions.  The Shapiro time-delay
    measurements using the Cassini spacecraft yielded an agreement with GR
    to $10^{-3}$ percent, and VLBI light deflection measurements have
    reached 0.02 percent.  Hipparcos denotes the optical astrometry
    satellite, which reached 0.1 percent.}
  \label{gammavalues}
\end{figure}}

However, the development of radio-interferometery, and later
of VLBI, very-long-baseline radio
interfer\-om\-etry, produced greatly improved determinations of
the deflection of light.  These techniques now have the capability
of measuring angular separations and changes in angles
to accuracies better than 
100 microarcseconds.  Early measurements took advantage
of a series of heavenly coincidences:
Each year, groups of strong quasistellar radio sources pass
very close to the Sun (as seen from the Earth), including the
group 3C273, 3C279, and 3C48, and the group 0111+02, 0119+11
and 0116+08.  As the Earth moves in its orbit, changing the
lines of sight of the quasars relative to the Sun, the angular
separation $\delta \theta$ between pairs of quasars
varies (Eq.~(\ref{E30})).  The time variation in the
quantities $d$, $d_{\rm r}$, $\chi$ and $\Phi_{\rm r}$ in Eq.~(\ref{E30}) is
determined using an accurate ephemeris for the Earth and initial
directions for the quasars, and the resulting prediction for
$\delta \theta$  as a function of time is used as a basis for a
least-squares fit of the measured $\delta \theta$,  with one of
the fitted parameters being the coefficient ${1 \over 2}(1+ \gamma )$.
A number of measurements of this kind over the period 1969--1975 yielded
an accurate determination of the coefficient ${1 \over 2}(1+ \gamma)$.
A 1995 VLBI measurement using 3C273 and
3C279 yielded $(1+\gamma)/2=0.9996 \pm 0.0017$~\cite{lebach}.

In recent years, transcontinental and intercontinental VLBI observations 
of quasars and
radio galaxies have been 
made primarily to monitor the Earth's rotation
(``VLBI'' in Figure~\ref{gammavalues}).  These measurements are
sensitive to the deflection of light over
almost the entire celestial sphere (at $90 ^\circ$ from the Sun, the
deflection is still 4 milli\-arcseconds).
A 2004 analysis of  almost 2 million VLBI observations
of 541 radio sources, made by 87 VLBI sites
yielded 
$(1+\gamma)/2=0.99992 \pm 0.00023$, or equivalently,
$\gamma-1= (-1.7 \pm 4.5) \times 10^{-4}$~\cite{sshapiro04}.

Analysis of observations made by the Hipparcos optical astrometry
satellite yielded a test at the level of 0.3
percent~\cite{hipparcos}.
A VLBI
measurement of the deflection of light by Jupiter was
reported; the predicted deflection of about 300
microarcseconds was seen with about 50 percent accuracy~\cite{treuhaft}.
The results of light-deflection measurements are summarized in
Figure~\ref{gammavalues}.

%%%%%%%%%%%%%%%%%%%%%%%%%%%%%%%%%%%%%%%%%%%%%%%%%%%%%%%%%%%%%%%%%%%%%%%%%%%%%%%%%%%

\subsubsection{The time delay of light}
\label{timedelay}

A radar signal sent across the solar system past the Sun to a
planet or satellite and returned to the Earth suffers an
additional non-Newtonian delay in its round-trip travel time,
given by (see Figure~\ref{deflectiongeom})
\begin{equation}\label{E31}
      \delta t = 2(1+ \gamma ) m_\odot
      \ln [(r_\oplus
  + {\bf x}_\oplus \cdot {\bf n} )
      (r_{\rm e} - {\bf x}_{\rm e} \cdot {\bf n} )/d^2],
\end{equation}
where $ {\bf x}_{\rm e} $ ($ {\bf x}_\oplus $) are the vectors, and
$ r_{\rm e} $ ($ r_\oplus $) are the distances from the Sun
to the source (Earth), respectively (TEGP 7.2~\cite{tegp}).  For a ray
which passes close to the Sun,
\begin{equation} \label{E32}
    \delta t \approx {1 \over 2} (1+ \gamma )
    [240-20 \ln (d^2 /r)] {\rm\ } \mu {\rm s},
\end{equation}
where $d$ is the distance of closest approach of the ray in solar
radii, and $r$ is the distance of the planet or satellite from the
Sun, in astronomical units.

In the two decades following Irwin Shapiro's 1964 discovery of
this effect as a theoretical consequence of GR,
several high-precision measurements were made
using radar ranging to targets passing through superior
conjunction.  Since one does not have access to a ``Newtonian''
signal against which to compare the round-trip travel time of the
observed signal, it is necessary to do a differential measurement
of the variations in round-trip travel times as the target passes
through superior conjunction, and to look for the logarithmic
behavior of Eq.~(\ref{E32}).  In order to do this accurately however,
one must take into account the variations in round-trip travel
time due to the orbital motion of the target relative to the
Earth. This is done by using radar-ranging (and possibly other)
data on the target taken when it is far from superior conjunction
(i.e.\ when the time-delay term is negligible) to determine
an accurate ephemeris for the target, using the ephemeris to
predict the PPN  coordinate trajectory ${\bf x}_e (t)$ near
superior conjunction, then combining that trajectory with the
trajectory of the Earth ${\bf x}_\oplus (t)$ to determine the
Newtonian round-trip time and the logarithmic term in Eq.~(\ref{E32}).
The resulting predicted round-trip travel times in terms of the
unknown coefficient ${1 \over 2}(1+ \gamma)$
are then fit to the measured travel times using the method
of least-squares, and an estimate obtained for
${1 \over 2}(1+ \gamma)$.

The targets employed included
planets, such as Mercury or Venus, used as passive reflectors of
the radar signals (``passive radar''), and
artificial satellites, such as Mariners~6 and 7, Voyager~2,
the Viking
Mars landers and orbiters, and the Cassini spacecraft to Saturn,
used as
active retransmitters of the radar signals (``active radar'').

The results for the coefficient ${1 \over 2}(1+ \gamma)$
of all radar time-delay measurements
performed to date (including a measurement of the one-way time delay
of signals from the millisecond pulsar PSR 1937+21)
are shown in Figure~\ref{gammavalues} (see TEGP 7.2~\cite{tegp} 
for discussion and references).  The 1976 Viking experiment resulted in a
0.1 percent measurement~\cite{reasenberg}.  

A significant improvement was reported in 2003 
from Doppler tracking of the Cassini
spacecraft while it was on its way to Saturn~\cite{bertotti03}, with a
result $\gamma -1 = (2.1 \pm 2.3) \times 10^{-5}$.  This was made possible
by the ability to do Doppler measurements using both X-band (7175 MHz) and
Ka-band (34316 MHz) radar, thereby significantly reducing the dispersive
effects of the
solar corona.  In addition, the 2002 superior conjunction of Cassini was
particularly favorable: with the spacecraft at 8.43 astronomical units from
the Sun, the distance of closest approach of the radar signals to the Sun
was only $1.6 \, R_\odot$.

  From the results of the Cassini experiment, we
can conclude that the coefficient
${1 \over 2}(1+ \gamma)$
must be within at most 0.0012~percent of unity.
Scalar-tensor theories must have $\omega > 40000$ to be compatible with
this constraint.

\subsubsection{Shapiro time delay and the speed of gravity}
\label{gravityspeed}

In 2001, Kopeikin~\cite{kopeikin01}
suggested that a measurement of the time delay of
light from a quasar as the light passed
by the planet Jupiter could be used to measure the speed
of the gravitational interaction. He argued that, since Jupiter is moving
relative to the solar system, and since gravity propagates with a finite
speed, the gravitational field experienced by the light ray should be
affected by gravity's speed, since the field experienced at one time depends on
the location of the source a short time earlier, depending on how fast
gravity propagates. According to his calculations, there should be a
post$^{1/2}$-Newtonian
correction to the normal Shapiro time-delay formula (\ref{E31})
which depends on the velocity of Jupiter and on the velocity of gravity.
On September 8, 2002, Jupiter passed almost in
front of a quasar, and Kopeikin and Fomalont  made precise
measurements of the Shapiro
delay with picosecond timing accuracy, and claimed to have measured the
correction term
to about 20 percent~\cite{fomalont03,kopeikin02,kopeikin03,kopeikin04}. 

However, several authors pointed out that this 1.5PN effect does {\em not}
depend on the speed of propagation of gravity, but rather only depends on
the speed of light~\cite{asada02,willspeed03,samuel03,carlip04,samuel04}.  
Intuitively, if one is working to only first order in $v/c$, 
then all that counts is
the uniform motion of the planet, Jupiter (its acceleration about the Sun
contributes a higher-order, unmeasurably small effect). But if that is the
case, then the principle of relativity says that one can view things from the
rest frame of Jupiter. In this frame, Jupiter's gravitational field is
static, and the speed of propagation of gravity is irrelevant.
A detailed post-Newtonian
calculation of the effect was done 
using a variant of the PPN framework, 
in a class of theories
in which the speed of gravity could be different from that
of light~\cite{willspeed03}, 
and found explicitly that, at first order in $v/c$, the effect
depends on the speed of light, not the speed of gravity, in line with
intuition.  Effects dependent upon the speed of gravity show up only at
higher order in $v/c$.  Kopeikin gave a number of arguments in opposition to
this interpretation~\cite{kopeikin04,kopeikin05a,kopeikin05b,kopeikin05c}.   
On the other hand, the $v/c$ correction term {\it does}
show a dependence on the PPN parameter $\alpha_1$, which could be non-zero
in theories of gravity with a differing speed $c_g$ of gravity (see Eq. (7)
of \cite{willspeed03}).   But existing tight bounds on $\alpha_1$ from other
experiments (Table
\ref{ppnlimits}) already far exceed the capability of the Jupiter VLBI
experiment.

\begin{table}[t]
  \begin{center}
    \begin{tabular}{@{}clcl@{}}
      \hline \hline
      Parameter & Effect & Limit & Remarks \\
      \hline\hline
      $\gamma-1$ & time delay & $2.3 \times 10^{-5}$ & Cassini tracking \\
       & light deflection & $4 \times 10^{-4}$ & VLBI \\
      $\beta-1$ & perihelion shift & $3 \times 10^{-3}$ &
      $J_2=10^{-7}$ from helioseismology \\
      & Nordtvedt effect & $2.3 \times 10^{-4}$ & $\eta_{\rm N}=4\beta-\gamma-3$
      assumed \\
      $\xi$ & Earth tides & $10^{-3}$ & gravimeter data \\
      $\alpha_1$ & orbital polarization & $10^{-4}$ & Lunar laser
      ranging \\
      & & $2 \times 10^{-4}$ & PSR J2317+1439\\
      $\alpha_2$ & spin precession & $4 \times 10^{-7}$ & solar alignment
      with ecliptic \\
      $\alpha_3$ & pulsar acceleration & $4 \times 10^{-20}$
      & pulsar $\dot P$ statistics \\
      $\eta_{\rm N}$ & Nordtvedt effect & $9 \times 10^{-4}$ & lunar laser ranging \\
      $\zeta_1$ & --  & $2 \times 10^{-2}$ & combined PPN  bounds \\
      $\zeta_2$ & binary acceleration & $4 \times 10^{-5}$ &
      $\ddot P_{\rm p}$ for PSR 1913+16 \\
      $\zeta_3$ & Newton's 3rd law & $10^{-8}$ & Lunar acceleration \\
      $\zeta_4$ & --  & --  & not independent [Eq. (\ref{E37})]\\
      \hline \hline
    \end{tabular}
    \caption{\it Current limits on the PPN parameters. Here $\eta_{\rm N}$ is
      a combination of other parameters given by $ \eta_{\rm N} = 4\beta
      -\gamma - 3 - 10 \xi /3 -\alpha_1 +2\alpha_2 /3 - 2\zeta_1 /3 -
      \zeta_2 /3 $.}
    \label{ppnlimits}
  \end{center}
\end{table}

%%%%%%%%%%%%%%%%%%%%%%%%%%%%%%%%%%%%%%%%%%%%%%%%%%%%%%%%%%%%%%%%%%%%%%%%%%%%%%%%%%%
%%%%%%%%%%%%%%%%%%%%%%%%%%%%%%%%%%%%%%%%%%%%%%%%%%%%%%%%%%%%%%%%%%%%%%%%%%%%%%%%%%%

\subsection{The perihelion shift of Mercury}
\label{perihelion}

The explanation of the anomalous perihelion shift of Mercury's
orbit was another of the triumphs of GR.  This had
been an unsolved problem in celestial mechanics for over half a
century, since the announcement by Le Verrier in 1859 that, after
the perturbing effects of the planets on Mercury's orbit had been
accounted for, and after the effect of the precession of the
equinoxes on the astronomical coordinate system had been
subtracted, there remained in the data an unexplained advance
in the perihelion of Mercury.  The modern value for this
discrepancy is 43 arcseconds per century.  A number of {\it ad
hoc} proposals were made in an attempt to account for this
excess, including, among others, the existence of a new planet
Vulcan near the Sun, a ring of planetoids, a solar quadrupole
moment and a deviation from the inverse-square law of gravitation,
but none was successful.  General relativity accounted
for the anomalous shift in a natural way without disturbing the
agreement with other planetary observations.

The predicted advance per orbit $\Delta \tilde \omega$, including
both relativistic PPN  contributions and the Newtonian
contribution resulting from a
possible solar quadru\-pole moment, is given by
\begin{equation}\label{E33}
 \Delta \tilde \omega
 = (6 \pi m/p)\left[ {1 \over 3} (2+2 \gamma - \beta )
 + {1 \over 6}~
     (2 \alpha_1 - \alpha_2 + \alpha_3 +2 \zeta_2 ) \mu /m
 + J_2 (R^2 /2mp)\right],
\end{equation}
where $m \equiv m_1 + m_2$ and $\mu  \equiv m_1 m_2 /m$
are the total mass and reduced mass of the two-body system
respectively; $p \equiv a(1-e^2 )$ is the semi-latus rectum of
the orbit, with the semi-major axis $a$ and the eccentricity $e$; $R$ is
the mean radius of the oblate body; and $J_2$ is a
dimensionless measure of its quadrupole moment, given by
$J_2 = (C-A)/m_1 R^2$, where $C$ and $A$ are the
moments of inertia about the body's rotation and equatorial
axes, respectively (for details of the derivation see TEGP 7.3~\cite{tegp}).
We have ignored preferred-frame and galaxy-induced
contributions to  $\Delta \tilde \omega$; these are discussed in
TEGP 8.3~\cite{tegp}.

The first term in Eq.~(\ref{E33}) is the classical relativistic
perihelion shift, which depends upon the PPN  parameters $\gamma$
and $\beta$.  The second term depends upon the ratio of the masses
of the two bodies; it is zero in any fully conservative theory of
gravity
($ \alpha_1 \equiv \alpha_2 \equiv \alpha_3 \equiv \zeta_2 \equiv 0$);
it is also negligible for Mercury, since
$\mu /m \approx m_{\rm Merc} /m_\odot \approx 2 \times 10^{-7}$.
We shall drop this term henceforth.  

The third term depends upon
the solar quadrupole moment $J_2$.  For a Sun that rotates
uniformly with its observed surface angular velocity,  so that
the quadrupole moment is produced by centrifugal flattening, one
may estimate $J_2$ to be
$\sim 1 \times 10^{-7}$.  This actually agrees reasonably well with
values inferred from rotating solar models that are in accord with
observations of the normal modes of solar oscillations
(helioseismology); the latest inversions of helioseismology data give $J_2 =
(2.2 \pm 0.1) \times 10^{-7}$~\cite{paterno96,pijpers98,roxburgh01,mecheri04}.
Substituting standard orbital elements and physical
constants for Mercury and the Sun we obtain the rate of
perihelion shift $\dot {\tilde \omega}$, in seconds of arc per century,
\begin{equation}\label{E34}
\dot {\tilde \omega}= 42.''98
\left[ {1 \over 3} (2+2 \gamma - \beta )
                 + 3 \times 10^{-4} (J_2 /10^{-7} ) \right].
\end{equation}
Now, the measured perihelion shift of Mercury is known
accurately:  after the perturbing effects of the other planets have been
accounted for, the excess shift is known to about 0.1 percent
from radar observations of Mercury between 1966 and 1990~\cite{shapiroGR12}.
Analysis of data taken since 1990 could improve the accuracy.
The solar oblateness
effect is smaller than the observational error, so we obtain the PPN
bound $| 2\gamma -\beta-1 | < 3 \times 10^{-3}$.

%%%%%%%%%%%%%%%%%%%%%%%%%%%%%%%%%%%%%%%%%%%%%%%%%%%%%%%%%%%%%%%%%%%%%%%%%%%%%%%%%%%
%%%%%%%%%%%%%%%%%%%%%%%%%%%%%%%%%%%%%%%%%%%%%%%%%%%%%%%%%%%%%%%%%%%%%%%%%%%%%%%%%%%

\subsection{Tests of the strong equivalence principle}
\label{septests}

The next class of solar-system experiments that test relativistic
gravitational effects may be called tests of the strong
equivalence principle (SEP).  In Sec.~\ref{sep} we pointed out that many metric
theories of gravity (perhaps all except GR) can be
expected to violate one or more aspects of SEP.  Among the
testable violations of SEP are a
violation of the weak equivalence principle for gravitating bodies
that leads to perturbations in the Earth-Moon
orbit; preferred-location and preferred-frame effects in the
locally measured gravitational constant that could produce
observable geophysical effects; and possible variations in the
gravitational constant over cosmological timescales.

%%%%%%%%%%%%%%%%%%%%%%%%%%%%%%%%%%%%%%%%%%%%%%%%%%%%%%%%%%%%%%%%%%%%%%%%%%%%%%%%%%%

\subsubsection{The Nordtvedt effect and the lunar E\"otv\"os experiment}
\label{Nordtvedteffect}

In a pioneering calculation using his early form of the PPN
formalism, Nord\-tvedt~\cite{nordtvedt1} showed that many metric theories
of gravity predict that massive bodies violate the weak
equivalence principle -- that is, fall with different
accelerations depending on their gravitational self-energy.  
Dicke~\cite{dicke_2} argued that such an effect would occur in theories with
a spatially
varying gravitational constant, such as scalar-tensor
gravity.  For a
spherically symmetric body, the acceleration from rest in an
external gravitational potential $U$ has the form
{\setlength{\arraycolsep}{0.14 em}
\begin{eqnarray}
{\bf a} &=& (m_{\rm p}/m) \nabla U, \nonumber \\
m_{\rm p}/m &=& 1- \eta_{\rm N} (E_{\rm g}/m), \label{E35} \\
    \eta_{\rm N} 
 &=& 4 \beta - \gamma -3 - {10 \over 3} \xi
        - \alpha_1 +  {2 \over 3} \alpha_2
        -  {2 \over 3} \zeta_1
        -  {1 \over 3} \zeta_2, \nonumber
\end{eqnarray}}
where $E_{\rm g}$ is the negative of the gravitational self-energy
of the body ($E_{\rm g} >0$).  This violation of the massive-body
equivalence principle is known as the ``Nordtvedt effect''.  The
effect is absent in GR ($ \eta_{\rm N} = 0$) but present
in scalar-tensor theory ($ \eta_{\rm N} = 1/(2+ \omega )+4\Lambda$).  The existence
of the Nordtvedt effect does not violate the results of laboratory
E\"otv\"os experiments, since for laboratory-sized objects,
$E_{\rm g} /m \le 10^{-27}$, far below the sensitivity of
current or future experiments.  However, for astronomical bodies,
$E_{\rm g} /m$ may be significant ($3.6 \times 10^{-6}$ for the Sun, $10^{-8}$
for Jupiter,
$4.6 \times 10^{-10}$  for the Earth, $0.2 \times 10^{-10}$
for the Moon).  If the Nordtvedt effect is present
($\eta_{\rm N} \ne 0$) then the Earth should fall toward the Sun with a
slightly different acceleration than the Moon.  This perturbation
in the Earth-Moon orbit leads to a polarization of the orbit that
is directed toward the Sun as it moves around the Earth-Moon
system, as seen from Earth.  This polarization represents a
perturbation in the Earth-Moon distance of the form
\begin{equation}\label{E36}
  \delta r = 13.1 \, \eta_{\rm N} \cos( \omega_0 -\omega_{\rm s} )t {\rm\ [m]},
\end{equation}
where $\omega_0$ and $\omega_{\rm s}$ are the angular frequencies
of the orbits of the Moon and Sun around the Earth (see
TEGP 8.1~\cite{tegp} for detailed derivations and references; for
improved
calculations of the numerical coefficient, see~\cite{Nordtvedt95,DamourVokrou96}).

Since August 1969, when the first successful acquisition was made
of a laser signal reflected from the Apollo~11 retroreflector on
the Moon, the lunar laser-ranging experiment (LLR) has made
regular measurements of the round-trip travel times of laser
pulses between a network of observatories
and the lunar retroreflectors, with accuracies that are
at the 50\ ps (1\ cm) level,
and that may soon
approach 5\ ps (1\ mm).  These measurements are fit
using the method of least-squares to a theoretical model for the
lunar motion that takes into account perturbations due to the Sun
and the other planets, tidal interactions, and post-Newtonian
gravitational effects.  The predicted round-trip travel times
between retroreflector and telescope also take into account the
librations of the Moon, the orientation of the Earth, the
location of the observatories, and atmospheric effects on the
signal propagation.  The ``Nordtvedt'' parameter $\eta_{\rm N}$ along with
several other important parameters of the model are then
estimated in the least-squares method.

Numerous ongoing analyses of the data find no evidence, within
experimental uncertainty, for the Nordtvedt 
effect~\cite{williams04,williams04ijmp} 
(for earlier results
see~\cite{Dickey,Williams,MullerMG}).  
These results represent a limit on a possible violation of WEP for
massive bodies of about 
1.4 parts in $10^{13}$ (compare Figure~\ref{wepfig}).  

However, 
at this level of precision, one cannot regard the results of lunar laser
ranging as a ``clean'' test of SEP until one eliminates the
possibility of a compensating violation of WEP for the two bodies,
because the chemical compositions of the Earth
and Moon differ in the relative fractions of iron and silicates.  To
this end, the E{\"o}t-Wash group carried out an improved test of WEP
for laboratory bodies whose chemical compositions mimic that of the
Earth and Moon.  The resulting bound of 1.4 parts 
in $10^{13}$~\cite{baessler99,adelberger01} from composition effects
reduces the ambiguity in 
the lunar laser ranging
bound, and establishes the firm SEP test
at the level of about 2 parts in $10^{13}$.
These results
can be summarized by the Nordtvedt parameter bound $| \eta_{\rm N} | = (4.4 \pm 4.5) \times 10^{-4}$.

In the future, the Apache Point Observatory for Lunar Laser-ranging
Operation (APOLLO) project, a joint effort by researchers from the
Universities of Washington, Seattle, and California, San Diego, plans
to
use enhanced laser and telescope technology, together with a good,
high-altitude site in New Mexico, to improve the lunar laser-ranging
bound by as much as an order of magnitude~\cite{williams04ijmp}.

In GR, the Nordtvedt effect vanishes; at the level of
several centimeters and below,
a number of non-null general relativistic effects
should be present~\cite{Nordtvedt95}.

Tests of the Nordtvedt effect for neutron stars
have also been carried out using
a class of systems known as wide-orbit binary millisecond pulsars (WBMSP),
which are 
pulsar-white-dwarf binary systems with small orbital eccentricities.
In the gravitational field
of the galaxy, a non-zero Nordtvedt effect can induce an apparent anomalous
eccentricity pointed toward the galactic center~\cite{DamourSchaefer91},
which can be bounded using statistical methods, given enough WBMSPs
(see~\cite{StairsLRR} for a review and references).  Using
data from 21 WBMSPs, including recently discovered highly circular systems,
Stairs {\it et al.}~\cite{stairs05} obtained the bound $\Delta < 5.6 \times
10^{-3}$, where $\Delta = \eta_{\rm N} (E_g/M)_{\rm NS}$.  Because 
$(E_g/M)_{\rm NS} \sim 0.1$ for typical neutron stars, this bound does not
compete with the bound on $\eta_{\rm N}$ from Lunar laser ranging; on the
other hand, it does test SEP in the strong-field regime because of the
presence of the neutron stars.

%%%%%%%%%%%%%%%%%%%%%%%%%%%%%%%%%%%%%%%%%%%%%%%%%%%%%%%%%%%%%%%%%%%%%%%%%%%%%%%%%%%

\subsubsection{Preferred-frame and preferred-location effects}
\label{preferred}

Some theories of gravity violate SEP by predicting that the
outcomes of local gravitational experiments may depend on the
velocity of the laboratory relative to the mean rest frame of the
universe (preferred-frame effects) or on the location of the
laboratory relative to a nearby gravitating body
(preferred-location effects).  In the post-Newtonian limit,
preferred-frame effects are governed by the values of the PPN
parameters $\alpha_1$, $\alpha_2$, and $\alpha_3$, and some
preferred-location effects are governed by $\xi$ (see Table~\ref{ppnmeaning}).

The most important such effects are variations and anisotropies
in the locally-measured value of the gravitational constant, which
lead to anomalous Earth tides and variations in the Earth's
rotation rate; anomalous contributions to the
orbital dynamics of planets and the Moon; self-accelerations of
pulsars, and anomalous torques on the Sun that
would cause its spin axis to be randomly oriented relative to the
ecliptic (see TEGP 8.2, 8.3, 9.3 and 14.3~(c)~\cite{tegp}).  
An bound on $\alpha_3$ of $4
\times 10^{-20}$ from the period derivatives of 21 millisecond pulsars
was reported in~\cite{Bell,stairs05};  improved bounds on
$\alpha_1$ were achieved using lunar laser ranging data~\cite{MullerPRD},
and using
observations of the circular binary
orbit of the pulsar J2317+1439 \cite{Bell2}.
Negative searches for these effects have
produced strong constraints on the PPN  parameters (Table~\ref{ppnlimits}).

%%%%%%%%%%%%%%%%%%%%%%%%%%%%%%%%%%%%%%%%%%%%%%%%%%%%%%%%%%%%%%%%%%%%%%%%%%%%%%%%%%%

\subsubsection{Constancy of the Newtonian gravitational constant}
\label{bigG}

Most theories of gravity that violate SEP predict that the locally
measured Newtonian gravitational constant may vary with time as
the universe evolves.  For the scalar-tensor theories listed in
Table~\ref{ppnvalues},
the predictions for $\dot G/G$ can be written in terms of time
derivatives of the asymptotic scalar field.
Where $G$
does change with cosmic evolution, its rate of variation should
be of the order of the expansion rate of the universe,
i.e.\ $\dot G/G \sim H_0$, where $H_0$ is the Hubble expansion parameter
and is given by
$H_0 = 100 \, h {\rm\ km\ s^{-1}\ Mpc^{-1}} = 1.02 \times 10^{-10}\,
h\, {\rm\ yr^{-1}}$,
where current observations of the expansion of the
universe give $h \approx 0.73 \pm 0.03$.

Several observational constraints can be placed on $\dot G/G$, one kind
coming from bounding the present rate of variation, another from bounding a
difference between the present value and a past value.
The first type of bound typically comes from lunar laser-ranging measurements,
planetary
radar-ranging measurements, and pulsar timing data.
The second type comes from
studies of the evolution of the Sun, stars and the Earth,
big-bang nucleosynthesis, and 
analyses of ancient
eclipse data.
Recent
results are shown in Table~\ref{Gdottable}.

\begin{table}[t]
  \begin{center}
    \begin{tabular}{lr@{\,}c@{\,}ll}
      \hline \hline
      Method &\multicolumn{3}{c}
      {$\rule{0 cm}{0.4 cm}\dot G/G (10^{-13} {\rm\ yr}^{-1})$} &Reference\\
      \hline \hline
      Lunar Laser Ranging & 4 & $\pm$ & 9 &\cite{williams04}\\
      Binary Pulsar $1913+16$& \,\qquad 40 & $\pm$ & 50 \qquad\,
      &\cite{kaspi94}\\
      Helioseismology & 0 & $\pm$ & 16 &\cite{guenther98}\\
      Big Bang nucleosynthesis & 0 & $\pm$ & 4 &\cite{copi04,bambi04}\\
      \hline \hline
    \end{tabular}
    \caption{\it Constancy of the gravitational constant. For binary
      pulsar data, the bounds are dependent upon the theory of gravity in
      the strong-field regime and on neutron star equation of state.
      Big-bang nucleosynthesis bounds assume specific form for time
      dependence of $G$.}
    \label{Gdottable}
  \end{center}
\end{table}

The best limits on a current $\dot G/G$ 
come from
lunar laser ranging measurements (for earlier results see
~\cite{Dickey,Williams,MullerMG}).
These have largely supplanted earlier bounds from ranging to the 1976 Viking
landers (see TEGP, 14.3~(c)~\cite{tegp}), which were limited 
by uncertain knowledge  of the masses and orbits of asteroids.  However,
improvements in knowledge of the asteroid belt, combined with continuing
radar observations of planets and spacecraft, notably the Mars Global
Surveyor (1998-2003) and Mars Odyssey (2002 - present), may enable a bound 
on $\dot G/G$ at the level of a part in $10^{13}$ per year.  For an initial
analysis along these lines, see~\cite{pitjeva05}.  
It has
been suggested that radar observations of the planned 2012 Bepi-Colombo 
Mercury orbiter mission over a
two-year integration with 6\ cm rms accuracy in range could yield
$\Delta (\dot G/G) < 10^{-13} {\rm\ yr}^{-1}$; an eight-year mission could
improve this by a factor 15~\cite{milani02,ashby05}.

Although bounds on $\dot G/G$ from solar-system measurements can be
correctly
obtained in a phenomenological manner through the simple expedient of
replacing $G$ by $G_0 + {\dot G}_0 (t - t_0 )$ in
Newton's equations of motion, the same does not hold true for pulsar
and binary pulsar timing measurements.  The reason is that, in theories
of gravity that violate SEP, such as scalar-tensor theories,
the ``mass'' and moment of inertia of a
gravitationally bound body may vary with variation in $G$.  Because
neutron stars are highly relativistic, the fractional variation in
these quantities can be comparable to $\Delta G/G$, the precise
variation depending both on the equation of state of neutron star
matter and on the theory of gravity in the strong-field regime.  The
variation in the moment of inertia affects the spin rate of the pulsar,
while the variation in the mass can affect the orbital period in a
manner that can subtract from the direct effect of a variation in $G$,
given by $\dot P_{\rm b} /P_{\rm b} =-{2} \dot G/G$~\cite{nordtvedt3}.  Thus,
the bounds quoted in Table~\ref{Gdottable}
for the binary pulsar PSR 1913+16 and others~\cite{kaspi94} 
(see also~\cite{DamourTaylor91})
are theory-dependent and must be treated as merely
suggestive.

In a similar manner, bounds from helioseismology and big-bang
nucleosynthesis (BBN) assume a model for the evolution of $G$ over the
multi-billion year time spans involved.  For example, the concordance of
predictions for 
light elements produced around 3 minutes after the big bang with the
abundances observed indicate that $G$ then was within 20 percent of $G$
today.  Assuming a power-law variation of $G \sim t^{-\alpha}$, then yields
a bound on $\dot G/G$ today shown in Table \ref{Gdottable}.

%%%%%%%%%%%%%%%%%%%%%%%%%%%%%%%%%%%%%%%%%%%%%%%%%%%%%%%%%%%%%%%%%%%%%%%%%%%%%%%%%%%
%%%%%%%%%%%%%%%%%%%%%%%%%%%%%%%%%%%%%%%%%%%%%%%%%%%%%%%%%%%%%%%%%%%%%%%%%%%%%%%%%%%

\subsection{Other tests of post-Newtonian gravity}
\label{othertests}

%%%%%%%%%%%%%%%%%%%%%%%%%%%%%%%%%%%%%%%%%%%%%%%%%%%%%%%%%%%%%%%%%%%%%%%%%%%%%%%%%%%

\subsubsection{Search for gravitomagnetism}
\label{gravitomagnetism}

According to GR, moving or rotating matter should
produce a contribution to the gravitational field that is the analogue
of the magnetic field of a moving charge or a magnetic dipole.  In
particular, one can view the $g_{0i}$ part of the PPN metric (Box 2) as an
analogue of the vector potential of electrodynamics.  In a suitable gauge,
and dropping the preferred-frame terms,
it can be written
\begin{equation}
g_{0i} = -\frac{1}{2} (4\gamma +4 + \alpha_1) V_i \,.
\label{gravitomag}
\end{equation}
At PN order, this contributes a Lorentz-type acceleration ${\bf v} \times
{\bf B}_g$ to the equation of motion, where the gravitomagnetic field 
$B_g$ is given by $B_g = \nabla \times (g_{0i}{\bf e}^i)$.

Gravitomagnetism plays a role in a variety of measured
relativistic effects involving moving material sources, such as the
Earth-Moon system and binary pulsar systems.
Nordtvedt~\cite{nordtvedt88a,nordtvedt88b} has argued that, if the 
gravitomagnetic potential (\ref{gravitomag}) were turned off, then there
would be anomalous orbital effects in LLR and binary pulsar data.  

Rotation also produces a gravitomagnetic effect, since for a rotating body,
${\bf V} = -\frac{1}{2} {\bf x} \times {\bf J}/r^3$, where ${\bf J}$ is the
angular momentum of the body.  The result is a ``dragging of inertial
frames'' around the body, 
also called the Lense-Thirring effect.  A consequence
is a precession of a gyroscope's spin $\bf S$ according to 
\begin{equation}\label{E42}
\frac{d {\bf S} }{d \tau} = \mbox{\boldmath$\Omega$}_{\rm LT} \times {\bf S} \,,
\qquad
\mbox{\boldmath$\Omega$}_{\rm LT} = -{1 \over 2}\left(1+\gamma + {1 \over 4}\alpha_1 \right)
          \frac{{\bf J}-3{\bf n}({\bf n} \cdot {\bf J})}{r^3} \,,
\end{equation}
where $\bf n$ is a
unit radial vector, and $r$ is the distance from the center of the
body (TEGP 9.1~\cite{tegp}).           

The Relativity
Gyroscope Experiment (Gravity Probe B or GPB)
at Stanford University, in collaboration with
NASA  and Lockheed-Martin Corporation~\cite{gpbwebsite}, recently completed
a space mission to detect this frame-dragging or Lense-Thirring precession,
along with the ``geodetic'' precession 
(see Sec. \ref{geodeticprecession}).
A set of four superconducting-niobium-coated,
spherical quartz gyroscopes were flown in a 
polar Earth orbit (642 km mean altitude, 0.0014 eccentricity),
and the precessions of the gyroscopes relative to a distant guide star (HR
8703, IM Pegasi)
were measured.  
For the given orbit, the predicted
secular angular precession of the gyroscopes is in a direction perpendicular
to the orbital plane at a rate
${1 \over 2}(1+\gamma + {1 \over 4}\alpha_1 ) \times
41 \times 10^{-3}$ arcsec/yr.
The accuracy goal of the experiment is about 0.5\ milliarcseconds
per year.  The spacecraft was launched on April 20, 2004, and the mission
ended in September 2005, as scheduled, when the remaining
liquid helium boiled off.  

It is too early to know whether the
relativistic precessions were measured in the amount predicted by GR,
because an important calibration of the instrument exploits the
effect of the aberration of starlight on the pointing of the on-board
telescope toward the guide star, and completing this
calibration required
the full mission data set.
In addition, part of the measured effect includes the motion of
the guide star relative to distant inertial frames.  This  was measured
before, during and after the mission
separately by radio astronomers at Harvard/SAO and elsewhere using
VLBI, and the results of those
measurements were to be strictly embargoed until the GPB team has completed its
analysis of the gyro data.  Final results from the experiment are expected
in 2006.

Another way to look for frame-dragging is to
measure the precession of orbital planes of bodies circling a rotating
body.  One implementation of this idea is to
measure the relative precession, at about 31 milliarcseconds per year,
of the line of nodes of a pair
of laser-ranged geodynamics satellites (LAGEOS), ideally with supplementary
inclination angles; the inclinations must be supplementary in order
to cancel the dominant (126 degrees per year)
nodal precession caused by the Earth's
Newtonian gravitational multipole moments.  Unfortunately, the two
existing LAGEOS satellites are not in appropriately inclined orbits,
and no concrete plans exist at present to launch a third satellite in a
supplementary orbit.  Nevertheless, Ciufolini and Pavlis~\cite{ciufolini04} 
combined 
nodal precession data
from LAGEOS I and II with improved models for the Earth's multipole moments
provided by 
two recent orbiting geodesy satellites, Europe's
CHAMP (Challenging Minisatellite Payload)
and NASA's GRACE (Gravity Recovery and Climate Experiment),
and reported a 
5 -- 10 percent confirmation of GR.  In earlier reports, Ciufolini {\em et
al.} had reported tests at the the 20 -- 30 percent level, without the
benefit of the GRACE/CHAMP data~\cite{ciufolini97,ciufolini98,ciufolini00}.
Some authors stressed the importance of adequately assessing systematic
errors in the LAGEOS data~\cite{ries,iorio05}.

%%%%%%%%%%%%%%%%%%%%%%%%%%%%%%%%%%%%%%%%%%%%%%%%%%%%

\subsubsection{Geodetic precession}
\label{geodeticprecession}

A gyroscope moving through curved spacetime suffers a precession of
its spin axis given by
\begin{equation}\label{E41}
d {\bf S} /d \tau = \mbox{\boldmath$\Omega$}_{\rm G} \times {\bf S},
 \qquad
\mbox{\boldmath$\Omega$}_{\rm G} = \left(\gamma + {1 \over 2} \right) {\bf v}
           \times \nabla U,
\end{equation}
where $\bf {v}$ is the velocity of the gyroscope, and $U$ is the
Newtonian gravitational potential of the source (TEGP 9.1~\cite{tegp}).
The Earth-Moon system can be considered as a ``gyroscope'', with its axis
perpendicular to the orbital plane.  The predicted precession is about
2 arcseconds per century, an effect first calculated by de Sitter.
This effect has been measured to about 0.6 percent using lunar laser
ranging data~\cite{Dickey,Williams,williams04}.

For the GPB gyroscopes orbiting the Earth, the precession is 6.6 arcseconds
per year.  A goal of GPB is to
measure this effect to $8 \times 10^{-5}$; if
achieved, this could bound the parameter
$\gamma$ to a part in $10^4$, not competitive with the Cassini bound.

%%%%%%%%%%%%%%%%%%%%%%%%%%%%%%%%%%%%%%%%%%%%%%%%%%%%%%%%%%%%%%%%%%%%%%%%%%%%%%%%%%%

\subsubsection{Tests of post-Newtonian conservation laws}
\label{conservation}

Of the five ``conservation law'' PPN  parameters
$\zeta_1$, $\zeta_2$, $\zeta_3$, $\zeta_4$,
and $\alpha_3$, only three, $\zeta_2$, $\zeta_3$ and $\alpha_3$,
have been constrained directly with any precision;
$\zeta_1$ is constrained
indirectly through its appearance in the Nordtvedt effect parameter
$\eta_{\rm N}$, Eq.~(\ref{E35}).  There is strong theoretical
evidence that $\zeta_4$, which is related to the gravity generated by
fluid pressure, is not really an independent parameter -- in any
reasonable theory of gravity there should be a connection between the
gravity produced by kinetic energy ($\rho v^2$), internal energy ($\rho
\Pi$), and pressure ($p$).
From such considerations, there follows~\cite{Will76}
the additional theoretical
constraint
\begin{equation}\label{E37}
6\zeta_4= 3\alpha_3+2\zeta_1-3\zeta_3.
\end{equation}

A non-zero value for any of these parameters would result in a
violation of conservation of momentum, or of Newton's third law
in gravitating systems.  An alternative statement of Newton's
third law for gravitating systems is that the ``active gravitational mass'',
that is the mass that determines the gravitational potential exhibited
by a body, should equal the ``passive gravitational mass'',
the mass that determines the force on a body in a gravitational field.
Such an equality guarantees the equality of action and reaction and
of conservation of momentum, at least in the Newtonian limit.

A classic test of Newton's third law for gravitating systems was
carried out in 1968 by Kreuzer, in which the gravitational attraction
of fluorine and bromine were compared to a precision of
5\ parts in $10^5$.

A remarkable planetary test was reported by
Bartlett and van Buren~\cite{bartlett}.  They noted that current understanding
of the structure of the Moon involves an iron-rich, aluminum-poor
mantle whose center of mass is offset about 10 km from the center of
mass of an aluminum-rich, iron-poor crust.  The direction of offset
is toward the Earth, about $14 ^\circ$ to the east of the Earth-Moon line.
Such a model accounts for the basaltic maria which face the Earth,
and the aluminum-rich highlands on the Moon's far side, and for a 2\ km
offset between the observed center of mass and center of figure for
the Moon.  Because of this asymmetry, a violation of Newton's third
law for aluminum and iron would result in a momentum non-conserving
self-force on the Moon, whose component along the orbital direction
would contribute to the secular acceleration of the lunar orbit.
Improved knowledge of the lunar orbit through lunar laser ranging, and
a better understanding of tidal effects in the Earth-Moon system
(which also contribute to the secular acceleration) through satellite
data, severely limit any anomalous secular acceleration, with
the resulting limit
\begin{equation}\label{E38}
\left| {{(m_{\rm A} / m_{\rm P} )_{\rm Al} - (m_{\rm A} /m_{\rm P})_{\rm Fe}}
       \over
       {(m_{\rm A} /m_{\rm P} )_{\rm Fe}}}
\right| <4 \times 10^{-12}.
\end{equation}
According to the PPN  formalism, in a theory of gravity that violates
conservation of momentum, but that obeys the constraint of Eq.~(\ref{E37}),
the electrostatic binding energy $E_{\rm e}$
of an atomic nucleus could make a contribution to the ratio of active to
passive mass of the form
\begin{equation}\label{E39}
     m_{\rm A} = m_{\rm P}
 + {1 \over 2} \zeta_3 E_{\rm e} \,.
\end{equation}
The resulting limit on $\zeta_3$ from the lunar experiment
is $\zeta_3 < 1 \times 10^{-8}$ (TEGP 9.2, 14.3~(d)~\cite{tegp}).
Nordtvedt~\cite{nordtvedt01} has examined whether this bound could be
improved by considering the asymmetric distribution of ocean water on
Earth.

Another consequence of a violation of conservation of momentum is a
self-accel\-er\-at\-ion of the center of mass of a binary stellar system,
given by
\begin{equation}\label{E40}
{\bf a}_{\rm CM} = - {1 \over 2} (\zeta_2+\alpha_3) {m \over a^2}{\mu
\over a}{{\delta m} \over m} {e \over {(1-e^2)^{3/2}}} {\bf n}_{\rm P},
\end{equation}
where $\delta m = m_1-m_2$, $a$ is the semi-major axis, and ${\bf n}_{\rm P}$
is a unit vector directed from the center of mass to the point of
periastron of $m_1$ (TEGP 9.3~\cite{tegp}).
A consequence of this acceleration would be non-vanishing values for
$d^2 P/dt^2$, where $P$ denotes the period of any intrinsic process in
the system (orbit, spectra,
pulsar periods).  The observed upper limit on $d^2 P_{\rm p}
/dt^2$ of the binary pulsar PSR 1913+16 places a
strong constraint on such an effect, resulting in the bound $|
\alpha_3 + \zeta_2 |<4 \times 10^{-5}$.  Since $\alpha_3$ has already
been constrained to be much less than this (Table~\ref{ppnlimits}),
we obtain a strong
solitary bound on $\zeta_2 < 4 \times 10^{-5}$~\cite{Will92c}.

%%%%%%%%%%%%%%%%%%%%%%%%%%%%%%%%%%%%%%%%%%%%%%%%%%%%%%%%%%%%%%%%%%%%%%%
\subsection{Prospects for improved PPN  parameter values}
\label{improvedPPN}

A number of advanced experiments or space missions are under development or
have been 
proposed which could lead to
significant improvements in values of the PPN  parameters,
of $J_2$ of the Sun, and of $\dot G/G$.  

Lunar laser ranging at the  Apache Point Observatory (APOLLO project) could
improve bounds on the Nordvedt parameter to the level $3 \times 10^{-5}$
and on $\dot G/G$ to better than 
$10^{-13} \, {\rm yr}^{-1}$~\cite{williams04ijmp}.

The proposed 2012 ESA Bepi-Columbo Mercury orbiter,
in a two-year experiment, with 6\ cm range capability, could yield
improvements in $\gamma$
to $3 \times 10^{-5}$, in $\beta$ to
$3 \times 10^{-4}$, in $\alpha_1$ to $10^{-5}$, 
in $\dot G/G$ to $10^{-13} {\rm\ yr}^{-1}$, and in
$J_2$ to $3 \times 10^{-8}$.  An eight-year mission could yield further
improvements by factors of 2 -- 5 in $\beta$, $\alpha_1$ and $J_2$, and a
further factor 15 in $\dot G/G$~\cite{milani02,ashby05}.  

GAIA is a high-precision astrometric orbiting telescope (a successor to
Hipparcos), which could
measure light-deflection and
$\gamma$ to the $10^{-6}$ level~\cite{gaia}.  It is planned for launch by
ESA in the 2011 time frame.

LATOR, Laser Astrometric Test of Relativity, is a concept for a NASA mission in
which two microsatellites orbit the Sun on Earth-like orbits near
superior conjunction, so that their lines of sight are close to the Sun.
Using optical tracking and an optical interferometer on the International
Space Station, it may be possible to measure the deflection of light with
sufficient accuracy to bound $\gamma$ to a part in $10^8$ and $J_2$ to a part
in $10^{8}$, and to measure the 
solar frame-dragging effect to one percent~\cite{turyshev04a,turyshev04b}.

Nordtvedt~\cite{Nordtvedt00} has argued that ``grand fits'' of large solar
system ranging data sets, including radar ranging to Mercury, Mars and
satellites, and laser ranging to the Moon,
could yield substantially improved measurements of PPN parameters.  A recent
contribution in that direction is~\cite{pitjeva05}.

%\newpage

%%%%%%%%%%%%%%%%%%%%%%%%%%%%%%%%%%%%%%%%%%%%%%%%%%%%%%%%%%%%%%%%%%%%%%%%%%%%%%%%%%%
%%%%%%%%%%%%%%%%%%%%%%%%%%%%%%%%%%%%%%%%%%%%%%%%%%%%%%%%%%%%%%%%%%%%%%%%%%%%%%%%%%%
%%%%%%%%%%%%%%%%%%%%%%%%%%%%%%%%%%%%%%%%%%%%%%%%%%%%%%%%%%%%%%%%%%%%%%%%%%%%%%%%%%%

\section{Strong Gravity and Gravitational Waves: A New Testing Ground}
\label{S4}

%%%%%%%%%%%%%%%%%%%%%%%%%%%%%%%%%%%%%%%%%%%%%%%%%%%%%%%%%%%%%%%%%%%%%%%%%%%%%%%%%%%
%%%%%%%%%%%%%%%%%%%%%%%%%%%%%%%%%%%%%%%%%%%%%%%%%%%%%%%%%%%%%%%%%%%%%%%%%%%%%%%%%%%

\subsection{Strong-field systems in general relativity}
\label{strong}

%%%%%%%%%%%%%%%%%%%%%%%%%%%%%%%%%%%%%%%%%%%%%%%%%%%%%%%%%%%%%%%%%%%%%%%%%%%%%%%%%%%

\subsubsection{Defining weak and strong gravity}
\label{strongvweak}

In the solar system, gravity is weak, in the sense that the  Newtonian
gravitational
potential and related variables ($ U ( {\bf x}, t) \sim v^2 \sim p/\rho \sim
\epsilon$)
are much smaller than unity everywhere.
This is the basis for the post-Newtonian expansion and for the
``parametrized
post-Newtonian'' framework
described in Sec.~\ref{ppn}.
``Strong-field'' systems are those for which
the simple 1PN approximation of the PPN framework
is no longer
appropriate.  This can occur in a number of situations:

\begin{itemize}

\item
The system may contain strongly relativistic objects, such as neutron
stars or black holes, near and inside which $\epsilon \sim 1$, and the
post-Newtonian approximation breaks down.  Nevertheless, under
some circumstances,
the orbital motion may be such that the interbody potential and
orbital velocities still satisfy $\epsilon \ll 1$ so that a kind of
post-Newtonian approximation for the orbital motion
might work; however, the strong-field
internal gravity of the bodies could (especially in alternative theories
of gravity) leave imprints on the orbital motion.

\item
The evolution of the system may be affected by the emission of
gravitational radiation.  The 1PN approximation
does not contain the effects of gravitational radiation back-reaction.
In the expression for the metric given in Box~\ref{box2},
radiation back-reaction effects do not occur until $O(\epsilon^{7/2})$
in $g_{00}$, $O(\epsilon^{3})$
in $g_{0i}$, and $O(\epsilon^{5/2})$
in $g_{ij}$.
Consequently, in order to describe such systems, one must carry out a
solution of the equations substantially beyond 1PN order,
sufficient to incorporate the leading
radiation damping terms at 2.5PN order.  In addition, the PPN metric
described in Sec. \ref{ppn} is valid in the {\em near zone} of the system,
i.e. within one gravitational wavelength of the system's center of mass.  As
such it cannot describe the gravitational waves seen by a detector.

\item
The system may be highly relativistic in its orbital motion, so that
$U \sim v^2 \sim 1$ even for the interbody field and orbital velocity.
Systems like this include the late stage of the inspiral of binary
systems of neutron stars or black holes, driven by gravitational
radiation damping, prior to a merger and collapse to a final
stationary state.  Binary inspiral is one of the leading candidate
sources for detection by a world-wide network of laser interferometric
gravitational wave observatories nearing completion.  A proper
description of such systems requires not only equations for the motion
of the binary carried to extraordinarily high PN orders (at least
3.5PN), but also requires equations for the far-zone
gravitational waveform measured at the detector, that are equally
accurate to high PN orders beyond the leading ``quadrupole''
approximation.

\end{itemize}

Of course, some systems cannot be properly described by any post-Newt\-onian
approximation because their behavior is fundamentally controlled by
strong gravity.  These include the imploding cores of supernovae, the
final merger of two compact objects, the quasinormal-mode vibrations
of neutron stars and black holes, the structure of rapidly rotating
neutron stars, and so on.  Phenomena such as these must be analysed
using different techniques.  Chief among these is the full solution of
Einstein's equations via numerical methods.  This field of ``numerical
relativity'' is a rapidly growing and maturing branch of gravitational
physics, whose description is beyond the scope of this review 
(see~\cite{Lehner01,BaumgarteShapiro03} for reviews).
Another is black hole perturbation theory 
(see~\cite{msstt97,KokkotasSchmidt99,SasakiTagoshi03} for  reviews).

%%%%%%%%%%%%%%%%%%%%%%%%%%%%%%%%%%%%%%%%%%%%%%%%%%%%%%%%%%%%%%%%%%%%%%%%%%%%%%%%%%%

\subsubsection{Compact bodies and the strong equivalence principle}
\label{compact-SEP}

When dealing with the motion and gravitational-wave generation by
orbiting bodies, one finds a remarkable simplification within GR. 
As long as the bodies are sufficiently well-separated
that one can ignore
tidal interactions and other effects that depend upon the finite extent of
the bodies (such as their quadrupole and higher multipole moments),
then all aspects of their orbital behavior and gravitational wave
generation
can be characterized by just two parameters:
mass and angular momentum.
Whether their internal structure is highly relativistic, as in black
holes or neutron stars, or non-relativistic as in the Earth and Sun,
only the mass and angular momentum are needed.  Furthermore, both
quantities are measurable in principle by examining the external
gravitational field of the bodies, and make no reference whatsoever to
their interiors.

Damour~\cite{Damour300} calls this the ``effacement'' of the bodies' internal
structure.
It is a consequence of the strong equivalence principle (SEP), described in
Section~\ref{sep}.

General relativity satisfies SEP because it contains one and only one
gravitational field, the spacetime metric $g_{\mu\nu}$.  Consider the
motion of a body in a binary system, whose size is small compared to
the binary separation.  Surround the body by a region that is large
compared to the size of the body, yet small compared to the
separation.  Because of the general covariance of the theory, one can
choose a freely-falling coordinate system which comoves with the body,
whose spacetime metric
takes the Minkowski form at its outer boundary (ignoring tidal
effects generated by the companion).
There is thus no evidence of the presence of the companion body,
and the structure of the chosen body can be obtained using the field
equations of GR in this coordinate system.  Far from the chosen body,
the metric is characterized by the mass and angular momentum (assuming
that one ignores quadrupole and higher multipole moments of the body) as
measured far from the body using orbiting test particles and gyroscopes.
These asymptotically measured quantities are oblivious to
the body's internal structure.  A black hole of mass $m$ and a planet
of mass $m$ would produce identical spacetimes in this outer region.

The geometry of this region surrounding the one body must be matched to
the geometry provided by the companion body.  Einstein's equations
provide consistency conditions for this matching that yield
constraints on the motion of the bodies.  These are the equations of
motion.  As a result the motion of two planets of mass and angular
momentum $m_1$, $m_2$, ${\bf J}_1$ and  ${\bf J}_2$ is
identical to that of two black holes of the same mass and angular
momentum (again, ignoring tidal effects).

This effacement does not occur in an alternative gravitional theory
like scalar-tensor gravity.  There, in addition to the spacetime
metric, a scalar field $\phi$ is generated by the masses of
the bodies, and controls the local value of the gravitational coupling
constant (i.e.\ $G$ is a function of $\phi$).   Now, in the local frame
surrounding one of the bodies in our binary system, while the metric
can still be made Minkowskian far away, the scalar field will take on
a value $\phi_0$ determined by the companion body.  This can affect
the value of $G$ inside the chosen body, alter its
internal structure (specifically its gravitational binding energy)
and hence alter its mass.
Effectively, each body can be characterized by several mass 
functions $m_A(\phi)$, which depend on
the value of the scalar field at its location, and several distinct masses
come into play, such as inertial mass, gravitational mass, ``radiation'' mass,
etc.  The precise nature of the
functions will depend on the body, specifically on its gravitational
binding energy, and as a result, the motion and
gravitational radiation may depend on the internal structure of each
body.  For compact bodies such as neutron stars and black holes these internal
structure effects could be large; for example, the gravitational binding energy
of a neutron star can be 10 -- 20 percent of its total mass.
At 1PN order, the leading manifestation of this phenomenon is
the Nordtvedt effect.

This is how the study of orbiting systems containing
compact objects provides strong-field tests of GR.
Even though the strong-field nature of the bodies is effaced in GR, it
is not in other theories, thus any result in agreement with the
predictions of GR constitutes a
kind of ``null'' test of strong-field gravity.

%%%%%%%%%%%%%%%%%%%%%%%%%%%%%%%%%%%%%%%%%%%%%%%%%%%%%%%%%%%%%%%%%%%%%%%%%%%%%%%%%%%
%%%%%%%%%%%%%%%%%%%%%%%%%%%%%%%%%%%%%%%%%%%%%%%%%%%%%%%%%%%%%%%%%%%%%%%%%%%%%%%%%%%

\subsection{Motion and gravitational radiation in general relativity}
\label{eomgw}

The motion of bodies and the
generation of gravitational radiation are long-standing problems
that date back to the first years following the publication of
GR, when Einstein calculated the gravitational
radiation emitted by a laboratory-scale object using the linearized
version of GR, and de Sitter calculated $N$-body equations of motion for
bodies in the 1PN approximation to GR.
It has at times been controversial, with disputes over such issues as
whether Einstein's equations alone imply equations of motion for
bodies (Einstein, Infeld and Hoffman demonstrated explicitly that they
do, using a matching procedure similar to the one described
above), whether gravitational waves are real or are artifacts of general
covariance (Einstein waffled; Bondi and colleagues proved their
reality rigorously in the 1950s), and even over algebraic errors
(Einstein erred by a factor of 2 in his first radiation calculation;
Eddington found the mistake).
Shortly after the discovery of the binary pulsar PSR
1913+16 in 1974, questions were raised about the foundations of the
``quadrupole formula'' for gravitational radiation damping
(and in
some quarters, even about its quantitative validity).
These questions were answered in part by theoretical
work designed to shore up the
foundations of the quadrupole approximation, and in
part
(perhaps mostly) by the
agreement between the predictions of the
quadrupole formula and the {\it observed}
rate of damping of the pulsar's orbit (see Section~\ref{binarypulsars}).
Damour~\cite{Damour300} gives a
thorough historical and technical review of this subject up to 1986.

The problem of motion and radiation in GR has received renewed interest
since 1990, with proposals for construction of
large-scale laser interferometric
grav\-it\-at\-ional-wave observatories, such as the LIGO project in the US,
VIRGO and GEO600 in Europe, and TAMA300 in Japan,
and the realization that a leading
candidate source of detectable waves would be the inspiral, driven
by grav\-it\-at\-ional radiation damping,
of a binary system of compact objects (neutron stars
or black holes)~\cite{LIGO,snowmass}.  The
analysis of signals from such systems
will require theoretical predictions from GR that are
extremely accurate, well beyond the leading-order prediction of
Newtonian or even post-Newtonian gravity for the orbits, and well beyond
the leading-order formulae for gravitational waves.

This presented a major theoretical challenge: to calculate the motion
and radiation of systems of compact objects
to very high PN order, a formidable algebraic task,
while addressing a number of issues of principle that have
historically plagued this subject, sufficiently well
to ensure that the results were physically meaningful.  This
challenge has been largely met, so that we
may soon see a remarkable convergence between observational
data and accurate predictions of gravitational theory that could
provide new, strong-field tests of GR.

Here we give a brief overview of the problem of motion
and gravitational radiation in GR.

%%%%%%%%%%%%%%%%%%%%%%%%%%%%%%%%%%%%%%%%%%%%%%%%%%%%%%%%%%%%%%%%%%%%%%%%%%%%%%%%%%%
%%%%%%%%%%%%%%%%%%%%%%%%%%%%%%%%%%%%%%%%%%%%%%%%%%%%%%%%%%%%%%%%%%%%%%%%%%%%%%%%%%%

\subsection{Einstein's equations in ``relaxed'' form}
\label{EErelaxed}

The Einstein equations $G_{\mu\nu} = 8\pi T_{\mu\nu}$ are elegant and
deceptively simple, showing geometry (in the form of the Einstein
tensor $G_{\mu\nu}$, which is a function of spacetime curvature)
being generated by matter (in the form of the material stress-energy tensor
$T_{\mu\nu}$).  However, this is not the most useful form for actual
calculations.  For post-Newtonian calculations, a far more useful form
is the so-called ``relaxed'' Einstein equations:
\begin{equation}
\Box h^{ \alpha \beta } = -16 \pi {\tau}^{ \alpha \beta },
\label{relaxed}
\end{equation}
where $\Box \equiv  -{\partial}^2 / \partial t^2 + {\nabla}^2 $
is the flat-spacetime wave operator,
$h^{ \alpha \beta }$ is a ``gravitational tensor potential''
related to the deviation of the spacetime metric from its Minkowski
form by the formula
$h^{\alpha \beta} \equiv \eta^{\alpha \beta} - (-g)^{1/2} g^{\alpha
\beta}$, $g$ is the determinant of $g_{\alpha
\beta}$, and a particular coordinate system has been specified
by the de~Donder
or harmonic gauge condition
$\partial h^{\alpha \beta} /\partial x^\beta =0$ (summation on
repeated indices is assumed).
This form of Einstein's equations bears a striking similarity to Maxwell's
equations for the vector potential $A^\alpha$ in Lorentz gauge: $\Box
A^\alpha = -4\pi J^\alpha$, $\partial A^\alpha /\partial
x^\alpha =0$.   There is a key difference, however: The source on the right
hand side of Eq.~(\ref{relaxed}) is given by the ``effective''
stress-energy pseudotensor
\begin{equation}
\tau^{\alpha\beta} = (-g)T^{\alpha\beta} + (16\pi)^{-1}
\Lambda^{\alpha\beta},
\label{effective}
\end{equation}
where $\Lambda^{\alpha\beta}$ is the non-linear ``field'' contribution
given by terms quadratic (and higher) in $h^{\alpha \beta}$ and its
derivatives (see~\cite{MTW}, Eqs.~(20.20, 20.21) for formulae).
In GR, the gravitational field itself generates
gravity, a reflection of the nonlinearity of Einstein's equations, and
in
contrast to the linearity of Maxwell's equations.

Equation~(\ref{relaxed}) is exact, and depends only on the assumption
that spacetime can be covered by harmonic coordinates.  It is called
``relaxed'' because it
can be solved formally as a functional of source variables without
specifying the motion of the source, in the form
\begin{equation}
h^{\alpha \beta} (t,{\bf x}) = 4 \int_{\cal C}
{ \tau^{\alpha \beta} (t -| {\bf x} - {\bf x'} |, {\bf x'}
)
\over | {\bf x} - {\bf x'} | } d^3x',
\label{nearintegral}
\end{equation}
where the integration is over the past flat-spacetime null cone $\cal
C$ of the field point $(t,{\bf x})$.
The motion of the source is then determined either by the equation
$\partial {\tau}^{\alpha \beta} /\partial x^\beta =0$ (which follows
from the harmonic gauge condition), or from the usual covariant
equation of motion ${T^{\alpha\beta}}_{;\beta}=0$, where the subscript
$;\beta$ denotes a covariant divergence.
This formal solution can then be iterated in a slow motion ($v<1$)
weak-field ($||h^{\alpha \beta}||<1$) approximation.  One begins by
substituting
$h_0^{\alpha \beta} =0$ into the source $\tau^{\alpha \beta}$ in Eq.
(\ref{nearintegral}), and
solving for the first iterate $h_1^{\alpha \beta}$, and then repeating the
procedure sufficiently many times to achieve a solution of the desired
accuracy.  For example, to obtain the 1PN equations of motion, {\it
two} iterations are needed (i.e.\ $h_2^{\alpha \beta}$ must be
calculated); likewise, to obtain the leading gravitational waveform
for a binary system, two iterations are needed.

At the same time, just as in electromagnetism, the formal integral
(\ref{nearintegral}) must be handled differently, depending on whether
the field point is in the far zone or the near zone.
For
field points in the far zone or radiation zone, $|{\bf x}| >
{\lambda\!\!\!{\scriptscriptstyle{{}^{-}}}} > |{\bf x}'|$
(${\lambda\!\!\!{\scriptscriptstyle{{}^{-}}}}$ is the gravitational
wavelength$/2\pi$), the field can be expanded in inverse powers of
$R=|{\bf x}|$ in a multipole expansion, evaluated at the ``retarded
time'' $t-R$.  The leading term in $1/R$ is
the gravitational waveform.  For field points in the near zone or
induction zone, $|{\bf x}| \sim |{\bf x}'| <
{\lambda\!\!\!{\scriptscriptstyle{{}^{-}}}}$, the field is expanded in
powers of $|{\bf x}-{\bf x}'|$ about the local time $t$,
yielding instantaneous potentials that go into the equations of
motion.

However, because the source ${\tau}^{\alpha \beta}$ contains
$h^{\alpha \beta}$ itself, it is not confined to a compact region, but
extends over all spacetime.  As a result, there is a danger that the
integrals involved in the various expansions will diverge or be
ill-defined.  This consequence of the non-linearity of Einstein's
equations has bedeviled the subject of gravitational radiation for
decades.  Numerous approaches have been developed to try to
handle this difficulty.  The ``post-Minkowskian'' method of Blanchet,
Damour and Iyer~\cite{bd86,bd88,bd89,di91,bd92,blanchet95}
solves Einstein's equations by two
different techniques, one in the near zone and one in the far zone,
and uses the method of singular asymptotic matching to join the
solutions in an overlap region.  The method provides a natural
``regularization'' technique to control potentially divergent
integrals (see~\cite{BlanchetLRR} for a thorough review).  
The ``Direct Integration of the Relaxed Einstein
Equations'' (DIRE) approach of Will, Wiseman and Pati~\cite{opus,DIRE} retains
Eq.~(\ref{nearintegral}) as the global solution, but splits the
integration into one over the near zone and another over the far zone,
and uses different
integration variables to carry out the explicit integrals over the two
zones.  In the DIRE method, all integrals are finite and convergent.
Itoh and Futamase have used an extension of the Einstein-Infeld-Hoffman
matching approach combined with a specific method for taking a
point-particle limit~\cite{ItohFutamase03}, while Damour, Jaranowski and
Sch\"afer have pioneered an ADM Hamiltonian approach that focuses on
the equations of 
motion~\cite{jaraschaefer98,jaranowski,damjaraschaefer,DJSequiv,DJSdim}.

These methods assume from the outset that gravity is
sufficiently weak that $||h^{\alpha\beta}||<1$ and harmonic
coordinates exists everywhere, including inside the bodies.  Thus, in
order to apply the results to cases where the bodies may be neutron
stars or black holes, one relies upon the strong equivalence principle
to argue that, if tidal forces are ignored, and equations are
expressed in terms of masses and spins,  one can simply
extrapolate the results unchanged
to the situation where the bodies are ultrarelativistic.
While no general proof of this exists, it has been shown to be valid
in specific circumstances, such as at 2PN order in the equations of
motion, and for black holes moving in a Newtonian 
background field~\cite{Damour300}.

Methods such as these
have resolved most of the issues that led to criticism of the
foundations of gravitational radiation theory during the 1970s.

%%%%%%%%%%%%%%%%%%%%%%%%%%%%%%%%%%%%%%%%%%%%%%%%%%%%%%%%%%%%%%%%%%%%%%%%%%%%%%%%%%%
%%%%%%%%%%%%%%%%%%%%%%%%%%%%%%%%%%%%%%%%%%%%%%%%%%%%%%%%%%%%%%%%%%%%%%%%%%%%%%%%%%%

\subsection{Equations of motion and gravitational waveform}
\label{eomwaveform}

Among the results of these approaches are formulae for the equations of
motion and gravitational waveform of binary systems of compact
objects, carried out to high orders in a PN expansion.  Here we shall
only state the key formulae that will be needed for this review.
For example,
the relative two-body equation of motion has the form
\begin{equation}
{\bf a} = {{d{\bf v}} \over dt} = {m \over r^2} \left\{- {\bf {\hat
      n}} + {\bf A}_{1\rm PN} + {\bf
A}_{2\rm PN} + {\bf A}_{2.5\rm PN} + {\bf A}_{3\rm PN} + {\bf
A}_{3.5\rm PN} + \dots
\right\},
\label{EOM}
\end{equation}
where $m=m_1+m_2$ is the total mass, $r= |{\bf x}_1 -{\bf x}_2|$,
${\bf v}={\bf v}_1-{\bf v}_2$, and
${\bf {\hat n}} = ({\bf x}_1 -{\bf x}_2)/r$.
The notation ${\bf A}_{n\rm PN}$ indicates that the term is
$O(\epsilon^n)$ relative to the Newtonian term $-{\bf {\hat n}}$.
Explicit and unambiguous
formulae for non-spinning bodies through 3.5PN order have been
calculated by
various authors (see~\cite{BlanchetLRR} for a review).
Here we quote only the first PN corrections and the
leading radiation-reaction terms at 2.5PN order:
{\setlength{\arraycolsep}{0.14 em}
\begin{eqnarray}
 {\bf A}_{1\rm PN} &=& \left\{ (4+2\eta){m \over r} - (1+3\eta)v^2 +
{3 \over 2} \eta {\dot r}^2 \right\}{\bf {\hat n}} + (4-2\eta) \dot r
{\bf v}, \label{APN} \\
{\bf A}_{2.5\rm PN} &=& -{8 \over 15} \eta {m \over r} \left\{ \left( 9v^2 +17{m
\over r} \right) \dot r {\bf {\hat n}} - \left( 3v^2 +9{m \over r}
\right) {\bf v} \right\}, \label{A2.5PN}
\end{eqnarray}}
where $\eta = m_1m_2/(m_1+m_2)^2$.
These terms are sufficient to analyse the orbit and evolution of the
binary pulsar (Sec.~\ref{binarypulsars}).  For example, the 1PN terms are
responsible for
the periastron advance of an eccentric orbit, given by $\dot \omega =
6\pi f_{\rm b} m/a(1-e^2)$, where $a$ and $e$ are the semi-major axis and
eccentricity, respectively, of the orbit, and $f_{\rm b}$ is the orbital
frequency, given to the needed order by Kepler's third law
$2 \pi f_{\rm b} = (m/a^3)^{1/2}$.

Another product is a formula for the gravitational field
far from the system, written schematically in the form
\begin{equation}
h^{ij} = {2m \over R} \left\{ Q^{ij} + Q_{0.5\rm PN}^{ij} +
Q_{1\rm PN}^{ij} + Q_{1.5\rm PN}^{ij} + Q_{2\rm PN}^{ij} + Q_{2.5\rm PN}^{ij} +
\dots \right\},
\label{waveform}
\end{equation}
where $R$ is the distance from the source, and the variables
are to be evaluated at retarded time $t-R$.  The leading term
is the so-called quadrupole formula
\begin{equation}
h^{ij}(t,{\bf x}) =  {2 \over R}{\ddot I}^{ij}(t-R),
\label{waveformquad}
\end{equation}
where $I^{ij}$ is the quadrupole moment of the source, and overdots
denote time derivatives.  For a binary system this leads to
\begin{equation}
Q^{ij} = 2\eta (v^iv^j - m{\hat n}^i{\hat n}^j/r).
\label{Qij}
\end{equation}
For binary systems, explicit
formulae for the waveform through 2PN order have been derived 
(see~\cite{bdiww} for a ready-to-use presentation of the waveform for
circular orbits; see~\cite{BlanchetLRR} for a full review).

Given the gravitational waveform, one can
compute the rate at which energy is carried off by the radiation
(schematically $\int \dot h \dot h d\Omega$,
the gravitational analog of the Poynting
flux).
The lowest-order quadrupole formula leads to the
gravitational wave energy flux
\begin{equation}
\dot E = {8 \over 15} \eta^2 {m^4 \over r^4} (12v^2-11 {\dot r}^2).
\label{EdotGR}
\end{equation}
This has been extended to 3.5PN order beyond the quadrupole formula 
(see~\cite{BlanchetLRR} for a
review).
Formulae for fluxes of angular and linear momentum can also be
derived.
The 2.5PN radiation-reaction terms in the equation of motion~(\ref{EOM})
result in
a damping of the orbital energy that precisely balances the energy
flux~(\ref{EdotGR})
determined from the waveform.  Averaged over one orbit, this
results in a rate of increase of the binary's orbital frequency  given by
\begin{eqnarray}
\dot f_{\rm b} &=& \frac{192\pi}{5} f_{\rm b}^2 (2\pi{\cal M}f_{\rm b})^{5/3} F(e),
\nonumber \\
F(e)&=&(1-e^2)^{-7/2}\left ( 1+ \frac{73}{24}e^2+\frac{37}{96} e^4 \right ) \,,
\label{fdotGR}
\end{eqnarray}
where ${\cal M}$ is the so-called ``chirp'' mass, given by ${\cal
M}=\eta^{3/5} m$.
Notice that by making precise measurements of the phase $\Phi (t) = 2\pi
\int^t
f(t') dt'$ of either the orbit or the gravitational waves
(for which $f =2f_{\rm b}$ for the dominant component) as a function of
the frequency, one in effect measures the ``chirp'' mass of the
system.

These formalisms have also been generalized to include the leading effects of
spin-orbit and spin-spin coupling between 
the bodies~\cite{kww93,kidder95,DIRE3}.

Another approach to gravitational radiation is applicable to the special
limit in which one mass is much smaller than the other.
This is the method of black-hole perturbation theory.  One begins with an
exact background spacetime of a black hole, either the non-rotating
Schwarzschild or the rotating Kerr solution, and perturbs it according to
$g_{\mu\nu}=g^{(0)}_{\mu\nu} + h_{\mu\nu}$.  The particle moves on a
geodesic of the background spacetime, and a suitably defined source
stress-energy tensor for the particle acts as a source for the gravitational
perturbation and wave field $h_{\mu\nu}$.  This method provides
numerical results that are exact in $v$, as well as analytical results
expressed as series in powers of $v$, both for
non-rotating and for rotating black holes.  For
non-rotating holes, the analytical expansions have been carried to
5.5PN order, or $\epsilon^{5.5}$ beyond the
quadrupole approximation.  All results of black hole
perturbation agree precisely with the $m_1 \to 0$ limit of the PN results,
up to the highest PN order where they can be compared (for reviews
see~\cite{msstt97,SasakiTagoshi03}).

%%%%%%%%%%%%%%%%%%%%%%%%%%%%%%%%%%%%%%%%%%%%%%%%%%%%%%%%%%%%%%%%%%%%%%%%%%%%%%%%%%%
%%%%%%%%%%%%%%%%%%%%%%%%%%%%%%%%%%%%%%%%%%%%%%%%%%%%%%%%%%%%%%%%%%%%%%%%%%%%%%%%%%%

\subsection{Gravitational wave detection}
\label{gwdetection}

A gravitational wave detector can be modelled as a body of mass $M$ at
a
distance $L$ from a fiducial laboratory point, connected to the point
by a spring of resonant frequency $\omega_0$ and quality factor $Q$.
From the equation of geodesic deviation, the infinitesimal
displacement $\xi$ of the mass along the line of separation from its
equilibrium position satisfies the equation of motion
\begin{equation}
\ddot \xi + 2{\omega_0 \over Q} \dot \xi + \omega_0^2 \xi
= {L \over 2} \left( F_+ (\theta,\phi,\psi) {\ddot h}_+ (t) + F_\times
(\theta,\phi,\psi) {\ddot h}_\times (t) \right),
\label{detector}
\end{equation}
where $F_+ (\theta,\phi,\psi)$ and $F_\times
(\theta,\phi,\psi)$ are ``beam-pattern'' factors, that depend on the
direction of the source $(\theta,\phi)$ and on a polarization  angle
$\psi$, and $h_+(t)$ and $h_\times (t)$ are gravitational waveforms
corresponding to the two polarizations of
the gravitational wave (for a review, see~\cite{Thorne300}).  In
a source coordinate system in which the $x-y$
plane is the plane of the sky and the $z$-direction points toward the
detector, these two modes are given by
\begin{equation}
h_+ (t) = {1 \over 2} (h^{xx}_{\rm TT} (t) - h^{yy}_{\rm TT} (t) ), \quad
h_\times  (t) = h^{xy}_{\rm TT} (t),
\label{modes}
\end{equation}
where $h^{ij}_{\rm TT}$ represent transverse-traceless (TT) projections of the
calculated waveform of Eq.~(\ref{waveform}), given by
\begin{equation}
h^{ij}_{\rm TT} = h^{kl} [(\delta^{ik}-{\hat N}^i{\hat N}^k)
(\delta^{jl}-{\hat N}^j{\hat N}^l ) - {1 \over 2} (\delta^{ij}-{\hat
N}^i{\hat N}^j)
(\delta^{kl}-{\hat N}^k{\hat N}^l ) ],
\label{TTprojection}
\end{equation}
where ${\hat N}^j$ is a unit vector pointing toward the detector.
The beam pattern factors depend on the orientation and nature of the
detector.  For a wave approaching along the laboratory
$z$-direction, and for a mass whose location on the $x-y$ plane
makes an angle $\phi$
with the $x$ axis, the beam pattern factors are given by $F_+ =
\cos 2 \phi$ and $F_\times = \sin 2 \phi$.
For a resonant cylinder oriented along the laboratory $z$
axis, and for source direction $(\theta,\phi)$,
they are given by $F_+ = \sin^2 \theta \cos 2 \psi $, $F_\times
= \sin^2 \theta \sin 2 \psi $ (the angle $\psi$ measures the relative
orientation of the laboratory and source $x$-axes).  For a laser interferometer with one
arm along the laboratory $x$-axis, the other along the $y$-axis, and
with $\xi$ defined as the {\it differential} displacement
along the two arms, the beam pattern functions are
$F_+ = {1 \over 2} (1+\cos^2 \theta )\cos 2 \phi \cos 2 \psi - \cos
\theta \sin 2 \phi \sin 2 \psi $ and
$F_\times = {1 \over 2} (1+\cos^2 \theta )\cos 2 \phi \sin 2 \psi + \cos
\theta \sin 2 \phi \cos 2 \psi $.

The waveforms $h_+ (t)$ and $h_\times (t)$ depend on the nature and
evolution of the source.  For example, for a binary system in a
circular orbit, with an inclination $i$ relative to the plane of the
sky, and the $x$-axis oriented along the major axis of the projected
orbit, the quadrupole approximation of Eq.~(\ref{Qij}) gives
\begin{eqnarray}
h_+ (t) &=& - \frac{2{\cal M}}{R} (2\pi {\cal M} f_{\rm b})^{2/3} (1+ \cos^2 i) \cos
2 \Phi_{\rm b} (t) \,, \\
h_\times (t) &=& - \frac{2{\cal M}}{R} (2\pi {\cal M} f_{\rm b})^{2/3} 2 \cos i
\cos 2 \Phi_{\rm b} (t) \,,
\end{eqnarray}
where $\Phi_{\rm b} (t) = 2\pi \int^t f_{\rm b} (t') dt'$ is the orbital
phase.

%\newpage

%%%%%%%%%%%%%%%%%%%%%%%%%%%%%%%%%%%%%%%%%%%%%%%%%%%%%%%%%%%%%%%%%%%%%%%%%%%%%%%%%%%
%%%%%%%%%%%%%%%%%%%%%%%%%%%%%%%%%%%%%%%%%%%%%%%%%%%%%%%%%%%%%%%%%%%%%%%%%%%%%%%%%%%
%%%%%%%%%%%%%%%%%%%%%%%%%%%%%%%%%%%%%%%%%%%%%%%%%%%%%%%%%%%%%%%%%%%%%%%%%%%%%%%%%%%

\section{Stellar System Tests of Gravitational Theory}
\label{stellar}

%%%%%%%%%%%%%%%%%%%%%%%%%%%%%%%%%%%%%%%%%%%%%%%%%%%%%%%%%%%%%%%%%%%%%%%%%%%%%%%%%%%
%%%%%%%%%%%%%%%%%%%%%%%%%%%%%%%%%%%%%%%%%%%%%%%%%%%%%%%%%%%%%%%%%%%%%%%%%%%%%%%%%%%

\subsection{The binary pulsar and general relativity}
\label{binarypulsars}

The 1974 discovery of the binary pulsar B1913+16 by Joseph Taylor
and Russell Hulse during a routine search for new pulsars
provided the first possibility of probing new aspects of gravitational
theory:  the effects of strong relativistic internal gravitational fields
on orbital dynamics, and the effects of gravitational radiation reaction.
For reviews of the discovery, see the published
Nobel Prize lectures by Hulse and Taylor~\cite{Hulse,Taylor94}. 
For  reviews of the current status of pulsars, 
including binary and millisecond
pulsars, see~\cite{lorimer,StairsLRR}.

The system consists of a pulsar of nominal period 59 ms in a close binary
orbit with an as yet unseen companion.  The orbital period is about
7.75 hours, and the eccentricity is 0.617.  From detailed analyses of the
arrival times of pulses (which amounts to an integrated version of the
Doppler-shift methods used in spectroscopic binary systems), extremely
accurate orbital and physical parameters for the system have been obtained
(Table~\ref{bpdata}).  Because the orbit is so close
($ \approx 1 R_\odot$)
and because there is no evidence of an eclipse of the pulsar signal or of
mass transfer from the companion, it is generally agreed that the companion
is compact.  Evolutionary arguments suggest that it is most likely
a dead pulsar, while B1913+16 is a ``recycled'' pulsar.  
Thus the orbital motion is
very clean, free from tidal or other
complicating effects.  Furthermore, the data acquisition is ``clean'' in
the sense that by exploiting the intrinsic stability of the pulsar
clock combined with the ability to maintain and transfer atomic time
accurately using GPS,
the observers can keep track of pulse time-of-arrival with
an accuracy of 13~$\mu$s, despite extended gaps between
observing sessions (including a several-year gap in the middle 1990s
for an upgrade of
the Arecibo radio telescope).  The pulsar has shown no evidence of ``glitches''
in its pulse period.

\begin{table}[t]
  \begin{center}
    \begin{tabular}{lll}
      \hline \hline
      & Symbol & \\
      Parameter & (units) & Value \\
      \hline \hline
      \multicolumn{3}{l}{\bf (i) ``Physical'' parameters} \\
      \quad Right Ascension & $\alpha$ &
      $19^{\rm h} 15^{\rm m}27.^{\rm s} 99999(2)$ \\
      \quad Declination & $\delta$ & $16 ^\circ 06' 27.'' 4034(4)$ \\
      \quad Pulsar Period & $P_{\rm p}$ (ms) & $59.0299983444181(5)$ \\
      \quad Derivative of Period & $\dot P_{\rm p}$ &
      $8.62713(8) \times 10^{-18}$ \\ \\
      \multicolumn{3}{l}{\bf (ii) ``Keplerian'' parameters} \\
      \quad Projected semimajor axis & $a_{\rm p} \sin i$ (s) &
      $2.341774(1)$ \\
      \quad Eccentricity & $e$ & $0.6171338(4)$ \\
      \quad Orbital Period & $P_{\rm b}$ (day) & $0.322997462727(5)$ \\
      \quad Longitude of periastron & $\omega_0$ ($^\circ$) &
      $226.57518(4)$ \\
      \quad Julian date of periastron & $T_0$ (MJD) &
      $46443.99588317(3)$ \\ \\
      \multicolumn{3}{l}{\bf (iii) ``Post-Keplerian'' parameters} \\
      \quad Mean rate of periastron advance &
      $\langle \dot\omega \rangle$ $(^\circ {\rm\ yr}^{-1})$ &
      $4.226595(5)$ \\
      \quad Redshift/time dilation & $\gamma'$ (ms) & $4.2919(8)$ \\
      \quad Orbital period derivative & $\dot P_{\rm b}$ $(10^{-12})$ &
      $-2.4184(9)$ \\
      \hline \hline
    \end{tabular}
    \caption{\it Parameters of the binary pulsar B1913+16. The
      numbers in parentheses denote errors in the last digit. Data
      taken from~\cite{ATNFpulsarcat,WeisbergTaylor05}.  Note that $\gamma'$
      is not the same as the PPN parameter $\gamma$ 
      [see Eq. (\ref{pkparameters})].}
    \label{bpdata}
  \end{center}
\end{table}

Three factors make this system an arena where relativistic celestial
mechanics must be used: the relatively large size of relativistic
effects
[$ v_{\rm orbit} \approx (m/r)^{1/2} \approx 10^{-3}$], a factor of 10
larger than the corresponding values for solar-system orbits;
the short orbital period, allowing secular effects to
build up rapidly; and the cleanliness of the system, allowing
accurate determinations of small effects.  Because the orbital
separation is large compared to the neutron stars' compact size, tidal
effects can be ignored.  Just as Newtonian gravity
is used as a tool for measuring astrophysical parameters of ordinary binary
systems, so GR is used as a tool for measuring
astrophysical parameters in the binary pulsar.

The observational parameters that are obtained from a least-squares
solution of the arrival-time data fall into three groups:  (i) non-orbital
parameters, such as the pulsar period and its rate of change (defined
at a given epoch), and the
position of the pulsar on the sky; (ii) five ``Keplerian'' parameters,
most closely related to those appropriate for standard Newtonian binary
systems,
such as the eccentricity $e$, the orbital period $P_{\rm b}$, and the
semi-major axis of the pulsar projected along the line of sight, 
$a_{\rm p} \sin \, i$;
and (iii) five ``post-Keplerian'' parameters.   The five post-Keplerian
parameters are: $\langle \dot \omega \rangle$, the average rate of periastron advance;
$\gamma'$, the amplitude of delays in arrival of pulses caused by the
varying effects of the gravitational redshift and time dilation as the
pulsar moves in its elliptical orbit at varying distances from the
companion and with varying speeds;
$\dot P_{\rm b}$, the rate of change of orbital period, caused
predominantly by gravitational radiation damping; and  $r$ and
$s = \sin i$, respectively the ``range'' and ``shape'' of the Shapiro
time delay of the pulsar signal as it propagates through the curved
spacetime region near the companion, where $i$ is the angle of
inclination of the orbit relative
to the plane of the sky.  An additional 14 relativistic parameters are
measurable in principle~\cite{DamourTaylor92}.

In GR, the five post-Keplerian parameters can be related
to the masses of the two bodies and to measured Keplerian parameters
by the equations (TEGP 12.1, 14.6~(a)~\cite{tegp})
{\setlength{\arraycolsep}{0.14 em}
\begin{eqnarray}
        \langle \dot \omega \rangle
&=& 6\pi f_{\rm b} (2\pi m f_{\rm b} )^{2/3} (1-e^2 )^{-1}, \nonumber \\
          \gamma'
 &=&e (2 \pi f_{\rm b} )^{-1} (2\pi m f_{\rm b})^{2/3}
          (m_2 /m) (1+m_2 /m ), \nonumber \\
          \dot P_{\rm b}
 &=&-(192 \pi /5)(2 \pi {\cal M}f_{\rm b} )^{5/3} F(e),
 \label{pkparameters} \\
          s &=& \sin i,  \nonumber \\
          r &=& m_2, \nonumber
\end{eqnarray}}%
where
$m_1$ and $m_2$ denote the pulsar and companion masses, respectively.
The formula for $\langle \dot \omega \rangle$ ignores
possible non-relativistic contributions to the periastron shift,
such as tidally or rotationally induced effects caused by the companion
(for discussion of these effects, see TEGP 12.1~(c)~\cite{tegp}).   The formula
for $\dot P_{\rm b}$ includes only quadrupole
gravitational radiation;
it ignores other sources of energy loss, such as tidal
dissipation (TEGP 12.1~(f)~\cite{tegp}).  Notice that, by virtue of Kepler's 
third
law, $(2\pi f_{\rm b})^2 = m/a^3$,  $(2\pi m f_{\rm b})^{2/3} = m/a \sim
\epsilon$, thus the first two post-Keplerian parameters can be seen
as $O(\epsilon)$, or 1PN corrections to the underlying variable, while the
third is an $O(\epsilon^{5/2})$, or 2.5PN correction.
The current observed values for the Keplerian
and post-Keplerian parameters are shown in Table~\ref{bpdata}.
The parameters $r$ and $s$ are not separately
measurable with interesting accuracy for B1913+16 because the
orbit's $47 ^\circ$ inclination does not lead to a substantial Shapiro
delay.

Because $f_{\rm b}$ and $e$ are separately measured parameters, the
measurement of the three post-Keplerian parameters provide three
constraints on the two unknown masses.  The periastron shift measures
the total mass of the system, $\dot P_{\rm b}$ measures the chirp mass, and
$\gamma'$ measures a complicated function of the masses.
GR passes the test if it provides a consistent solution to these
constraints, within the measurement errors.

From the intersection of the  $\langle \dot \omega \rangle$
and $\gamma' $ constraints we obtain the values
$m_1 = 1.4414 \pm 0.0002 M_\odot$ and
$m_2 = 1.3867 \pm 0.0002 M_\odot$.  The third of Eqs.~(\ref{pkparameters})
then predicts the value
$\dot P_{\rm b} = -2.40242 \pm 0.00002 \times 10^{-12}$.
In order to compare the predicted
value for $\dot P_{\rm b}$ with the observed value of Table~\ref{bpdata}, it
is necessary to
take into account the small effect of a relative acceleration between the
binary pulsar system and the solar system caused by the differential
rotation of the galaxy.  This effect was previously considered unimportant
when $\dot P_{\rm b}$ was known only to 10 percent accuracy.  Damour and
Taylor~\cite{DamourTaylor91}
carried out a careful estimate of this effect using data
on the location and proper motion of the pulsar, combined with the best
information available on galactic rotation; the current value of this effect
is
 $\dot P_{\rm b}^{ \rm GAL} \simeq -(0.0128 \pm 0.0050) \times 10^{-12}$.
Subtracting this from the observed $\dot P_{\rm b}$ (Table~\ref{bpdata})
gives the corrected 
$\dot P_{\rm b}^{\rm CORR} = -(2.4056 \pm 0.0051)
\times 10^{-12}$,
which agrees with the prediction within the errors.  In other words,
\begin{equation}
     \frac{\dot P_{\rm b}^{\rm CORR}}{\dot P_{\rm b}^{\rm GR}}
 = 1.0013 \pm 0.0021 \,.
\label{Pdotcompare}
\end{equation}
The consistency among the measurements is displayed in Figure~\ref{bpfigure1},
in which the regions allowed by the three most precise constraints
have a single common overlap.
Uncertainties in the parameters that go into the galactic correction are now
the limiting factor in the accuracy of the test of gravitational damping.

A third way to display the agreement with GR is by
comparing the observed phase of the orbit with a theoretical template
phase as a function of time.  If $f_{\rm b}$ varies slowly in time, then to
first order in a Taylor expansion, the orbital phase is given by
$\Phi_{\rm b} (t) = 2\pi f_{\rm b0} t + \pi {\dot f}_{\rm b0} t^2$.  The time of
periastron passage $t_{\rm P}$ is given by $\Phi (t_{\rm P})=2\pi N$, where $N$ is
an integer, and consequently, the periastron time will not grow
linearly with $N$.  Thus the cumulative difference between periastron
time $t_{\rm P}$ and $N/f_{\rm b0}$, the quantities actually measured in
practice,  should vary according to
$t_{\rm P} - N/f_{\rm b0} = -{\dot f}_{\rm b0} N^2/2 f_{\rm b0}^3 \approx - ({\dot
f}_{\rm b0}/2f_{\rm b0}) t^2$.  Figure~\ref{bpfigure2} shows the results: the
dots are the data points, while the curve is the predicted difference
using the measured masses and the quadrupole formula for ${\dot
f}_{\rm b0}$~\cite{WeisbergTaylor05}.

The consistency among the constraints
provides a test of the assumption that the two bodies
behave as ``point'' masses, without complicated tidal effects, obeying
the general relativistic equations of motion including
gravitational radiation.  It is also a test of strong gravity,
in that the highly relativistic internal structure of the
neutron stars does not influence their orbital motion, as predicted by
the strong equivalence principle of GR.

\epubtkImage{}{
\begin{figure}[hptb]
  \def\epsfsize#1#2{0.6#1}
  \centerline{\epsfbox{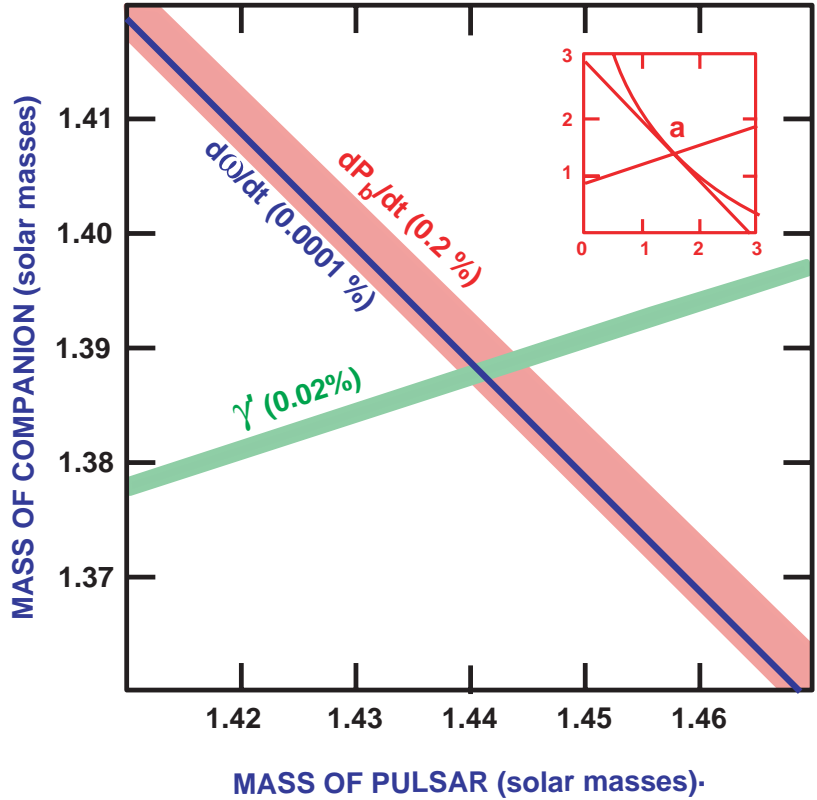}}
  \caption{\it Constraints on masses of the pulsar and its companion
    from data on B1913+16, assuming GR to be valid.  The width of
    each strip in the plane reflects observational accuracy, shown as
    a percentage.  An inset shows the three constraints on the full
    mass plane;  the intersection region (a) has been magnified 400
    times for the full figure.}
  \label{bpfigure1}
\end{figure}}

\epubtkImage{}{
\begin{figure}[hptb]
  \def\epsfsize#1#2{0.6#1}
  \centerline{\epsfbox{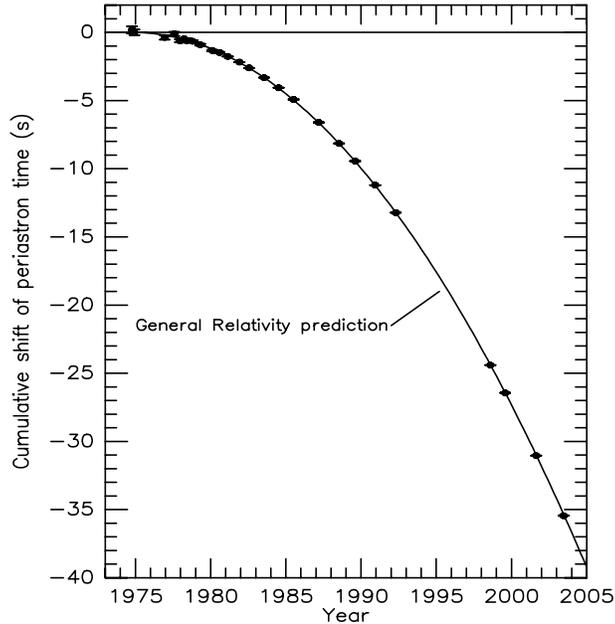}}
  \caption{\it Plot of the cumulative shift of the periastron time
    from 1975--2005. The points are data, the curve is the GR
    prediction.  The gap during the middle 1990s was caused by a
    closure of Arecibo for upgrading~\cite{WeisbergTaylor05}.}
  \label{bpfigure2}
\end{figure}}

Recent observations~\cite{kramer,WeisbergTaylor02} indicate
variations in the pulse profile, which suggests that the pulsar is
undergoing geodetic precession on a 300-year timescale
as it moves through the curved spacetime
generated by its companion (Sec. \ref{geodeticprecession}).
The amount is consistent with GR, assuming that the pulsar's
spin is suitably misaligned with the orbital angular momentum.
Unfortunately, the evidence suggests that the pulsar beam may precess
out of our line of sight by 2025.

%%%%%%%%%%%%%%%%%%%%%%%%%%%%%%%%%%%%%%%%%%%%%%%%%%%%%%%%%%%%%%%%%%%%%%%%%%%%%%%%%%%
%%%%%%%%%%%%%%%%%%%%%%%%%%%%%%%%%%%%%%%%%%%%%%%%%%%%%%%%%%%%%%%%%%%%%%%%%%%%%%%%%%%

\subsection{A zoo of binary pulsars}
\label{population}

Nine relativistic binary neutron star systems with orbital periods less than
a day are now known.  While some are less interesting for
testing relativity, some have yielded interesting tests, and others, notably
the recently discovered ``double pulsar'' are likely to produce significant
results in the future.  Here we describe some of the more interesting
or best studied cases; 
the parameters of the first four are listed in Table \ref{bpdata2}.

\begin{table}[t]
  \begin{center}
    \begin{tabular}{@{}l@{\,\,}l@{\,\,}l@{\,\,}l@{\,\,}l@{}}
      \hline \hline
      Parameter & B1534+12 & B2127+11C & J1141-6545 & J0737-3039(A,B)\\
      \hline \hline
      \multicolumn{3}{@{}l}{\bf (i) ``Keplerian'' parameters} \\
      \quad $a_{\rm p} \sin i$ (s) & \small 3.7294626(8)
      & \small 2.520(3) & \small 1.85894(1)& \small 1.41504(2)/1.513(3)\\
      \quad $e$ & \small 0.2736767(1) & \small 0.68141(2)
      & \small 0.171876(2)& \small 0.087779(5)\\
      \quad $P_{\rm b}$ (day) & \small 0.420737299153(4)
      & \small 0.335282052(6) & \small 0.1976509587(3)
      & \small 0.102251563(1)\\ \\
      \multicolumn{3}{@{}l}{\bf (ii) ``Post-Keplerian'' parameters} \\
      \quad $\langle \dot \omega \rangle$ $(^\circ {\rm\ yr}^{-1})$ &
      \small 1.755805(3) & \small 4.457(12) & \small 5.3084(9) &
      \small 16.90(1)\\
      \quad $\gamma'$ (ms) & \small 2.070(2) & \small 4.67 & \small 0.72(3)
      & \small 0.382(5)\\
      \quad ${\dot P}_{\rm b}$ $(10^{-12})$ & \small $-0.137(3)$ &$-3.94$ &
      -0.43(10)& \small $-1.21(6)$ \\
      \quad $r(\mu {\rm s})$ & \small 6.7(1.0) & & 
      & 6.2(5)\\
      \quad $s= \sin i$ & \small 0.975(7) & & &\small 0.9995(4) \\
      \hline \hline
    \end{tabular}
    \caption{\it Parameters of other binary
      pulsars.  References may be found in the text; for an
      online catalogue of pulsars with reasonably up-to-date parameters, 
      see~\cite{ATNFpulsarcat}.}
    \label{bpdata2}
  \end{center}
\end{table}

\begin{description}
\item[B1534+12.] This is a binary pulsar system in our 
galaxy~\cite{stairs2,StairsLRR,ATNFpulsarcat}.  Its
  pulses are significantly stronger and narrower than those of 
  B1913+16, so timing measurements are more precise, reaching 3~$\mu$s
  accuracy.  
  The orbital plane
  appears to be almost edge-on relative to the line  of sight
  ($i \simeq 80 ^\circ$); as a result the Shapiro delay is substantial,
  and separate values of the parameters $r$ and $s$ have been obtained
  with interesting accuracy.  Assuming GR, one infers
  that the two masses are  $m_1=1.335 \pm 0.002 M_\odot$ and
  $m_2=1.344 \pm 0.002 M_\odot$.  The rate of orbit decay
  $\dot P_{\rm b}$ agrees with GR to about 15 percent, but the precision
  is
  limited by the poorly known distance to the pulsar, which introduces
  a significant uncertainty into the subtraction of galactic
  acceleration.  Independently of $\dot P_{\rm b}$, measurement of the
  four other post-Keplerian parameters gives two tests of strong-field
  gravity in the non-radiative regime~\cite{TWDW92}.
\item[B2127+11C.]
  This system appears to be a clone of the Hulse-Taylor binary pulsar,
  with very similar values for orbital period and eccentricity (see
  Table~\ref{bpdata2}).  The inferred total mass of the system is
  $2.706 \pm 0.011 M_\odot$. But because the system is in the globular
  cluster M15 (NGC 7078), it suffers Doppler shifts resulting from
  local accelerations, either by the mean cluster gravitational field
  or by nearby stars, that are more difficult to estimate than was
  the case with the galactic system B1913+16.  This makes a
  separate, precision measurement of the relativistic contribution to
  $\dot P_{\rm b}$ essentially impossible.
\item[J0737-3039A,B.]
This binary pulsar system, discovered in 2003~\cite{burgay03}, was already
remarkable for its extraordinarily short orbital period (0.1 days) and large
periastron advance ($16.88^\circ \,{\rm yr}^{-1}$), but then the companion was
also discovered to be a pulsar~\cite{lyne04}.  Because two projected
semi-major axes can now be measured, one can obtain the mass ratio directly
from the
ratio of the two values of $a_{\rm p} \sin \,i$, and thereby obtain the two
masses by combining that ratio with the periastron advance, assuming GR.  The
results are $m_A = 1.337 \pm 0.005 \,M_\odot$ and 
$m_B = 1.250 \pm 0.005 \,M_\odot$, where $A$ denotes the primary (first)
pulsar.  From these values, one finds that the orbit is nearly edge-on, with
$\sin \,i = 0.9991$, a value which is completely consistent with that inferred
from the Shapiro delay parameter (Table~\ref{bpdata2}).  In fact, the five
measured post-Keplerian parameters plus the ratio of the projected
semi-major axes give six constraints on the masses (assuming GR): all six
overlap within their measurement errors.  This system provides a unique
opportunity for tight tests of strong-field and radiative effects in
GR.
Furthermore, it is likely that galactic proper motion effects will
play a significantly
smaller role in the interpretation of $\dot P_{\rm b}$ measurements 
than they did in B1913+16.
\item[J1141-6545.]  
This is a case where the companion is probably a white 
dwarf~\cite{bailes03,ATNFpulsarcat}.  
The mass of the pulsar and companion are $1.30 \pm 0.02$
and $0.986 \pm 0.02\, M_\odot$, respectively.
${\dot P}_{\rm b}$ has been measured to about 25 percent, consistent with the GR
prediction.  But because of the asymmetry in sensitivities ($s_{\rm NS} \sim
0.2$, $s_{\rm WD} \sim 10^{-4}$), there is the possibility, absent in
the double neutron-star systems, 
to place a strong bound on scalar-tensor gravity (see Sec.
\ref{binarypulsarsscalar}).  
\item[J1756-2251.]
Discovered in 2004, this pulsar is in a binary system with a probable
neutron star companion, with $P_{\rm b} = 7.67 \,{\rm hr}$, $e= 0.18$
and $\dot{\omega} = 2.585 \pm 0.002 \, {\rm deg. \, yr}^{-1}$~\cite{faulkner05}.
\item[J1906+0746.]
The discovery of this system was reported in late 2005~\cite{lorimer05}.  
It is a young, 144-ms pulsar in a relativistic orbit with $P_{\rm b} = 3.98
\,{\rm hr}$, $e= 0.085$, and $\dot{\omega} = 7.57 \pm 0.03 \, {\rm deg. \,
yr}^{-1}$.  
\end{description}

%%%%%%%%%%%%%%%%%%%%%%%%%%%%%%%%%%%%%%%%%%%%%%%%%%%%%%%%%%%%%%%%%%%%%%%%%%%%%%%%%%%
%%%%%%%%%%%%%%%%%%%%%%%%%%%%%%%%%%%%%%%%%%%%%%%%%%%%%%%%%%%%%%%%%%%%%%%%%%%%%%%%%%%

\subsection{Binary pulsars and alternative theories}
\label{binarypulsarsalt}

Soon after the discovery of the binary pulsar it was widely hailed as
a new testing ground for relativistic gravitational effects.
As we have seen in the case of GR, in most respects,
the system has lived up to, indeed exceeded, the early expectations.

In another respect, however, the system has only partially lived up to its
promise, namely as a direct testing ground for alternative theories of
gravity.  The origin of this promise was the discovery
that alternative theories of gravity generically predict the emission
of dipole gravitational radiation from binary star systems.
In GR, there is no dipole radiation because the
``dipole moment'' (center of mass) of isolated systems is
uniform in time (conservation
of momentum), and because the ``inertial mass'' that determines the
dipole moment is the same as the mass that generates gravitational
waves (SEP).  In other theories, while the
inertial dipole moment may remain uniform, the ``gravity wave'' dipole
moment need not, because the mass that generates gravitational waves
depends differently on the internal
gravitational binding energy of each body than does the inertial mass
(violation of SEP).
Schematically, in a coordinate system in which the center of inertial
mass is at the origin, so that $m_{\rm I,1} {\bf x}_1 + m_{\rm I,2} {\bf x}_2 =0$,
the dipole part of the retarded gravitational field would be given by
\begin{equation}
h \sim \frac{1}{R} \frac{d}{dt} (m_{\rm GW,1} {\bf x}_1 + m_{\rm GW,2} {\bf x}_2 )
\sim \frac{\eta m}{R} {\bf v} \left( \frac{m_{\rm GW,1}}{m_{\rm I,1}} -
\frac{m_{\rm GW,2}}{m_{\rm I,2}} \right),
\label{hdipole}
\end{equation}
where ${\bf v} = {\bf v}_1 -{\bf v}_2$ and $\eta$ and $m$ are defined
using inertial masses.  In theories that violate SEP,
the difference between gravitational-wave mass and inertial mass is a
function of the internal gravitational binding energy of the bodies.
This additional form of gravitational radiation damping could,
at least in principle, be significantly stronger than the usual quadrupole
damping, because it depends on fewer powers of the orbital velocity $v$,
and it depends on the gravitational binding energy per unit mass of
the bodies, which, for neutron stars, could be as large as 20 percent
(see TEGP 10~\cite{tegp} for further details).
As one fulfillment of this promise, Will and Eardley worked out in
detail the effects of dipole gravitational radiation in the bimetric theory
of Rosen, and, when the first observation of the decrease of
the orbital period was announced in 1979, the Rosen theory
suffered a terminal blow.  A wide
class of alternative theories also fails the binary pulsar test because
of dipole gravitational radiation (TEGP 12.3~\cite{tegp}).

On the other hand, the early observations of PSR 1913+16
already indicated that, in
GR, the masses of the two bodies were nearly equal, so
that, in theories of gravity that are in some sense ``close'' to
GR, dipole gravitational radiation would not be a
strong effect, because of the apparent symmetry of the system.
The Rosen theory, and others like it, are not ``close'' to GR,
except in their predictions for the weak-field, slow-motion
regime of the solar system.  When relativistic neutron stars are present,
theories like these can predict strong effects on the motion of the bodies
resulting from their internal highly relativistic gravitational structure
(violations of SEP).  As a consequence,
the masses inferred from observations of the periastron shift
and $\gamma'$ may
be significantly different from those inferred using GR,
and may be different from each other, leading to strong
dipole gravitational radiation damping.  By contrast, the Brans-Dicke
theory is ``close'' to GR, roughly
speaking within $1/ \omega_{\rm BD}$ of the predictions of the latter,
for large values of the coupling constant $\omega_{\rm BD}$. 
Thus, despite the presence of dipole gravitational radiation,
the binary pulsar provides at present only a weak test of Brans-Dicke
theory, not competitive with solar-system tests.

%%%%%%%%%%%%%%%%%%%%%%%%%%%%%%%%%%%%%%%%%%%%%%%%%%%%%%%%%%%%%%%%%%%%%%%%%%%%%%%%%%%
%%%%%%%%%%%%%%%%%%%%%%%%%%%%%%%%%%%%%%%%%%%%%%%%%%%%%%%%%%%%%%%%%%%%%%%%%%%%%%%%%%%

\subsection{Binary pulsars and scalar-tensor gravity}
\label{binarypulsarsscalar}

Making the usual assumption that both members of the system are neutron
stars, and using the methods summarized in TEGP 10--12~\cite{tegp},
one can obtain
formulas for the periastron shift, the gravitational redshift/second-order
Doppler shift parameter, and the rate of change of orbital period,
analogous to Eqs.~(\ref{pkparameters}).  These formulas depend on the
masses of the two neutron stars, on their self-gravitational binding
energy, represented by ``sensitivities'' $s$ and $\kappa^*$, and on
the Brans-Dicke coupling constant $\omega_{\rm BD}$.  First, there is
a modification of Kepler's third law, given by
\begin{equation}
    2 \pi f_{\rm b} = ( {\cal G} m /a^3)^{1/2}.
\label{KeplerBD}
\end{equation}
Then, the predictions for $\langle \dot \omega \rangle$, $\gamma^\prime$
and $\dot P_{\rm b}$
are
\begin{equation}
       \langle \dot \omega \rangle
= 6 \pi f_{\rm b} (2\pi m f_{\rm b})^{2/3} (1-e^2 )^{-1} {\cal P}{\cal G}^{-4/3}
, \label{periastronBD}
\end{equation}
\begin{equation}
          \gamma^\prime
= e (2 \pi f_{\rm b} )^{-1} (2\pi m f_{\rm b})^{2/3}
          (m_2 /m) {\cal G}^{-1/3}
          ( \alpha_2^* + {\cal G} m_2 /m
      + \kappa_1^* \eta_2^* ),\label{gammaBD}
\end{equation}
\begin{equation}
          \dot P_{\rm b}
= -(192 \pi /5)(2 \pi {\cal M} f_{\rm b} )^{5/3} F^\prime (e)
     - 4 \pi (2 \pi \mu f_{\rm b} ) \xi {\cal S}^2 G(e), \label{PdotBD}
\end{equation}
where ${\cal M} \equiv \chi^{3/5} {\cal G}^{-4/5} \eta^{3/5} m$, and,
to first order in $\xi \equiv (2 + \omega_{\rm BD} )^{-1}$, we
have
\begin{equation}
          F^\prime (e)
= F(e) + \frac{5}{144} \xi (\Gamma + 3 \Gamma^\prime )^2
       \left(  \frac{1}{2} e^2
       +   \frac{1}{8} e^4 \right) (1-e^2 )^{-7/2} \,, \nonumber 
\end{equation}
\begin{equation}
          G(e)
= (1-e^2 )^{-5/2} \left(1+ \frac{1}{2} e^2 \right), \nonumber
\end{equation}
\begin{equation}
          {\cal S}
= s_1  -  s_2 \,, \quad
          {\cal G}
=  1  -  \xi (s_1 + s_2 - 2s_1 s_2 ) \,, \quad
          {\cal P}
= {\cal G} [1 -  \frac{2}{3} \xi
       +  \frac{1}{3} \xi
          (s_1 + s_2 -2s_1 s_2 )] \,, \nonumber
\end{equation}
\begin{equation}
          \alpha_2^*
= 1 -  \xi s_2 \,, \quad
          \eta_2^*
= (1-2s_2 ) \xi \,, \quad
          \chi = {\cal G}^2
\left[1 -   \frac{1}{2} \xi  +  \frac{1}{12} \xi \Gamma^2 \right] \,,
\nonumber
\end{equation}
\begin{equation}
          \Gamma
= 1 - 2(m_1 s_2  +  m_2 s_1 )/m \,, \quad
          \Gamma^\prime
= 1 - s_1 - s_2 \,,
\label{BDcoefficients}
\end{equation}
where $F(e)$ is defined in Eq (\ref{fdotGR}).
The quantities $s_{\rm a}$ and $\kappa_{\rm a}^*$ are defined by
\begin{equation}
       s_{\rm a} = -
\left( \frac{\partial (\ln m_{\rm a} )}{\partial (\ln G)} \right)_N \,, \qquad
       \kappa_{\rm a}^*  =  -
\left( \frac{\partial (\ln I_{\rm a} )}{\partial (\ln G)} \right)_N \,, 
\label{sensitivities}
\end{equation}
and measure the ``sensitivity'' of the mass $m_{\rm a}$ and moment
of inertia $I_{\rm a}$ of each body to changes in the scalar field
(reflected in changes in $G$) for a fixed baryon number $N$ (see
TEGP 11, 12 and 14.6~(c)~\cite{tegp} for further details).  The
quantity $s_{\rm a}$ is
related to the gravitational binding energy.  These sensitivities will
depend on the neutron-star equation of state.  Notice how the violation of
SEP in Brans-Dicke theory introduces complex structure-dependent
effects in everything from the Newtonian limit (modification of the
effective coupling constant in Kepler's third law) to
gravitational radiation.  In the limit $\xi \to 0$, we recover GR, and
all structure dependence disappears.
The first term in $\dot P_{\rm b}$ (Eq.~(\ref{PdotBD}))
is the combined effect of quadrupole and
monopole gravitational radiation, while the second term is the
effect of dipole radiation.

Unfortunately, because of the near equality of the neutron star
masses in the binary pulsar, dipole radiation is suppressed, and the
bounds obtained are not competitive with the 
Cassini bound on $\gamma$~\cite{zaglauer}, 
except for those generalized scalar-tensor theories,
with $\beta_0 < 0$~\cite{DamourEspo98}.  
Bounds on the parameters $\alpha_0$ and $\beta_0$ from solar-system,
binary-pulsar and gravitational wave observations
(see Secs.~\ref{binarypulsars} and~\ref{backreaction})
are found in~\cite{DamourEspo98}.  

Alternatively, a binary pulsar system with dissimilar objects, such as
a white dwarf or black hole companion, would provide potentially more
promising tests of dipole radiation.  In this regard, the recently
discovered binary pulsar J1141+6545, with an apparent white dwarf
companion, may play an important role.  Here one can treat $s_{\rm WD}
\sim 10^{-4}$ as negligible.
Then, from
Eq. (\ref{PdotBD}), it is straightforward to show that, if the timing reaches
sufficient accuracy to determine ${\dot P}_{\rm b}$ to an accuracy $\sigma$ 
{\em in agreement with the prediction of GR}, 
then the resulting lower bound on $\omega_{\rm BD}$
would be 
\begin{equation}
\omega_{\rm BD} > 4 \times 10^4 s_{\rm NS}^2/\sigma \,.
\label{omegabound}
\end{equation}
Thus, for 
$s_{\rm NS} \sim 0.2$, a 4 percent measurement would already compete with
the Cassini bound (for further details, see~\cite{GerardWiaux02,Esposito03}).

%\newpage

%%%%%%%%%%%%%%%%%%%%%%%%%%%%%%%%%%%%%%%%%%%%%%%%%%%%%%%%%%%%%%%%%%%%%%%%%%%%%%%%%%%
%%%%%%%%%%%%%%%%%%%%%%%%%%%%%%%%%%%%%%%%%%%%%%%%%%%%%%%%%%%%%%%%%%%%%%%%%%%%%%%%%%%
%%%%%%%%%%%%%%%%%%%%%%%%%%%%%%%%%%%%%%%%%%%%%%%%%%%%%%%%%%%%%%%%%%%%%%%%%%%%%%%%%%%

\section{Gravitational Wave Tests of Gravitational Theory}
\label{gwaves}

%%%%%%%%%%%%%%%%%%%%%%%%%%%%%%%%%%%%%%%%%%%%%%%%%%%%%%%%%%%%%%%%%%%%%%%%%%%%%%%%%%%
%%%%%%%%%%%%%%%%%%%%%%%%%%%%%%%%%%%%%%%%%%%%%%%%%%%%%%%%%%%%%%%%%%%%%%%%%%%%%%%%%%%

\subsection{Gravitational wave observatories}
\label{gwobservatories}

Some time in the next decade, a new opportunity for testing
relativistic gravity will be realized, 
when a worldwide network of kilometer-scale, laser interferometric
gravitational wave observatories in the U.S.\ (LIGO pro\-ject), Europe
(VIRGO and GEO600 projects) and Japan (TAMA300 project) begins regular
detection and analysis of gravitational-wave signals from astrophysical
sources.
These
broad-band antennas will have the capability of detecting and
measuring the gravitational waveforms from astronomical sources in a
frequency band between about 10 Hz (the seismic noise cutoff) and
500 Hz (the photon counting noise cutoff), with a maximum
sensitivity to strain at around 100 Hz of $h \sim \Delta l/l \sim 10^{-22}$
(rms), for the kilometer-scale LIGO/VIRGO projects.  
The most promising source for detection and study of
the gravitational wave signal is the ``inspiralling compact binary''
-- a binary system of neutron stars or black holes (or one of each) in
the final minutes of a death spiral leading to a violent merger.
Such is the fate, for example, of the Hulse-Taylor binary pulsar 
B1913+16 in about 300 Myr, or the ``double pulsar'' J0737-3039
in about 85 Myr.  Given the expected sensitivity of the
``advanced LIGO'' (around 2010), which could see such sources out to
many hundreds of megaparsecs, it has been estimated that from 40 to
several hundred annual inspiral events could be detectable.
Other sources, such as supernova core collapse events, instabilities
in rapidly rotating newborn neutron stars, signals from
non-axisymmetric pulsars, and a stochastic background of waves, may be
detectable (for reviews, see~\cite{LIGO,snowmass}; for updates on
the status of various projects, see~\cite{fritschel,brillet}).

A similar
network of cryogenic resonant-mass gravitational antennas have  been  in
operation for many years, albeit at lower levels of sensitivity
($h \sim 10^{-19}$).  While modest improvements in sensitivity may be
expected in the future, these resonant detectors are not expected to
be competitive with the large interferometers, unless new designs
involving masses of spherical, or nearly spherical shape come to
fruition.  These systems are primarily sensitive to waves in
relatively narrow bands about frequencies in the hundreds to thousands
of Hz range~\cite{rome,allegro,niobe,auriga}, although future improvements
in sensitivity and increases in bandwidth may be possible~\cite{coccia03}.

In addition, plans are being developed for an orbiting laser
interferometer space antenna (LISA for short).  Such a system,
consisting of three spacecraft orbiting the sun in a triangular
formation separated from each other by five million kilometers,
would be sensitive primarily in the very low frequency band between
$10^{-4}$ and $10^{-1}$ Hz, with peak strain sensitivity of order $h \sim
10^{-23}$~\cite{danzmann}.

In addition to opening a new astronomical window, the
detailed observation of gravitational waves by such observatories may
provide the means to test general relativistic predictions for the
polarization and speed of the waves, for gravitational radiation
damping and for strong-field gravity.

%%%%%%%%%%%%%%%%%%%%%%%%%%%%%%%%%%%%%%%%%%%%%%%%%%%%%%%%%%%%%%%%%%%%%%%%%%%%%%%%%%%
%%%%%%%%%%%%%%%%%%%%%%%%%%%%%%%%%%%%%%%%%%%%%%%%%%%%%%%%%%%%%%%%%%%%%%%%%%%%%%%%%%%

\subsection{Polarization of gravitational waves}

A laser interferometric or resonant bar gravitational wave detector
whose scale is small compared to the gravitational wavelength
measures the local components of a symmetric $3\times3$ tensor which
is composed of the ``electric'' components of the Riemann curvature tensor,
$R_{0i0j}$, via the equation of geodesic deviation, given, for a pair
of freely falling particles by
$ {\ddot x}^i =  - R_{0i0j} x^j $,
where $x^i$ denotes the spatial separation.
In general there are six independent components, which can be expressed in
terms of polarizations (modes with specific transformation properties
under rotations and boosts).  Three are transverse to the direction of
propagation, with two representing quadrupolar deformations and one
representing a monopolar ``breathing'' deformation.  Three modes are
longitudinal, with one an axially symmetric
stretching mode in the propagation direction,
and one quadrupolar mode in each of the two orthogonal planes containing the
propagation direction.  Figure~\ref{wavemodes} shows the displacements
induced on a ring of freely falling test particles by each of these
modes.
General relativity predicts only the first two
transverse quadrupolar modes (a) and (b) independently of the source; these
correspond to the waveforms $h_+$
and $h_\times$
discussed earlier (note the $\cos 2 \phi $ and $\sin 2 \phi $
dependences of the displacements).
Massless scalar-tensor gravitational waves
can in addition contain the transverse breathing mode (c).  In massive
scalar-tensor theories, the longitudinal mode (d) can also be present, but
is
suppressed relative to (c) by a factor $(\lambda/\lambda_c)^2$, where $\lambda$
is the wavelength of the radiation, and $\lambda_c$ is the Compton
wavelength of the massive scalar.  More general
metric theories predict additional longitudinal modes,
up to the full complement of
six (TEGP 10.2~\cite{tegp}).

A suitable array of gravitational antennas could delineate or limit
the number of modes present in a given wave.  The strategy depends on
whether or not the source direction is known.
In general there are eight unknowns (six polarizations and two direction
cosines), but only six measurables ($R_{0i0j}$).  If the direction can
be established by either association of the waves with optical or
other observations, or by time-of-flight measurements between
separated detectors, then six suitably oriented detectors suffice to
determine all six components.  If the direction cannot be established,
then the system is underdetermined, and no unique solution can be
found.  However, if one assumes that only transverse waves are
present, then there are only three unknowns if the source direction is
known, or five unknowns otherwise.  Then the corresponding number
(three or five) of detectors can determine the polarization.  If
distinct evidence were found of any mode other than the two
transverse quadrupolar modes of GR, the result would be disastrous for
GR.  On the other hand, the absence of a breathing mode would not
necessarily rule out scalar-tensor gravity, because the strength
of that mode depends on the nature of the source.

\epubtkImage{}{
\begin{figure}[hptb]
  \def\epsfsize#1#2{0.4#1}
  \centerline{\epsfbox{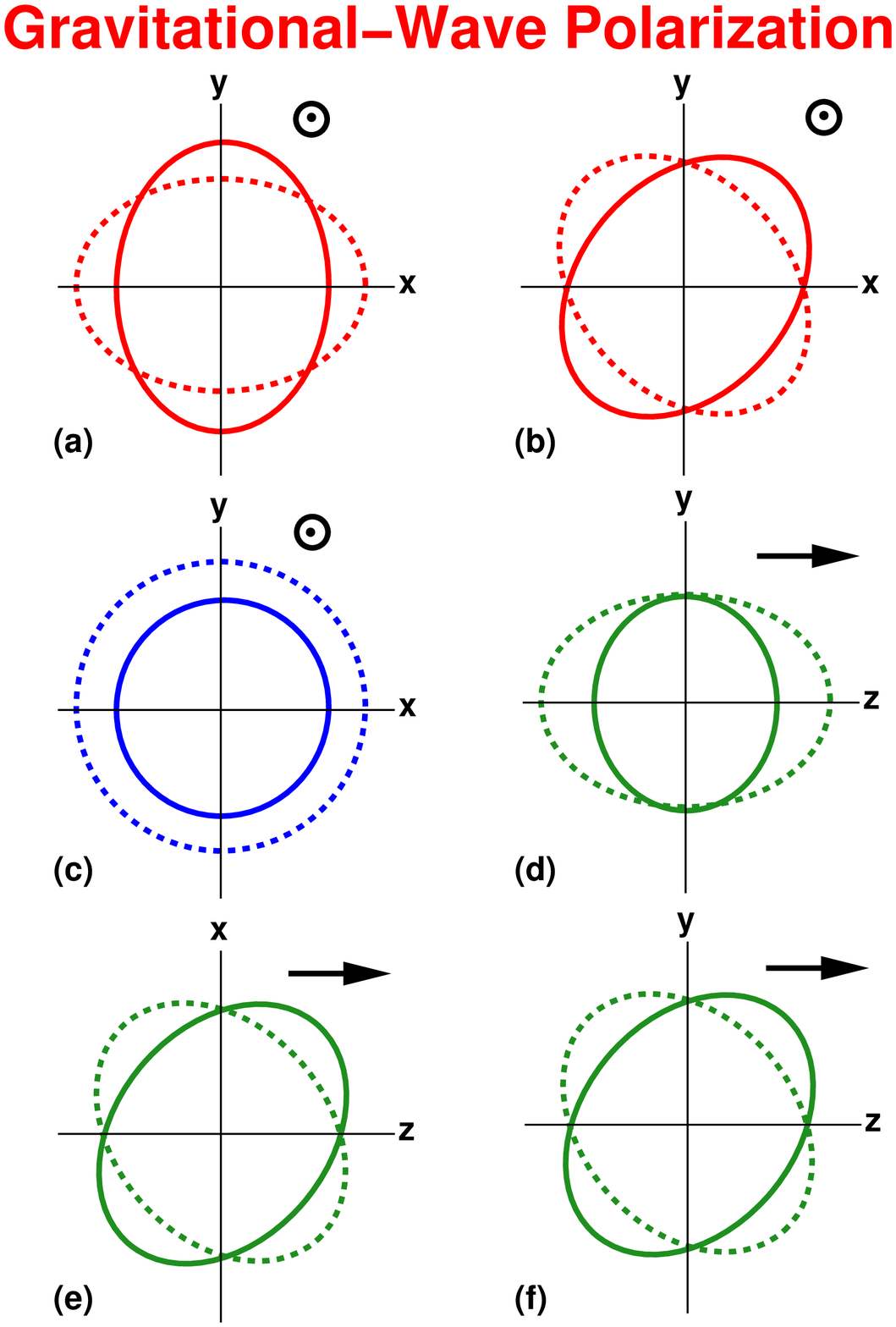}}
  \caption{\it The six polarization modes for gravitational waves
    permitted in any metric theory of gravity.  Shown is the
    displacement that each mode induces on a ring of test particles.
    The wave propagates in the $+z$ direction.  There is no
    displacement out of the plane of the picture.  In (a), (b) and
    (c), the wave propagates out of the plane; in (d), (e), and (f),
    the wave propagates in the plane.  In GR, only (a)
    and (b) are present; in massless scalar-tensor gravity, (c) may also be
    present.}
  \label{wavemodes}
\end{figure}}

Some
of the details of implementing such polarization observations have
been worked out for arrays of resonant cylindrical, disk-shaped,
spherical and truncated icosahedral
detectors (TEGP 10.2~\cite{tegp}, for recent reviews see~\cite{lobo,wagoner});
initial work has been done to assess whether
the ground-based or space-based
laser interferometers (or combinations of the two types) could perform
interesting polarization 
measurements~\cite{wagoner2,brunetti,maggiore,gasperini,WenSchutz05}.
Unfortunately for this purpose, the two LIGO observatories (in Washington
and Louisiana states, respectively) have been constructed to have their
respective arms as parallel as possible, apart from the curvature of the
Earth; while this maximizes the joint sensitivity of the two detectors to
gravitational waves, it minimizes their ability to detect two modes of
polarization.

%%%%%%%%%%%%%%%%%%%%%%%%%%%%%%%%%%%%%%%%%%%%%%%%%%%%%%%%%%%%%%%%%%%%%%%%%%%%%%%%%%%
%%%%%%%%%%%%%%%%%%%%%%%%%%%%%%%%%%%%%%%%%%%%%%%%%%%%%%%%%%%%%%%%%%%%%%%%%%%%%%%%%%%

\subsection{Gravitational radiation back-reaction}
\label{backreaction}

In the binary pulsar, a test of GR was made possible by measuring at
least three relativistic effects that depended upon only two unknown
masses.  The evolution of the orbital phase under the damping effect
of gravitational radiation played a crucial role.  Another situation
in which measurement of orbital phase can lead to tests of GR is that
of the inspiralling compact binary system.  The key differences are
that here gravitational radiation itself is the detected signal,
rather than radio pulses, and the phase evolution alone carries all
the information.  In the binary pulsar, the first derivative of the
binary frequency $\dot f_{\rm b}$, was measured; here the full nonlinear
variation of $f_{\rm b}$ as a function of time is measured.

Broad-band laser interferometers
are especially sensitive to the phase evolution of the gravitational
waves, which carry the information about the orbital phase evolution.
The analysis of gravitational wave data from such sources will involve
some form of matched filtering of the noisy detector output against an
ensemble of theoretical ``template'' waveforms which depend on the
intrinsic parameters of the inspiralling binary, such as the component
masses, spins, and so on, and on its inspiral evolution.
How accurate must a template be in order to
``match'' the waveform from a given source (where by a match we mean
maximizing the cross-correlation or the
signal-to-noise ratio)?  In the total accumulated phase
of the wave detected in the sensitive bandwidth, the template must
match the signal to a fraction of a cycle.  For two inspiralling
neutron stars, around 16,000 cycles should be detected during the final few
minutes of inspiral; this implies a
phasing accuracy of $10^{-5}$ or better.  Since $v \sim 1/10$ during
the late inspiral, this means that correction terms in the phasing
at the level of
$v^5$ or higher are needed.  More formal analyses confirm this
intuition~\cite{3min,finnchern,cutlerflan,poissonwill}.

Because it is a slow-motion system ($v \sim 10^{-3}$), the binary
pulsar is sensitive only to the lowest-order effects of gravitational
radiation as predicted by the quadrupole formula.  Nevertheless, the
first correction terms of order $v$ and $v^2$ to the quadrupole formula
were
calculated as early as 1976~\cite{wagwill} (see TEGP 10.3~\cite{tegp}).

But for laser interferometric observations of gravitational waves,
the bottom line is that, in order to measure the astrophysical
parameters of the source and to test the properties of the
gravitational waves, it is necessary to derive the
gravitational waveform and the resulting radiation back-reaction on the orbit
phasing at least to 2PN
order beyond the quadrupole
approximation, and preferably to 3PN
order.

For the special case of
non-spinning bodies moving on quasi-circular orbits
(i.e.\ circular apart from a slow inspiral), the evolution of the
gravitational wave frequency $f = 2f_{\rm b}$ through 2PN
order has the
form
{\setlength{\arraycolsep}{0.14 em}
\begin{eqnarray}
\dot f &=&{96\pi \over 5} f^2
(\pi {\cal M} f)^{5/3}
\biggl [ 1
- \left( {743 \over 336} + {11 \over 4} \eta \right) (\pi mf)^{2/3}
+ 4\pi(\pi mf)  \nonumber \\
&&  + \left( {34103 \over 18144} + {13661 \over 2016} \eta + {59 \over
18} \eta^2 \right)  (\pi mf)^{4/3}
+ O[(\pi mf)^{5/3}] \biggr ],
\label{fdot2PN}
\end{eqnarray}}%
where $\eta= m_1m_2/m^2$.  The first term is the quadrupole
contribution (compare Eq.~(\ref{fdotGR})),
the second term is the 1PN contribution, the
third term, with the coefficient $4\pi$,
is the ``tail'' contribution,
and the fourth term is the 2PN
contribution, first reported jointly by Blanchet {\em et al.}~\cite{bdiww,bdi2pn,opus}.
The 2.5PN, 3PN and 3.5PN contributions have also been calculated
(see~\cite{BlanchetLRR} for a review).

Similar expressions can be derived for the loss of angular momentum
and linear momentum.  Expressions for non-circular
orbits have also been derived~\cite{gopu,damgopuiyer04}. These losses react
back on the orbit to circularize it and cause it to inspiral.  The
result is that the orbital phase (and consequently the
gravitational wave
phase) evolves non-linearly with time.  It is the sensitivity of the
broad-band laser interferometric detectors to phase that makes the
higher-order contributions to $df/dt$ so observationally relevant.

If the coefficients of each of the powers of $f$ in Eq.~(\ref{fdot2PN})
can be measured, then one again obtains more than two constraints on
the two unknowns $m_1$ and $m_2$, leading to the possibility to test
GR.  For example,
Blanchet and Sathyaprakash~\cite{lucsathya,lucsathya2} have shown that, by
observing a source with a sufficiently strong signal, an interesting
test of the $4\pi$ coefficient of the ``tail'' term could be
performed.

Another possibility involves gravitational waves from a small mass
orbiting and inspiralling into a (possibly supermassive) spinning black hole.
A general non-circular, non-equatorial orbit will precess around the
hole, both in periastron and in orbital plane,
leading to a complex gravitational waveform that carries
information about the non-spherical, strong-field spacetime around the hole.
According to GR, this spacetime must be the Kerr spacetime of a
rotating black hole, uniquely specified by its mass and angular
momentum, and consequently, observation of the waves could test this
fundamental hypothesis of GR~\cite{ryan,poissonBH}.

Thirdly, the dipole gravitational radiation predicted by scalar-tensor
theories will result in a modification of the gravitational radiation
back-reaction, and thereby of the phase evolution.
Including only the leading quadrupole and dipole contributions, one
obtains, in Brans-Dicke theory,
\begin{equation}
\dot f ={96\pi \over 5} f^2
(\pi {\cal M} f)^{5/3}
\biggl [ 1 + b (\pi mf)^{-2/3}  \biggr ],
\label{fdotBD}
\end{equation}
where ${\cal M} = (\chi^{3/5} {\cal G}^{-4/5})\eta^{3/5} m$,
and $b$ is the coefficient of the dipole term, given by
$b= (5/48)(\chi^{-1} {\cal G}^{4/3} )\xi {\cal S}^2$, where
$\chi$, $\cal G$, $\cal S$ are given by Eqs.~(\ref {BDcoefficients}),
and $\xi = 1/(2+\omega_{\rm BD})$.
Double neutron star systems are not
promising because the small range of masses available near
$1.4 M_\odot$ results in suppression of dipole radiation by symmetry
For black holes, $s=0.5$ identically, consequently
double black-hole systems turn out to be observationally
identical in the two theories.
Thus mixed systems involving a neutron star and a
black hole are preferred.  However, a number of analyses of the
capabilities of both ground-based
and space-based (LISA) observatories have shown that observing waves from
neutron-star black-hole inspirals is not likely to bound scalar-tensor
gravity at a level competitive with the Cassini bound or with future
solar-system improvements~\cite{willbd,krolak95,scharrewill,willyunes,bbw1,bbw2}.

%%%%%%%%%%%%%%%%%%%%%%%%%%%%%%%%%%%%%%%%%%%%%%%%%%%%%%%%%%%%%%%%%%%%%%%%%%%%%%%%%%%
%%%%%%%%%%%%%%%%%%%%%%%%%%%%%%%%%%%%%%%%%%%%%%%%%%%%%%%%%%%%%%%%%%%%%%%%%%%%%%%%%%%

\subsection{Speed of gravitational waves}

According to GR, in the limit in which the wavelength of gravitational
waves is small compared to the radius of curvature of the background
spacetime, the waves propagate along null geodesics of the background
spacetime,i.e.\ they have the same speed $c$ as light (in this
section, we do not set $c=1$).  In other
theories, the speed could differ from $c$ because of coupling of
gravitation to ``background'' gravitational fields.  For example, in
the Rosen bimetric theory with a flat background metric
{\boldmath $\eta$},
gravitational waves follow null geodesics of {\boldmath $\eta$},
while light follows null geodesics of ${\bf {g}}$ (TEGP 10.1~\cite{tegp}).

Another way in which the speed of gravitational waves could differ
from $c$ is if gravitation were propagated by a massive field (a
massive graviton), in which case $v_{\rm g}$ would
be given by, in a local inertial frame,
\begin{equation}
{v_{\rm g}^2 \over c^2} = 1- {m_{\rm g}^2c^4 \over E^2},
\label{eq1}
\end{equation}
where $m_{\rm g}$ and $E$ are the graviton rest mass and energy,
respectively.  

The simplest attempt to incorporate a massive graviton into
general relativity in a ghost-free manner 
suffers from the so-called van Dam-Veltman-Zakharov
(vDVZ) discontinuity~\cite{vdv70,zakharov70}.  Because of the 3 additional
helicity states available to the massive spin-2 graviton, the limit of small
graviton mass does not coincide with pure GR, and the predicted
perihelion advance,
for example, violates experiment.
A model theory by Visser~\cite{visser} attempts to circumvent the vDVZ
problem by
introducing a non-dynamical flat-background metric.  This theory is truly
continuous with GR in the limit of vanishing graviton mass; on the other
hand, its observational implications have been only partially explored.
Braneworld scenarios predict a tower or a continuum of massive gravitons,
and may avoid the vDVZ discontinuity, although the full details are
still a work in progress~\cite{deffayet02,creminelli05}.

The most obvious way to test this is to
compare the arrival times of a gravitational wave and an
electromagnetic
wave from the same event, e.g.\ a supernova.
For a source at a distance $D$, the
resulting value of the difference $1-v_{\rm g}/c$ is
\begin{equation}
1- {v_{\rm g} \over c}= 5 \times 10^{-17} \left(
{{200 {\rm\ Mpc}} \over D} \right) \left( {{\Delta t}
\over {1 {\rm\ s}}} \right),
\label{eq2}
\end{equation}
where
$\Delta t \equiv \Delta t_{\rm a} - (1+Z) \Delta t_{\rm e} $ is the ``time difference'',
where $\Delta t_{\rm a}$ and $\Delta t_{\rm e}$ are the differences
in arrival time and emission time, respectively, of the
two signals, and $Z$ is the redshift of the source.
In many cases, $\Delta t_{\rm e}$ is unknown,
so that the best one can do is employ an upper bound on
$\Delta t_{\rm e}$ based on observation or modelling.
The result will then be a bound on $1-v_{\rm g}/c$.

For a massive graviton, if
the frequency of the gravitational waves is such that $hf \gg
m_{\rm g}c^2$,
where $h$ is Planck's constant, then $v_{\rm g}/c \approx 1- {1 \over 2}
(c/\lambda_{\rm g}
f)^{2}$, where $\lambda_{\rm g}= h/m_{\rm g}c$ is the graviton Compton wavelength,
and
the bound on $1-v_{\rm g}/c$ can be converted to a bound on $\lambda_{\rm g}$,
given
by
\begin{equation}
\lambda_{\rm g} > 3 \times 10^{12} {\rm\ km} \left( {D \over {200
      {\rm\ Mpc}}}
{{100 {\rm\ Hz}} \over f} \right)^{1/2} \left({1 \over
{f\Delta t}} \right)^{1/2}.
\label{eq4}
\end{equation}

The foregoing discussion assumes that the source emits {\it both}
gravitational and electromagnetic radiation in detectable amounts, and
that the relative time of emission can be established
to sufficient accuracy, or can be shown to be sufficiently
small.

However, there is a situation in which a bound on the graviton mass
can be set using gravitational radiation alone~\cite{graviton}.
That is the case of
the inspiralling compact binary.  Because the frequency of the
gravitational radiation sweeps from low frequency at the initial
moment of observation to higher frequency at the final moment, the
speed of the gravitons emitted will vary, from lower speeds initially
to higher speeds (closer to $c$) at the end.  This will cause a
distortion of the observed phasing of the waves and result in a
shorter than expected
overall time $\Delta t_{\rm a}$ of passage of a given number of cycles.
Furthermore, through the technique of matched filtering, the
parameters of the compact binary can be measured accurately,
(assuming that GR is a good approximation to the orbital evolution, even in
the presence of a massive graviton), and
thereby the emission time $\Delta t_{\rm e}$ can be determined accurately.
Roughly speaking, the ``phase interval'' $f\Delta t$ in Eq.
(\ref{eq4}) can be
measured to an accuracy $1/\rho$, where $\rho$ is the
signal-\-to-\-noise
ratio.

Thus one can estimate the bounds on $\lambda_{\rm g}$ achievable for various
compact inspiral systems, and for various detectors.   For
stellar-mass inspiral (neutron stars or black holes) observed by the
LIGO/VIRGO class of ground-based interferometers, $D \approx$
200~Mpc, $f \approx$ 100~Hz, and $f\Delta t \sim
\rho^{-1} \approx 1/10$.
The result is $\lambda_{\rm g} > 10^{13}$~km.  For supermassive binary
black holes ($10^4$ to $10^7 M_\odot$) observed by the proposed laser interferometer space antenna
(LISA), $D \approx$ 3~Gpc, $f \approx 10^{-3}$~Hz,
and $f\Delta t \sim
\rho^{-1} \approx 1/1000$. The result is $\lambda_{\rm g} >
10^{17}$~km.

A full noise analysis using proposed noise curves for the advanced
LIGO and for LISA weakens these crude bounds by factors between two and 
10~\cite{graviton,willyunes,bbw1,bbw2}.  For example, for the inspiral of
two $10^6 \, M_\odot$ black holes with aligned spins at a distance of 3 Gpc
observed by LISA,
a bound of $2 \times 10^{16}$ km could be placed~\cite{bbw1}.  Other
possibilities include using binary pulsar data to bound modifications of
gravitational radiation damping by a massive graviton~\cite{FinnSutton02},
and using LISA observations of the phasing of waves from compact white-dwarf
binaries, eccentric galactic binaries, and eccentric inspiral 
binaries~\cite{CutlerLarHis03,Jones05}.

Bounds obtainable from gravitational radiation effects
should be compared 
with the solid bound 
$\lambda_{\rm g} > 2.8 \times 10^{12} \, {\rm km}$,~\cite{talmadge}
derived from solar system dynamics, which limit
the presence of a Yukawa modification of Newtonian gravity of the form
\begin{equation}
V(r) =(GM/r)\exp(-r/\lambda_{\rm g}),
\end{equation}
and with the model-dependent bound
$\lambda_{\rm g} > 6\times 10^{19} \,{\rm km}$ from consideration of
galactic and cluster dynamics~\cite{visser}.

%%%%%%%%%%%%%%%%%%%%%%%%%%%%%%%%%%%%%%%%%%%%%%%%%%%%%%%%%%%%%%%%%%%%%%%%%%%%%%%%%%%
%%%%%%%%%%%%%%%%%%%%%%%%%%%%%%%%%%%%%%%%%%%%%%%%%%%%%%%%%%%%%%%%%%%%%%%%%%%%%%%%%%%

\subsection{Strong-gravity tests}
\label{other}

One of the central difficulties of testing GR in the
strong-field regime is the possibility of contamination by uncertain
or complex physics.  In the solar system, weak-field gravitational effects
could
in most cases be measured cleanly and separately from
non-gravitational effects.  The remarkable cleanliness of the binary
pulsar permitted precise measurements of gravitational phenomena
in a strong-field context.

Unfortunately, nature is rarely so kind.
Still, under suitable conditions, qualitative and even quantitative
strong-field tests of GR could be carried out.

One example is in cosmology.  From a few seconds after the big bang
until the present, the underlying physics of the universe is well
understood, in terms of a Standard Model of a nearly
spatially flat universe, 13.6 Byr old, 
dominated by dark matter and dark energy.
Some alternative theories of gravity that
are qualitatively different from GR fail to produce cosmologies that
meet even the minimum requirements of agreeing qualitatively with big-bang
nucleosynthesis (BBN) or the properties of the cosmic microwave background
(TEGP 13.2~\cite{tegp}).
Others, such as Brans-Dicke theory, are sufficiently close to GR (for
large enough $\omega_{\rm BD}$) that they conform to all cosmological
observations, given the underlying uncertainties.  The generalized
scalar-tensor theories, however, could have small $\omega$ at
early times, while evolving through the attractor mechanism to large
$\omega$ today.  One way to test such theories is through
big-bang nucleosynthesis, since the abundances of the light elements
produced when the temperature of the universe was about 1 MeV are
sensitive to the rate of expansion at that epoch, which in turn
depends on the strength of interaction between geometry and the
scalar field.  Because the universe is radiation-dominated at that
epoch, uncertainties in the amount of cold dark matter or of the
cosmological constant are unimportant.  The nuclear reaction rates are
reasonably well understood from laboratory experiments and theory, and
the number of light neutrino families (3) conforms to evidence from
particle accelerators.  Thus, within modest uncertainties, one can
assess the quantitative difference between the BBN predictions of GR and
scalar-tensor gravity under strong-field conditions and compare with
observations.
For recent analyses, see~\cite{Santiago97,damourpichon,Clifton05,coc06}.

Another example is the exploration of the spacetime near black holes
and neutron stars
via accreting matter.  
Studies of certain kinds of accretion known as advection-dominated
accretion flow (ADAF) in low-luminosity binary X-ray sources
may yield the signature of the black hole event
horizon~\cite{narayan}.
The spectrum of frequencies of quasi-periodic oscillations (QPO) from galactic
black-hole binaries may permit measurement of the spins of the black
holes~\cite{Psaltis04}.
Aspects of strong-field gravity and frame-dragging may be revealed in
spectral shapes of iron fluorescence lines from the inner regions of
accretion disks~\cite{reynolds04,reynolds05}.
Because of uncertainties in the detailed models, the results to date of 
studies like these are
suggestive at best, but the combination of higher-resolution
observations and better modelling could lead to striking tests of
strong-field predictions of GR.

%\newpage

%%%%%%%%%%%%%%%%%%%%%%%%%%%%%%%%%%%%%%%%%%%%%%%%%%%%%%%%%%%%%%%%%%%%%%%%%%%%%%%%%%%
%%%%%%%%%%%%%%%%%%%%%%%%%%%%%%%%%%%%%%%%%%%%%%%%%%%%%%%%%%%%%%%%%%%%%%%%%%%%%%%%%%%
%%%%%%%%%%%%%%%%%%%%%%%%%%%%%%%%%%%%%%%%%%%%%%%%%%%%%%%%%%%%%%%%%%%%%%%%%%%%%%%%%%%

\section{Conclusions}
\label{S5}

We find
that general relativity has held up under extensive experimental scrutiny.
The question then arises, why bother to continue to test it?  One
reason is that gravity is a fundamental interaction of nature, and as
such requires the most solid empirical underpinning we can provide.
Another is that all attempts to quantize gravity and to unify it with
the other forces suggest that the standard general relativity of Einstein is
not likely to be the last word.
Furthermore, the predictions of general relativity are fixed;
the theory contains
no adjustable constants so nothing can be changed.  Thus every test
of the theory is either a potentially deadly test or a possible probe for
new physics.  Although it is remarkable
that this theory, born 90 years ago out of almost pure thought,
has managed to survive every test, the possibility of finding
a discrepancy will continue to drive experiments for years to come.

%\newpage

%%%%%%%%%%%%%%%%%%%%%%%%%%%%%%%%%%%%%%%%%%%%%%%%%%%%%%%%%%%%%%%%%%%%%%%%%%%%%%%%%%%
%%%%%%%%%%%%%%%%%%%%%%%%%%%%%%%%%%%%%%%%%%%%%%%%%%%%%%%%%%%%%%%%%%%%%%%%%%%%%%%%%%%
%%%%%%%%%%%%%%%%%%%%%%%%%%%%%%%%%%%%%%%%%%%%%%%%%%%%%%%%%%%%%%%%%%%%%%%%%%%%%%%%%%%

\section{Acknowledgments}

This work has been supported since the initial version 
in part by the National Science Foundation,
Grant Numbers PHY 96-00049, 00-96522, and 03-53180, and by the National
Aeronautics and Space Administration, Grant Number NAG5-10186.  We also
gratefully acknowledge the support of the Centre National de la Recherche
Scientifique, and the hospitality of the Institut d'Astrophysique de Paris,
where this update was completed.  Comments from referees were particularly
helpful in improving this update.

\newpage

%%%%%%%%%%%%%%%%%%%%%%%%%%%%%%%%%%%%%%%%%%%%%%%%%%%%%%%%%%%%%%%%%%%%%%%%%%%%%%%%%%%
%%%%%%%%%%%%%%%%%%%%%%%%%%%%%%%%%%%%%%%%%%%%%%%%%%%%%%%%%%%%%%%%%%%%%%%%%%%%%%%%%%%
%%%%%%%%%%%%%%%%%%%%%%%%%%%%%%%%%%%%%%%%%%%%%%%%%%%%%%%%%%%%%%%%%%%%%%%%%%%%%%%%%%%

\bibliography{refs}

\begin{thebibliography}{100}

\bibitem{gg}
``GG Small Mission Project'', project homepage, (2005). URL (cited on 15 July
  2005): \newline\url{http://tycho.dm.unipi.it/~nobili/ggproject.html}.
  \epubtkKeywords{equivalence principle tests}

\bibitem{gaia}
``Taking The Galactic Census'', project homepage, (2005). URL (cited on 15 July
  2005): \newline\url{http://astro.estec.esa.nl/GAIA/}. \epubtkKeywords{tests
  of relativistic gravity}

\bibitem{LIGO}
Abramovici, A., Althouse, W.E., Drever, R.W.P., G{\"{u}}rsel, Y., Kawamura, S.,
  Raab, F.J., Shoemaker, D., Siewers, L., Spero, R.E., Thorne, K.S., Vogt,
  R.E., Weiss, R., Whitcomb, S.E., and Zucker, M.E., ``LIGO: The laser
  interferometer gravitational-wave observatory'', {\em Science}, {\bf 256},
  325--333, (1992). \epubtkKeywords{interferometric gravitational wave
  detectors}

\bibitem{adelberger01}
Adelberger, E.G., ``New tests of Einstein's equivalence principle and Newton's
  inverse-square law'', {\em Class. Quantum Grav.}, {\bf 18}, 2397--2405,
  (2001). \epubtkKeywords{equivalence principle tests, inverse square law}

\bibitem{adelberger03}
Adelberger, E.G., Heckel, B.R., and Nelson, A.E., ``Tests of the gravitational
  inverse-square law'', {\em Annu. Rev. Nucl. Sci.}, {\bf 53}, 77--121, (dec,
  2003). Related online version (cited on 15 July 2005):
  \newline\url{http://arXiv.org/abs/hep-ph/0307284}. \epubtkKeywords{inverse
  square law}

\bibitem{Adelberger91}
Adelberger, E.G., Heckel, B.R., Stubbs, C.W., and Rogers, W.F., ``Searches for
  new macroscopic forces'', {\em Annu. Rev. Nucl. Sci.}, {\bf 41}, 269--320,
  (1991). \epubtkKeywords{equivalence principle tests}

\bibitem{alvager}
Alv\"ager, T., Farley, F. J.~M., Kjellman, J., and Wallin, I., ``Test of the
  second postulate of special relativity in the GeV region'', {\em Phys.
  Lett.}, {\bf 12}, 260--262, (1977). \epubtkKeywords{tests of lorentz
  invariance}

\bibitem{AlvarezMann96b}
Alvarez, C., and Mann, R.B., ``The equivalence principle and anomalous magnetic
  moment experiments'', {\em Phys. Rev. D}, {\bf 54}, 7097--7107, (1996).
  Related online version (cited on 15 January 2001):
  \newline\url{http://arXiv.org/abs/gr-qc/9511028}. \epubtkKeywords{equivalence
  principle}

\bibitem{AlvarezMann96a}
Alvarez, C., and Mann, R.B., ``Testing the equivalence principle by Lamb shift
  energies'', {\em Phys. Rev. D}, {\bf 54}, 5954--5974, (1996). Related online
  version (cited on 15 January 2001):
  \newline\url{http://arXiv.org/abs/gr-qc/9507040}. \epubtkKeywords{equivalence
  principle}

\bibitem{AlvarezMann97a}
Alvarez, C., and Mann, R.B., ``The equivalence principle and g-2 experiments'',
  {\em Phys. Lett. B}, {\bf 409}, 83--87, (1997). Related online version (cited
  on 15 January 2001): \newline\url{http://arXiv.org/abs/gr-qc/9510070}.
  \epubtkKeywords{equivalence principle}

\bibitem{AlvarezMann97b}
Alvarez, C., and Mann, R.B., ``The equivalence principle in the non-baryonic
  regime'', {\em Phys. Rev. D}, {\bf 55}, 1732--1740, (1997). Related online
  version (cited on 15 January 2001):
  \newline\url{http://arXiv.org/abs/gr-qc/9609039}. \epubtkKeywords{equivalence
  principle}

\bibitem{AlvarezMann97c}
Alvarez, C., and Mann, R.B., ``Testing the equivalence principle using atomic
  vacuum energy shifts'', {\em Mod. Phys. Lett. A}, {\bf 11}, 1757--1763,
  (1997). Related online version (cited on 15 January 2001):
  \newline\url{http://arXiv.org/abs/gr-qc/9612031}. \epubtkKeywords{equivalence
  principle}

\bibitem{antoniadis98}
{Antoniadis}, I., {Arkani-Hamed}, N., {Dimopoulos}, S., and {Dvali}, G., ``{New
  dimensions at a millimeter to a fermi and superstrings at a TeV}'', {\em
  Physics Letters B}, {\bf 436}, 257--263, (sep, 1998). Related online version
  (cited on 15 February 2006):
  \newline\url{http://arXiv.org/abs/hep-ph/9804398}. \epubtkKeywords{extra
  dimensions, inverse square law}

\bibitem{antonini05}
Antonini, P., Okhapkin, M., Goklu, E., and Schiller, S., ``Test of constancy of
  speed of light with rotating cryogenic optical resonators'', {\em Phys. Rev.
  A}, {\bf 71}, 050101, (2005). \epubtkKeywords{tests of Lorentz invariance}

\bibitem{add98}
Arkani-Hamed, N., Dimopoulos, S., and Dvali, G., ``The hierarchy problem and
  new dimensions at a millimeter'', {\em Phys. Lett. B}, {\bf 429}, 263--272,
  (1998). Related online version (cited on 15 July 2005):
  \newline\url{http://arXiv.org/abs/hep-ph/9803315}. \epubtkKeywords{extra
  dimensions, inverse square law}

\bibitem{asada02}
{Asada}, H., ``The light cone effect on the Shapiro time delay'', {\em
  Astrophys. J. Lett.}, {\bf 574}, L69--L70, (2002). Related online version
  (cited on 15 July 2005): \newline\url{http://arXiv.org/abs/astro-ph/0206266}.
  \epubtkKeywords{speed of gravity, Shapiro time delay}

\bibitem{ashby1}
Ashby, N., ``Relativistic effects in the Global Positioning System'', in
  Dadhich, N., and Narlikar, J.V., eds., {\em Gravitation and Relativity: At
  the Turn of the Millenium}, Proceedings of the 15th International Conference
  on General Relativity and Gravitation (GR-15), held at IUCAA, Pune, India,
  December 16--21, 1997,  231--258, (Inter-University Center for Astronomy and
  Astrophysics, Pune, India, 1998). \epubtkKeywords{equivalence principle,
  global positioning system}

\bibitem{ashby2}
Ashby, N., ``Relativity in the Global Positioning System'', {\em Living Rev.
  Relativity}, {\bf 6}(1), lrr-2003-1, (2003). URL (cited on 15 July 2005):
  \newline\url{http://www.livingreviews.org/lrr-2003-1}. \epubtkKeywords{global
  positioning system}

\bibitem{ashby05}
Ashby, N., Bender, P.L., and Wahr, J.M., ``Gravitational physics tests from
  ranging to a Mercury orbiter'', unknown status, (2005). \epubtkKeywords{radar
  ranging}

\bibitem{ATNFpulsarcat}
ATNF/CSIRO, ``ATNF Pulsar Catalogue'', web interface to database. URL (cited on
  15 July 2005):
  \newline\url{http://www.atnf.csiro.au/research/pulsar/psrcat/}.
  \epubtkKeywords{pulsars, neutron stars}

\bibitem{baessler99}
Baessler, S., Heckel, B.R., Adelberger, E.G., Gundlach, J.H., Schmidt, U., and
  Swanson, H.E., ``Improved test of the equivalence principle for gravitational
  self-energy'', {\em Phys. Rev. Lett.}, {\bf 83}, 3585--3588, (1999).
  \epubtkKeywords{equivalence principle tests}

\bibitem{bailes03}
Bailes, M., Ord, S.M., Knight, H.S., and Hotan, A.W., ``Self-consistency of
  relativistic observables with general relativity in the white dwarf-neutron
  star binary PSR J1141-6545'', {\em Astrophys. J. Lett.}, {\bf 595}, L49--L52,
  (2003). Related online version (cited on 15 July 2005):
  \newline\url{http://arXiv.org/abs/astro-ph/0307468}. \epubtkKeywords{binary
  pulsars}

\bibitem{bambi04}
Bambi, C., Giannotti, M., and Villante, F.L., ``Response of primordial
  abundances to a general modification of $G_{N}$ and/or of the early universe
  expansion rate'', {\em Phys. Rev. D}, {\bf 71}, 123524, (2005). Related
  online version (cited on 15 July 2005):
  \newline\url{http://arXiv.org/abs/astro-ph/0503502}. \epubtkKeywords{big bang
  nucleosynthesis}

\bibitem{bartlett}
Bartlett, D.F., and van Buren, D., ``Equivalence of active and passive
  gravitational mass using the moon'', {\em Phys. Rev. Lett.}, {\bf 57},
  21--24, (1986). \epubtkKeywords{Tests of relativistic gravity, lunar laser
  ranging}

\bibitem{bauch02}
Bauch, A., and Weyers, S., ``New experimental limit on the validity of local
  position invariance'', {\em Phys. Rev. D}, {\bf 65}, 081101, (2002).
  \epubtkKeywords{equivalence principle tests}

\bibitem{BaumgarteShapiro03}
Baumgarte, T.W., and Shapiro, S.L., ``Numerical relativity and compact
  binaries'', {\em Phys. Rep.}, {\bf 376}, 41--131, (2003). Related online
  version (cited on 15 July 2005):
  \newline\url{http://arXiv.org/abs/gr-qc/0211028}. \epubtkKeywords{numerical
  relativity}

\bibitem{Bell2}
Bell, J.F., Camilo, F., and Damour, T., ``A tighter test of local Lorentz
  invariance using PSR J2317+1439'', {\em Astrophys. J.}, {\bf 464}, 857,
  (1996). Related online version (cited on 15 January 2001):
  \newline\url{http://arXiv.org/abs/astro-ph/9512100}. \epubtkKeywords{tests of
  relativistic gravity, binary pulsars}

\bibitem{Bell}
Bell, J.F., and Damour, T., ``A new test of conservation laws and Lorentz
  invariance in relativistic gravity'', {\em Class. Quantum Grav.}, {\bf 13},
  3121--3128, (1996). Related online version (cited on 15 January 2001):
  \newline\url{http://arXiv.org/abs/gr-qc/9606062}. \epubtkKeywords{tests of
  relativistic gravity, binary pulsars}

\bibitem{bbw1}
Berti, E., Buonanno, A., and Will, C.M., ``Estimating spinning binary
  parameters and testing alternative theories of gravity with LISA'', {\em
  Phys. Rev. D}, {\bf 71}, 084025, (2005). Related online version (cited on 15
  July 2005): \newline\url{http://arXiv.org/abs/gr-qc/0411129}.
  \epubtkKeywords{gravitational wave data analysis, parameter estimation}

\bibitem{bbw2}
Berti, E., Buonanno, A., and Will, C.M., ``Testing general relativity and
  probing the merger history of massive black holes with LISA'', {\em Class.
  Quantum Grav.}, {\bf 22}, S943--S954, (2005). Related online version (cited
  on 15 July 2005): \newline\url{http://arXiv.org/abs/gr-qc/0504017}.
  \epubtkKeywords{gravitational wave data analysis, parameter estimation}

\bibitem{bertotti03}
Bertotti, B., Iess, L., and Tortora, P., ``A test of general relativity using
  radio links with the Cassini spacecraft'', {\em Nature}, {\bf 425}, 374--376,
  (September, 2003). \epubtkKeywords{tests of relativistic gravity, Shapiro
  time delay}

\bibitem{biller}
Biller, S.D., Breslin, A.C., Buckley, J., Catanese, M., Carson, M.,
  Carter-Lewis, D.A., Cawley, M.F., Fegan, D.J., Finley, J.P., Gaidos, J.A.,
  Hillas, A.M., Krennrich, F., Lamb, R.C., Lessard, R., Masterson, C., McEnery,
  J.E., McKernan, B., Moriarty, P., Quinn, J., Rose, H.J., Samuelson, F.,
  Sembroski, G., Skelton, P., and Weekes, T.C., ``Limits to quantum gravity
  effects on energy dependence of the speed of light from observations of TeV
  flares in active galaxies'', {\em Phys. Rev. Lett.}, {\bf 82}, 2108--2111,
  (1999). Related online version (cited on 15 July 2005):
  \newline\url{http://arXiv.org/abs/gr-qc/9810044}. \epubtkKeywords{tests of
  Lorentz invariance, quantum gravity}

\bibitem{bize03}
Bize, S., Diddams, S.~A., Tanaka, U., Tanner, C.~E., Oskay, W.~H., Drullinger,
  R.~E., Parker, T.~E., Heavner, T.~P., Jefferts, S.~R., Hollberg, L., Itano,
  W.~M., and Bergquist, J.~C., ``Testing the stability of Fundamental Constants
  with the Hg$^+$ single-ion optical clock'', {\em Phys. Rev. Lett.}, {\bf 90},
  150802--1--4, (2003). Related online version (cited on 15 July 2005):
  \newline\url{http://arXiv.org/abs/physics/0212109}.
  \epubtkKeywords{equivalence principle tests}

\bibitem{niobe}
Blair, D.G., Heng, I.S., Ivanov, E.N., and Tobar, M.E., ``Present status of the
  resonant-mass gravitational-wave antenna NIOBE'', in Coccia, E., Veneziano,
  G., and Pizzella, G., eds., {\em Second Edoardo Amaldi Conference on
  Gravitational Waves}, Proceedings of the conference, held at CERN,
  Switzerland, 1--4 July, 1997, Edoardo Amaldi Foundation Series,  127--147,
  (World Scientific, Singapore; River Edge, U.S.A., 1998).
  \epubtkKeywords{resonant gravitational wave detectors}

\bibitem{blanchet95}
Blanchet, L., ``Second-post-Newtonian generation of gravitational radiation'',
  {\em Phys. Rev. D}, {\bf 51}, 2559--2583, (1995). Related online version
  (cited on 15 January 2001): \newline\url{http://arXiv.org/abs/gr-qc/9501030}.
  \epubtkKeywords{gravitational radiation, post-Newtonian approximations,
  binary systems}

\bibitem{BlanchetLRR}
Blanchet, L., ``Gravitational Radiation from Post-Newtonian Sources and
  Inspiralling Compact Binaries'', {\em Living Rev. Relativity}, {\bf 5},
  lrr-2002-3, (2002). URL (cited on 15 July 2005):
  \newline\url{http://www.livingreviews.org/lrr-2002-3}.
  \epubtkKeywords{gravitational radiation, post-Newtonian approximations,
  binary systems}

\bibitem{bd86}
Blanchet, L., and Damour, T., ``Radiative gravitational fields in general
  relativity. I. General structure of the field outside the source'', {\em
  Philos. Trans. R. Soc. London, Ser. A}, {\bf 320}, 379--430, (1986).
  \epubtkKeywords{Gravitational radiation, Post-Newtonian approximations,
  binary systems}

\bibitem{bd88}
Blanchet, L., and Damour, T., ``Tail-transported temporal correlations in the
  dynamics of a gravitating system'', {\em Phys. Rev. D}, {\bf 37}, 1410--1435,
  (1988). \epubtkKeywords{Gravitational radiation, Post-Newtonian
  approximations}

\bibitem{bd89}
Blanchet, L., and Damour, T., ``Post-Newtonian generation of gravitational
  waves'', {\em Ann. Inst. Henri Poincare A}, {\bf 50}, 377--408, (1989).
  \epubtkKeywords{Gravitational radiation, Post-Newtonian approximations,
  binary systems}

\bibitem{bd92}
Blanchet, L., and Damour, T., ``Hereditary effects in gravitational
  radiation'', {\em Phys. Rev. D}, {\bf 46}, 4304--4319, (1992).
  \epubtkKeywords{gravitational radiation, post-Newtonian approximations}

\bibitem{bdi2pn}
Blanchet, L., Damour, T., and Iyer, B.R., ``Gravitational waves from
  inspiralling compact binaries: Energy loss and waveform to
  second-post-Newtonian order'', {\em Phys. Rev. D}, {\bf 51}, 5360--5386,
  (1995). Related online version (cited on 15 January 2001):
  \newline\url{http://arXiv.org/abs/gr-qc/9501029}. Erratum: Phys. Rev. D 54
  (1996) 1860. \epubtkKeywords{gravitational radiation, post-Newtonian
  approximations binary systems}

\bibitem{bdiww}
Blanchet, L., Damour, T., Iyer, B.R., Will, C.M., and Wiseman, A.G.,
  ``Gravitational-radiation damping of compact binary systems to second
  post-Newtonian order'', {\em Phys. Rev. Lett.}, {\bf 74}, 3515--3518, (1995).
  Related online version (cited on 15 January 2001):
  \newline\url{http://arXiv.org/abs/gr-qc/9501027}.
  \epubtkKeywords{gravitational radiation, post-Newtonian approximations,
  binary systems}

\bibitem{lucsathya2}
Blanchet, L., and Sathyaprakash, B.S., ``Signal analysis of gravitational wave
  tails'', {\em Class. Quantum Grav.}, {\bf 11}, 2807--2831, (1994).
  \epubtkKeywords{Gravitational radiation, Parameter estimation}

\bibitem{lucsathya}
Blanchet, L., and Sathyaprakash, B.S., ``Detecting a tail effect in
  gravitational-wave experiments'', {\em Phys. Rev. Lett.}, {\bf 74},
  1067--1070, (1995). \epubtkKeywords{Gravitational radiation, parameter
  estimation}

\bibitem{braginsky}
Braginsky, V.B., and Panov, V.I., ``Verification of the equivalence of inertial
  and gravitational mass'', {\em Sov. Phys. JETP}, {\bf 34}, 463--466, (1972).
  \epubtkKeywords{equivalence principle tests}

\bibitem{brecher}
Brecher, K., ``Is the speed of light independent of the velocity of the
  source?'', {\em Phys. Rev. Lett.}, {\bf 39}, 1051--1054, (1977).
  \epubtkKeywords{tests of lorentz invariance}

\bibitem{brillet}
Brillet, A., ``VIRGO -- Status Report, November 1997'', in Coccia, E.,
  Veneziano, G., and Pizzella, G., eds., {\em Second Edoardo Amaldi Conference
  on Gravitational Waves}, Proceedings of the conference, held at CERN,
  Switzerland, 1--4 July, 1997, vol.~4 of Edoardo Amaldi Foundation Series,
  86--96, (World Scientific, Singapore; River Edge, U.S.A., 1998).
  \epubtkKeywords{interferometric gravitational wave detectors}

\bibitem{brillethall}
Brillet, A., and Hall, J.L., ``Improved laser test of the isotropy of space'',
  {\em Phys. Rev. Lett.}, {\bf 42}, 549--552, (1979). \epubtkKeywords{tests of
  Lorentz invariance}

\bibitem{brunetti}
Brunetti, M., Coccia, E., Fafone, V., and Fucito, F., ``Gravitational-wave
  radiation from compact binary systems in the Jordan-Brans-Dicke theory'',
  {\em Phys. Rev. D}, {\bf 59}, 044027, (1999). Related online version (cited
  on 15 January 2001): \newline\url{http://arXiv.org/abs/gr-qc/9805056}.
  \epubtkKeywords{gravitational radiation, scalar-tensor gravity}

\bibitem{burgay03}
Burgay, M., D'Amico, N., Possenti, A., Manchester, R.N., Lyne, A.G., Joshi,
  B.C., McLaughlin, M.A., Kramer, M., Sarkissian, J.M., Camilo, F., Kalogera,
  V., Kim, C., and Lorimer, D.R., ``An increased estimate of the merger rate of
  double neutron stars from observations of a highly relativistic system'',
  {\em Nature}, {\bf 426}, 531--533, (December, 2003). Related online version
  (cited on 15 July 2005): \newline\url{http://arXiv.org/abs/astro-ph/0312071}.
  \epubtkKeywords{binary pulsars}

\bibitem{carlip04}
Carlip, S., ``Model-dependence of Shapiro time delay and the ``speed of
  gravity/speed of light'' controversy'', {\em Class. Quantum Grav.}, {\bf 21},
  3803--3812, (2004). Related online version (cited on 15 July 2005):
  \newline\url{http://arXiv.org/abs/gr-qc/0403060}. \epubtkKeywords{Shapiro
  time delay, speed of gravity}

\bibitem{Champeney}
Champeney, D.~C., Isaak, G.~R., and Khan, A.~M., ``An ``aether drift''
  experiment based on the M\"ossbauer effect'', {\em Phys. Lett.}, {\bf 7},
  241--243, (1963). \epubtkKeywords{tests of Lorentz invariance}

\bibitem{petitjean2}
Chand, H., Petitjean, P., Srianand, R., and Aracil, B., ``Probing the
  time-variation of the fine-structure constant: Results based on Si IV
  doublets from a UVES sample'', {\em Astron. Astrophys.}, {\bf 430}, 47--58,
  (2005). Related online version (cited on 15 July 2005):
  \newline\url{http://arXiv.org/abs/astro-ph/0408200}.
  \epubtkKeywords{equivalence principle tests}

\bibitem{kapitulnik}
Chiaverini, J., Smullin, S.J., Geraci, A.A., Weld, D.M., and Kapitulnik, A.,
  ``New experimental constraints on non-Newtonian forces below 100 {$\mu$}m'',
  {\em Phys. Rev. Lett.}, {\bf 90}, 151101, (2003). Related online version
  (cited on 15 July 2005): \newline\url{http://arXiv.org/abs/hep-ph/0209325}.
  \epubtkKeywords{inverse square law}

\bibitem{chupp}
Chupp, T.~E., Hoare, R.~J., Loveman, R.~A., Oteiza, E.~R., Richardson, J.~M.,
  Wagshul, M.~E., and Thompson, A.~K., ``Results of a new test of local Lorentz
  invariance: A search for mass anisotropy in $^{21}$Ne'', {\em Phys. Rev.
  Lett.}, {\bf 63}, 1541--1545, (1989). \epubtkKeywords{tests of Lorentz
  invariance}

\bibitem{ciufolini00}
Ciufolini, I., ``The 1995-99 measurements of the Lense-Thirring effect using
  laser-ranged satellites'', {\em Class. Quantum Grav.}, {\bf 17}, 2369--2380,
  (2000). \epubtkKeywords{Lense-Thirring effect}

\bibitem{ciufolini97}
Ciufolini, I., Chieppa, F., Lucchesi, D., and Vespe, F., ``Test of Lense -
  Thirring orbital shift due to spin'', {\em Class. Quantum Grav.}, {\bf 14},
  2701--2726, (1997). \epubtkKeywords{Lense-Thirring effect}

\bibitem{ciufolini98}
Ciufolini, I., Pavlis, E., Chieppa, F., Fernandes-Vieira, E., and
  P{\'{e}}rez-Mercader, J., ``Test of general relativity and measurement of the
  Lense-Thirring effect with two Earth satellites'', {\em Science}, {\bf 279},
  2100--2103, (1998). \epubtkKeywords{Lense-Thirring effect}

\bibitem{ciufolini04}
Ciufolini, I., and Pavlis, E.C., ``A confirmation of the general relativistic
  prediction of the Lense--Thirring effect'', {\em Nature}, {\bf 431},
  958--960, (October, 2004). \epubtkKeywords{Lense-Thirring effect}

\bibitem{Clifton05}
Clifton, T., Barrow, J.D., and Scherrer, R.J., ``Constraints on the variation
  of G from primordial nucleosynthesis'', {\em Phys. Rev. D}, {\bf 71},
  123526--1--11, (2005). Related online version (cited on 15 July 2005):
  \newline\url{http://arXiv.org/abs/astro-ph/0504418}. \epubtkKeywords{big bang
  nucleosynthesis}

\bibitem{microscope}
CNES, ``MICROSCOPE (MICRO-Satellite \`a Tra\^in\'ee Compens\'ee pour
  l'Observation du Principe d'Equivalence)'', project homepage. URL (cited on
  15 July 2005): \newline\url{http://smsc.cnes.fr/MICROSCOPE/}.
  \epubtkKeywords{equivalence principle tests}

\bibitem{coc06}
Coc, A., Olive, K.~A., Uzan, J.-P., and Vangioni, E., ``{Big bang
  nucleosynthesis constraints on scalar-tensor theories of gravity}'', (2006).
  URL (cited on 15 February 2006):
  \newline\url{http://arXiv.org/abs/gr-qc/0601299}. \epubtkKeywords{big bang
  nucleosynthesis}

\bibitem{coccia03}
{Coccia}, E., ``{Resonant-mass detectors of gravitational waves in the short-
  and medium-term future}'', {\em Class. Quantum Grav.}, {\bf 20}, 135--+,
  (may, 2003). \epubtkKeywords{resonant gravitational wave detectors}

\bibitem{Coley82}
Coley, A., ``{Schiff's Conjecture on Gravitation}'', {\em Phys. Rev. Lett.},
  {\bf 49}, 853--855, (1982). \epubtkKeywords{equivalence principle}

\bibitem{colladay97}
Colladay, D., and Kosteleck\'y, V.A., ``CPT violation and the standard model'',
  {\em Phys. Rev. D}, {\bf 55}, 6760--6774, (1997). Related online version
  (cited on 15 July 2005): \newline\url{http://arXiv.org/abs/hep-ph/9703464}.
  \epubtkKeywords{tests of Lorentz invariance}

\bibitem{colladay98}
Colladay, D., and Kosteleck\'y, V.A., ``Lorentz-violating extension of the
  standard model'', {\em Phys. Rev. D}, {\bf 58}, 116002, (1998). Related
  online version (cited on 15 July 2005):
  \newline\url{http://arXiv.org/abs/hep-ph/9809521}. \epubtkKeywords{tests of
  Lorentz invariance}

\bibitem{copi04}
Copi, C.J., Davis, A.N., and Krauss, L.M., ``New nucleosynthesis constraint on
  the variation of G'', {\em Phys. Rev. Lett.}, {\bf 92}(17), 171301, (2004).
  Related online version (cited on 15 July 2005):
  \newline\url{http://arXiv.org/abs/astro-ph/0311334}. \epubtkKeywords{big bang
  nucleosynthesis}

\bibitem{creminelli05}
{Creminelli}, P., {Nicolis}, A., {Papucci}, M., and {Trincherini}, E.,
  ``{Ghosts in massive gravity}'', {\em Journal of High Energy Physics}, {\bf
  9}, 3, (sep, 2005). Related online version (cited on 15 February 2006):
  \newline\url{http://arXiv.org/abs/hep-th/0505147}. \epubtkKeywords{gravitons}

\bibitem{3min}
Cutler, C., Apostolatos, T.A., Bildsten, L., Finn, L.S., Flanagan,
  {\'{E}}.{\'{E}}., Kennefick, D., Markovi{\'{c}}, D.M., Ori, A., Poisson, E.,
  Sussman, G.J., and Thorne, K.S., ``The last three minutes: Issues in
  gravitational wave measurements of coalescing compact binaries'', {\em Phys.
  Rev. Lett.}, {\bf 70}, 2984--2987, (1993). Related online version (cited on
  15 January 2001): \newline\url{http://arXiv.org/abs/astro-ph/9208005}.
  \epubtkKeywords{gravitational radiation, parameter estimation}

\bibitem{cutlerflan}
Cutler, C., and Flanagan, {\'{E}}.{\'{E}}., ``Gravitational waves from merging
  compact binaries: How accurately can one extract the binary's parameters from
  the inspiral waveform?'', {\em Phys. Rev. D}, {\bf 49}, 2658--2697, (1994).
  Related online version (cited on 15 January 2001):
  \newline\url{http://arXiv.org/abs/gr-qc/9402014}.
  \epubtkKeywords{gravitational radiation, parameter estimation}

\bibitem{CutlerLarHis03}
Cutler, C., Hiscock, W.A., and Larson, S.L., ``LISA, binary stars, and the mass
  of the graviton'', {\em Phys. Rev. D}, {\bf 67}, 024015, (2003). Related
  online version (cited on 15 July 2005):
  \newline\url{http://arXiv.org/abs/gr-qc/0209101}. \epubtkKeywords{graviton,
  parameter estimation}

\bibitem{Damour300}
Damour, T., ``The problem of motion in Newtonian and Einsteinian gravity'', in
  Hawking, S.W., and Israel, W., eds., {\em Three Hundred Years of
  Gravitation},  128--198, (Cambridge University Press, Cambridge, U.K.; New
  York, U.S.A., 1987). \epubtkKeywords{equations of motion, post-Newtonian
  approximations, gravitational radiation}

\bibitem{damourreview}
Damour, T., ``Gravitation, experiment and cosmology'', in Gazis, E.N.,
  Koutsoumbas, G., Tracas, N.D., and Zoupanos, G., eds., {\em Proceedings of
  the 5th Hellenic School and Workshops on Elementary Particle Physics},
  Proceedings of the workshops, held at Corfu, Greece, 3--24 September 1995,
  332--368. Corfu Summer Institute, (1995). Related online version (cited on 15
  January 2001): \newline\url{http://arXiv.org/abs/gr-qc/9606079}.
  \epubtkKeywords{tests of relativistic gravity}

\bibitem{DamourPDG}
Damour, T., ``Experimental tests of gravitational theory'', {\em Phys. Lett.},
  {\bf 592B}, 1, (2004). \epubtkKeywords{tests of relativistic gravity}

\bibitem{damourdyson}
Damour, T., and Dyson, F., ``The Oklo bound on the time variation of the
  fine-structure constant revisited'', {\em Nucl. Phys. B}, {\bf 480}, 37--54,
  (1996). Related online version (cited on 15 January 2001):
  \newline\url{http://arXiv.org/abs/hep-ph/9606486}. \epubtkKeywords{tests of
  relativistic gravity}

\bibitem{DamourEspo92}
Damour, T., and Esposito-Far{\`{e}}se, G., ``Tensor-multi-scalar theories of
  gravitation'', {\em Class. Quantum Grav.}, {\bf 9}, 2093--2176, (1992).
  \epubtkKeywords{theories of gravity, scalar-tensor gravity}

\bibitem{DamourEspo98}
Damour, T., and Esposito-Far{\`{e}}se, G., ``Gravitational-wave versus
  binary-pulsar tests of strong-field gravity'', {\em Phys. Rev. D}, {\bf 58},
  042001, (1998). Related online version (cited on 15 January 2001):
  \newline\url{http://arXiv.org/abs/gr-qc/9803031}. \epubtkKeywords{Theories of
  gravity, scalar-tensor gravity, binary pulsars}

\bibitem{damgopuiyer04}
{Damour}, T., {Gopakumar}, A., and {Iyer}, B.~R., ``{Phasing of gravitational
  waves from inspiralling eccentric binaries}'', {\em Phys. Rev. D}, {\bf
  70}(6), 064028, (sep, 2004). Related online version (cited on 15 February,
  2006): \newline\url{http://arXiv.org/abs/gr-qc/0404128}.
  \epubtkKeywords{gravitational radiation, parameter estimation, binary
  systems}

\bibitem{di91}
Damour, T., and Iyer, B.R., ``Post-Newtonian generation of gravitational waves.
  II. The spin moments'', {\em Ann. Inst. Henri Poincare A}, {\bf 54},
  115--164, (1991). \epubtkKeywords{gravitational radiation, post-Newtonian
  approximations}

\bibitem{damjaraschaefer}
Damour, T., Jaranowski, P., and Sch{\"{a}}fer, G., ``Poincar\'e invariance in
  the ADM Hamiltonian approach to the general relativistic two-body problem'',
  {\em Phys. Rev. D}, {\bf 62}, 021501--1--5, (2000). Related online version
  (cited on 15 January 2001): \newline\url{http://arXiv.org/abs/gr-qc/0003051}.
  Erratum: Phys.Rev. D 63 (2001) 029903. \epubtkKeywords{equations of motion,
  post-Newtonian approximations, binary systems}

\bibitem{DJSdim}
Damour, T., Jaranowski, P., and Sch\"afer, G., ``Dimensional regularization of
  the gravitational interaction of point masses'', {\em Phys. Lett. B}, {\bf
  513}, 147--155, (2001). \epubtkKeywords{equations of motion, post-Newtonian
  approximations, binary systems}

\bibitem{DJSequiv}
Damour, T., Jaranowski, P., and Sch\"afer, G., ``Equivalence between the
  ADM-Hamiltonian and the harmonic-coordinates approaches to the third
  post-Newtonian dynamics of compact binaries'', {\em Phys. Rev. D}, {\bf 63},
  044021, (2001). Erratum Phys. Rev. D {\bf 66}, 029901(E) (2002).
  \epubtkKeywords{equations of motion, post-Newtonian approximations, binary
  systems}

\bibitem{DamourNord93a}
Damour, T., and Nordtvedt, K., ``General relativity as a cosmological attractor
  of tensor-scalar theories'', {\em Phys. Rev. Lett.}, {\bf 70}, 2217--2219,
  (1993). \epubtkKeywords{theories of gravity, scalar-tensor gravity,
  cosmology}

\bibitem{DamourNord93b}
Damour, T., and Nordtvedt, K., ``Tensor-scalar cosmological models and their
  relaxation toward general relativity'', {\em Phys. Rev. D}, {\bf 48},
  3436--3450, (1993). \epubtkKeywords{theories of gravity, scalar-tensor
  gravity, cosmology}

\bibitem{damourpiazza02a}
Damour, T., Piazza, F., and Veneziano, G., ``Runaway dilaton and equivalence
  principle violations'', {\em Phys. Rev. Lett.}, {\bf 89}, 081601--1--4,
  (2002). Related online version (cited on 15 July 2005):
  \newline\url{http://arXiv.org/abs/gr-qc/0204094}. \epubtkKeywords{string
  theory, equivalence principle}

\bibitem{damourpiazza02b}
Damour, T., Piazza, F., and Veneziano, G., ``Violations of the equivalence
  principle in a dilaton-runaway scenario'', {\em Phys. Rev. D}, {\bf 66},
  046007, (2002). Related online version (cited on 15 July 2005):
  \newline\url{http://arXiv.org/abs/hep-th/0205111}. \epubtkKeywords{string
  theory, equivalence principle}

\bibitem{damourpichon}
Damour, T., and Pichon, B., ``Big bang nucleosynthesis and tensor-scalar
  gravity'', {\em Phys. Rev. D}, {\bf 59}, 123502, (1999). Related online
  version (cited on 15 January 2001):
  \newline\url{http://arXiv.org/abs/astro-ph/9807176}. \epubtkKeywords{theories
  of gravity, scalar-tensor gravity, big bang nucleosynthesis}

\bibitem{DamourPolyakov}
Damour, T., and Polyakov, A.M., ``The string dilaton and a least coupling
  principle'', {\em Nucl. Phys. B}, {\bf 423}, 532--558, (1994). Related online
  version (cited on 15 July 2005):
  \newline\url{http://arXiv.org/abs/hep-th/9401069}. \epubtkKeywords{string
  theory, equivalence principle}

\bibitem{DamourSchaefer91}
{Damour}, T., and {Schaefer}, G., ``{New tests of the strong equivalence
  principle using binary-pulsar data}'', {\em Physical Review Letters}, {\bf
  66}, 2549--2552, (May, 1991). \epubtkKeywords{equivalence principle, binary
  pulsars}

\bibitem{DamourTaylor92}
Damour, T., and Taylor, J.~H., ``Strong-field tests of relativistic gravity and
  binary pulsars'', {\em Phys. Rev. D}, {\bf 45}, 1840--1868, (1992).
  \epubtkKeywords{Tests of relativistic gravity, binary pulsars}

\bibitem{DamourTaylor91}
Damour, T., and Taylor, J.H., ``On the orbital period change of the binary
  pulsar PSR 1913+16'', {\em Astrophys. J.}, {\bf 366}, 501--511, (1991).
  \epubtkKeywords{gravitational radiation, binary pulsars}

\bibitem{DamourVokrou96}
Damour, T., and Vokrouhlick{\'{y}}, D., ``Equivalence principle and the Moon'',
  {\em Phys. Rev. D}, {\bf 53}, 4177--4201, (1996). Related online version
  (cited on 15 January 2001): \newline\url{http://arXiv.org/abs/gr-qc/9507016}.
  \epubtkKeywords{tests of relativistic gravity, equivalence principle, lunar
  laser ranging}

\bibitem{danzmann}
Danzmann, K., ``LISA -- An ESA cornerstone mission for a gravitational-wave
  observatory'', {\em Class. Quantum Grav.}, {\bf 14}, 1399--1404, (1997).
  \epubtkKeywords{interferometric gravitational wave detectors}

\bibitem{deffayet02}
Deffayet, C., Dvali, G., Gabadadze, G., and Vainshtein, A., ``Nonperturbative
  continuity in graviton mass versus perturbative discontinuity'', {\em Phys.
  Rev. D}, {\bf 65}, 044026, (2002). Related online version (cited on 15
  February 2006): \newline\url{http://arxiv.org/abs/hep-th/0106001}.
  \epubtkKeywords{graviton}

\bibitem{dick98}
{Dick}, R., ``{Inequivalence of Jordan and Einstein frame: What is the low
  energy gravity in string theory?}'', {\em J. Gen. Rel. Grav.}, {\bf 30},
  435--444, (mar, 1998). \epubtkKeywords{string theory}

\bibitem{dicke64}
Dicke, R.~H., ``Experimental relativity'', in DeWitt, C., and DeWitt, B., eds.,
  {\em Relativity, Groups and Toplogy},  165--313, (Gordon and Breach, New
  York, U.S.A., 1964). \epubtkKeywords{equivalence principle, tests of
  relativistic gravity}

\bibitem{dicke1}
Dicke, R.H., {\em Gravitation and the Universe}, vol.~78 of Memoirs of the
  American Philosophical Society. Jayne Lecture for 1969, (American
  Philosophical Society, Philadelphia, U.S.A., 1970). \epubtkKeywords{theories
  of gravity, equivalence principle, tests of relativistic gravity}

\bibitem{Dickey}
Dickey, J.O., Bender, P.L., Faller, J.E., Newhall, X.X., Ricklefs, R.L., Ries,
  J.G., Shelus, P.J., Veillet, C., Whipple, A.L., Wiant, J.R., Williams, J.G.,
  and Yoder, C.F., ``Lunar laser ranging: A continuing legacy of the Apollo
  program'', {\em Science}, {\bf 265}, 482--490, (1994). \epubtkKeywords{Tests
  of relativistic gravity, lunar laser ranging}

\bibitem{drever}
Drever, R.W.P., ``A search for anisotropy of inertial mass using a free
  precession technique'', {\em Phil. Mag.}, {\bf 6}, 683--687, (1961).
  \epubtkKeywords{tests of Lorentz invariance}

\bibitem{dyson72}
Dyson, F.~J., ``The fundamental constants and their time variation'', in Salam,
  A., and Wigner, E.~P., eds., {\em Aspects of quantum theory},  213--236,
  (Cambridge University Press, Cambridge, U.K., New York, U.S.A., 1972).
  \epubtkKeywords{equivalence principle tests}

\bibitem{eling04}
{Eling}, C., and {Jacobson}, T., ``Static post-Newtonian equivalence of general
  relativity and gravity with a dynamical preferred frame'', {\em Phys. Rev.
  D}, {\bf 69}, 064005, (2004). Related online version (cited on 15 July 2005):
  \newline\url{http://arXiv.org/abs/gr-qc/0310044}. \epubtkKeywords{theories of
  gravity}

\bibitem{eotvos}
E{\"{o}}tv{\"{o}}s, R.V., Pek{\'{a}}r, V., and Fekete, E., ``Beitrage zum
  Gesetze der Proportionalit\"at von Tr\"agheit und Gravit\"at'', {\em Ann.
  Phys. (Leipzig)}, {\bf 68}, 11--66, (1922). \epubtkKeywords{equivalence
  principle tests}

\bibitem{Esposito03}
Esposito-Far{\`e}se, G., ``Binary-pulsar tests of strong-field gravity and
  gravitational radiation damping'', in Novello, M., Perez-Bergliaffa, S., and
  Ruffini, R., eds., {\em The Tenth Marcel Grossmann Meeting on Recent
  Developments in Theoretical and Experimental General Relativity, Gravitation
  and Relativistic Field Theories}, Proceedings of the meeting held in Rio de
  Janeiro, Brazil, 20--26 July 2003, (World Scientific, Singapore, 2005).
  Related online version (cited on 15 July 2005):
  \newline\url{http://arXiv.org/abs/gr-qc/0402007}. in press.
  \epubtkKeywords{binary pulsars}

\bibitem{farley}
Farley, F.J.M., Bailey, J., Brown, R.C.A., Giesch, M., J\"ostlein, H., van~der
  Meer, S., Picasso, E., and Tannenbaum, M., {\em Nuovo Cim.}, {\bf 45}, 281,
  (1966). \epubtkKeywords{tests of Lorentz invariance}

\bibitem{faulkner05}
Faulkner, A.J., Kramer, M., Lyne, A.G., Manchester, R.N., McLaughlin, M.A.,
  Stairs, I.H., Hobbs, G., Possenti, A., Lorimer, D.R., D'Amico, N., Camilo,
  F., and Burgay, M., ``PSR J1756-2251: A new relativistic double neutron star
  system'', {\em Astrophys. J. Lett.}, {\bf 618}, L119--L122, (2005). Related
  online version (cited on 15 July 2005):
  \newline\url{http://arXiv.org/abs/astro-ph/0411796}. \epubtkKeywords{binary
  pulsars}

\bibitem{finnchern}
Finn, L.S., and Chernoff, D.F., ``Observing binary inspiral in gravitational
  radiation: One interferometer'', {\em Phys. Rev. D}, {\bf 47}, 2198--2219,
  (1993). Related online version (cited on 15 January 2001):
  \newline\url{http://arXiv.org/abs/gr-qc/9301003}.
  \epubtkKeywords{gravitational radiation, parameter estimation}

\bibitem{FinnSutton02}
Finn, L.S., and Sutton, P.J., ``Bounding the mass of the graviton using binary
  pulsar observations'', {\em Phys. Rev. D}, {\bf 65}, 044022, (2002). Related
  online version (cited on 15 July 2005):
  \newline\url{http://arXiv.org/abs/gr-qc/0109049}. \epubtkKeywords{binary
  pulsars, graviton}

\bibitem{fischbach92}
Fischbach, E., Gillies, G.T., Krause, D.E., Schwan, J.G., and Talmadge, C.L.,
  ``Non-Newtonian gravity and new weak forces: An index of measurements and
  theory'', {\em Metrologia}, {\bf 29}, 213--260, (1992).
  \epubtkKeywords{equivalence principle tests}

\bibitem{fischbach5}
Fischbach, E., Sudarsky, D., Szafer, A., Talmadge, C.L., and Aronson, S.H.,
  ``Reanalysis of the E\"otv\"os experiment'', {\em Phys. Rev. Lett.}, {\bf
  56}, 3--6, (1986). Erratum: Phys. Rev. Lett. 56 (1986) 1427.
  \epubtkKeywords{equivalence principle tests}

\bibitem{FischbachTalmadge}
Fischbach, E., and Talmadge, C.L., ``Six years of the fifth force'', {\em
  Nature}, {\bf 356}, 207--215, (1992). \epubtkKeywords{equivalence principle
  tests}

\bibitem{FischbachTalmadge2}
Fischbach, E., and Talmadge, C.L., {\em The Search for Non-Newtonian Gravity},
  (Springer, New York, U.S.A., 1998). \epubtkKeywords{equivalence principle
  tests}

\bibitem{fischer04}
Fischer, M., Kolachevsky, N., Zimmermann, M., Holzwarth, R., Udem, T.,
  H{\"a}nsch, T.W., Abgrall, M., Grunert, J., Maksimovic, I., Bize, S., Marion,
  H., Pereira Dos~Santos, F., Lemonde, P., Santarelli, G., Laurent, P.,
  Clairon, A., Salomon, C., Haas, M., Jentschura, U.D., and Keitel, C.H., ``New
  limits on the drift of fundamental constants from laboratory measurements'',
  {\em Phys. Rev. Lett.}, {\bf 92}, 230802--1--4, (2004). Related online
  version (cited on 15 July 2005):
  \newline\url{http://arXiv.org/abs/physics/0312086}.
  \epubtkKeywords{equivalence principle tests}

\bibitem{fomalont03}
Fomalont, E.B., and Kopeikin, S.M., ``The measurement of the light deflection
  from Jupiter: experimental results'', {\em Astrophys. J.}, {\bf 598},
  704--711, (2003). Related online version (cited on 15 July 2005):
  \newline\url{http://arXiv.org/abs/astro-ph/0302294}. \epubtkKeywords{Shapiro
  time delay, speed of gravity}

\bibitem{foster05}
Foster, B.~Z., and Jacobson, T., ``Post-Newtonian parameters and constraints on
  Einstein-{\AE}ther theory'', (2005). URL (cited on 15 September 2005):
  \newline\url{http://arXiv.org/abs/gr-qc/0509083}. \epubtkKeywords{theories of
  gravity}

\bibitem{fritschel}
Fritschel, P., ``The LIGO project: Progress and plans'', in Coccia, E.,
  Veneziano, G., and Pizzella, G., eds., {\em Second Edoardo Amaldi Conference
  on Gravitational Waves}, Proceedings of the conference, held at CERN,
  Switzerland, 1-4 July, 1997, Edoardo Amaldi Foundation Series,  74--85,
  (World Scientific, Singapore; River Edge, U.S.A., 1998).
  \epubtkKeywords{interferometric gravitational wave detectors}

\bibitem{hipparcos}
Froeschl{\'{e}}, M., Mignard, F., and Arenou, F., ``Determination of the PPN
  parameter $\gamma$ with the Hipparcos data'', in {\em Proceedings from the
  Hipparcos Venice '97 Symposium}, Proceedings of the symposium held on 13-16
  May 1997, (ESA, Noordwijk, Netherlands, 1997). URL (cited on 15 January
  2001): \newline\url{http://astro.estec.esa.nl/Hipparcos/venice.html}.
  \epubtkKeywords{tests of relativistic gravity}

\bibitem{fujii04}
Fujii, Y., ``Oklo constraint on the time-variability of the fine-structure
  constant'', in Karshenboim, S.G., and Peik, E., eds., {\em Astrophysics,
  Clocks and Fundamental Constants}, vol. 648 of Lecture Notes in Physics,
  167--185. Springer, (2004). Related online version (cited on 15 July 2005):
  \newline\url{http://arXiv.org/abs/hep-ph/0311026}.
  \epubtkKeywords{equivalence principle tests}

\bibitem{gasperini}
Gasperini, M., ``On the response of gravitational antennas to dilatonic
  waves'', {\em Phys. Lett. B}, {\bf 470}, 67--72, (1999). Related online
  version (cited on 15 January 2001):
  \newline\url{http://arXiv.org/abs/gr-qc/9910019}.
  \epubtkKeywords{Gravitational wave detectors, scalar-tensor gravity}

\bibitem{GerardWiaux02}
{G{\'e}rard}, J.-M., and {Wiaux}, Y., ``{Gravitational dipole radiations from
  binary systems}'', {\em Phys. Rev. D}, {\bf 66}(2), 024040, (jul, 2002).
  Related online version (cited on 15 February 2006):
  \newline\url{http://arXiv.org/abs/gr-qc/0109062}.
  \epubtkKeywords{gravitational radiation, scalar-tensor gravity}

\bibitem{gleiser}
Gleiser, R.J., and Kozameh, C.N., ``Astrophysical limits on quantum gravity
  motivated birefringence'', {\em Phys. Rev. D}, {\bf 64}, 083007, (2001).
  Related online version (cited on 15 July 2005):
  \newline\url{http://arXiv.org/abs/gr-qc/0102093}. \epubtkKeywords{tests of
  relativistic gravity, quantum gravity}

\bibitem{Godone}
Godone, A., Novero, C., and Tavella, P., ``Null gravitational redshift
  experiment with nonidentical atomic clocks'', {\em Phys. Rev. D}, {\bf 51},
  319--323, (1995). \epubtkKeywords{equivalence principle tests}

\bibitem{gopu}
Gopakumar, A., and Iyer, B.R., ``Gravitational waves from inspiraling compact
  binaries: Angular momentum flux, evolution of the orbital elements and the
  waveform to the second post-Newtonian order'', {\em Phys. Rev. D}, {\bf 56},
  7708--7731, (1997). Related online version (cited on 15 January 2001):
  \newline\url{http://arXiv.org/abs/gr-qc/9710075}.
  \epubtkKeywords{gravitational radiation, post-Newtonian approximations,
  binary systems}

\bibitem{guenther98}
Guenther, D.B., Krauss, L.M., and Demarque, P., ``Testing the constancy of the
  gravitational constant using helioseismology'', {\em Astrophys. J.}, {\bf
  498}, 871--876, (1998). \epubtkKeywords{tests of relativistic gravity}

\bibitem{allegro}
Hamilton, W.O., ``The ALLEGRO detector and the future of resonant detectors in
  the USA'', in Coccia, E., Veneziano, G., and Pizzella, G., eds., {\em Second
  Edoardo Amaldi Conference on Gravitational Waves}, Proceedings of the
  conference, held at CERN, Switzerland, 1--4 July, 1997, Edoardo Amaldi
  Foundation Series,  115--126, (World Scientific, Singapore; River Edge,
  U.S.A., 1998). \epubtkKeywords{resonant gravitational wave detectors}

\bibitem{haugan}
Haugan, M.P., ``Energy conservation and the principle of equivalence'', {\em
  Ann. Phys. (N.Y.)}, {\bf 118}, 156--186, (1979). \epubtkKeywords{equivalence
  principle}

\bibitem{hauganlammer1}
Haugan, M.P., and L\"ammerzahl, C.L., ``On the interpretation of
  Michelson--Morley experiments'', {\em Phys. Lett. A}, {\bf 282}, 223--229,
  (2001). Related online version (cited on 15 July 2005):
  \newline\url{http://arXiv.org/abs/gr-qc/0103052}. \epubtkKeywords{tests of
  Lorentz invariance}

\bibitem{hauganlammer2}
Haugan, M.P., and L\"ammerzahl, C.L., ``Principles of equivalence: Their role
  in gravitation physics and experiments that test them'', in L\"ammerzahl,
  C.L., Everitt, C.W.F., and Hehl, F.W., eds., {\em Gyros, Clocks, and
  Interferometers: Testing Relativistic Gravity in Space}, Proceedings of the
  meeting held at Bad Honnef, Germany, 21--27 August 1999, vol. 562 of Lecture
  Notes in Physics,  195--212, (Springer, Berlin, Germany, 2001). Related
  online version (cited on 15 July 2005):
  \newline\url{http://arXiv.org/abs/gr-qc/0103067}. \epubtkKeywords{equivalence
  principle tests}

\bibitem{hauganwill}
Haugan, M.P., and Will, C.M., ``Modern tests of special relativity'', {\em
  Phys. Today}, {\bf 40}, 69--76, (1987). \epubtkKeywords{tests of Lorentz
  invariance}

\bibitem{hellings73}
Hellings, R.W., and Nordtvedt, K., ``Vector-Metric Theory of Gravity'', {\em
  Phys. Rev. D}, {\bf 7}, 3593--3602, (1973). \epubtkKeywords{theories of
  gravity}

\bibitem{hoyle04}
Hoyle, C.D., Kapner, D.J., Heckel, B.R., Adelberger, E.G., Gundlach, J.H.,
  Schmidt, U., and Swanson, H.E., ``Submillimeter tests of the gravitational
  inverse-square law'', {\em Phys. Rev. D}, {\bf 70}(4), 042004, (2004).
  Related online version (cited on 15 July 2005):
  \newline\url{http://arXiv.org/abs/hep-ph/0405262}. \epubtkKeywords{inverse
  square law}

\bibitem{hoyle01}
Hoyle, C.D., Schmidt, U., Heckel, B.R., Adelberger, E.G., Gundlach, J.H.,
  Kapner, D.J., and Swanson, H.E., ``Submillimeter Test of the Gravitational
  Inverse-Square Law: A Search for ``Large'' Extra Dimensions'', {\em Phys.
  Rev. Lett.}, {\bf 86}, 1418--1421, (2001). Related online version (cited on
  15 July 2005): \newline\url{http://arXiv.org/abs/hep-ph/0011014}.
  \epubtkKeywords{inverse square law}

\bibitem{hughes}
Hughes, V.W., Robinson, H.G., and Beltran-Lopez, V., ``Upper limit for the
  anisotropy of inertial mass from nuclear resonance experiments'', {\em Phys.
  Rev. Lett.}, {\bf 4}, 342--344, (1960). \epubtkKeywords{tests of Lorentz
  invariance}

\bibitem{Hulse}
Hulse, R.A., ``Nobel Lecture: The discovery of the binary pulsar'', {\em Rev.
  Mod. Phys.}, {\bf 66}, 699--710, (1994). \epubtkKeywords{binary pulsars}

\bibitem{iorio05}
Iorio, L., ``On the reliability of the so-far performed tests for measuring the
  Lense-Thirring effect with the LAGEOS satellites'', {\em New Astron.}, {\bf
  10}, 603--615, (2005). Related online version (cited on 15 July 2005):
  \newline\url{http://arXiv.org/abs/gr-qc/0411024}.
  \epubtkKeywords{Lense-Thirring effect}

\bibitem{ItohFutamase03}
Itoh, Y., and Futamase, T., ``New derivation of a third post-Newtonian equation
  of motion for relativistic compact binaries without ambiguity'', {\em Phys.
  Rev. D}, {\bf 68}, 121501, (2003). URL (cited on 15 July 2005):
  \newline\url{http://arXiv.org/abs/gr-qc/0310028}.
  \epubtkKeywords{post-Newtonian approximations, equations of motion, binary
  systems}

\bibitem{ivanchik05}
Ivanchik, A., Petitjean, P., Varshalovich, D., Aracil, B., Srianand, R., Chand,
  H., Ledoux, C., and Boisse, P., ``A new constraint on the time dependence of
  the proton-to-electron mass ratio: Analysis of the Q 0347-383 and Q 0405-443
  spectra'', (2005). URL (cited on 15 July 2005):
  \newline\url{http://arXiv.org/abs/astro-ph/0507174}.
  \epubtkKeywords{equivalence principle tests}

\bibitem{ives}
Ives, H.~E., and Stilwell, G.~R., ``An experimental study of the rate of a
  moving atomic clock'', {\em J. Opt. Soc. Am.}, {\bf 28}, 215--226, (1938).
  \epubtkKeywords{tests of Lorentz invariance}

\bibitem{jacobson01}
Jacobson, T., and Mattingly, D., ``Gravity with a dynamical preferred frame'',
  {\em Phys. Rev. D}, {\bf 64}, 024028, (2001). Related online version (cited
  on 15 July 2005): \newline\url{http://arXiv.org/abs/gr-qc/0007031}.
  \epubtkKeywords{theories of gravity}

\bibitem{jacobson04}
Jacobson, T., and Mattingly, D., ``Einstein-aether waves'', {\em Phys. Rev. D},
  {\bf 70}, 024003, (2004). Related online version (cited on 15 July 2005):
  \newline\url{http://arXiv.org/abs/gr-qc/0402005}. \epubtkKeywords{theories of
  gravity}

\bibitem{jaraschaefer98}
Jaranowski, P., and Sch{\"{a}}fer, G., ``3rd post-Newtonian higher order
  Hamilton dynamics for two-body point-mass systems'', {\em Phys. Rev. D}, {\bf
  57}, 7274--7291, (1998). Related online version (cited on 15 January 2001):
  \newline\url{http://arXiv.org/abs/gr-qc/9712075}. Erratum: Phys. Rev. D 63
  (2001) 029902. \epubtkKeywords{equations of motion, post-Newtonian
  approximations, binary systems}

\bibitem{jaranowski}
Jaranowski, P., and Sch{\"{a}}fer, G., ``The binary black-hole problem at the
  third post-Newtonian approximation in the orbital motion: Static part'', {\em
  Phys. Rev. D}, {\bf 60}, 124003--1--7, (1999). Related online version (cited
  on 15 January 2001): \newline\url{http://arXiv.org/abs/gr-qc/9906092}.
  \epubtkKeywords{equations of motion, post-Newtonian approximations, binary
  systems}

\bibitem{jaseja}
Jaseja, T.S., Javan, A., Murray, J., and Townes, C.H., ``Test of special
  relativity or of the isotropy of space by use of infrared masers'', {\em
  Phys. Rev.}, {\bf 133}, A1221--A1225, (1964). \epubtkKeywords{tests of
  Lorentz invariance}

\bibitem{Jones05}
Jones, D.I., ``Bounding the mass of the graviton using eccentric binaries'',
  {\em Astrophys. J. Lett.}, {\bf 618}, L115--L118, (2005). Related online
  version (cited on 15 July 2005):
  \newline\url{http://arXiv.org/abs/gr-qc/0411123}. \epubtkKeywords{graviton}

\bibitem{kaspi94}
Kaspi, V.M., Taylor, J.H., and Ryba, M.F., ``High-precision timing of
  millisecond pulsars. 3: Long-term monitoring of PSRs B1855+09 and B1937+21'',
  {\em Astrophys. J.}, {\bf 428}, 713--728, (1994). \epubtkKeywords{binary
  pulsars}

\bibitem{kidder95}
Kidder, L.E., ``Coalescing binary systems of compact objects to
  (post)$^{5/2}$-Newtonian order. V. Spin effects'', {\em Phys. Rev. D}, {\bf
  52}, 821--847, (1995). Related online version (cited on 15 January 2001):
  \newline\url{http://arXiv.org/abs/gr-qc/9506022}.
  \epubtkKeywords{gravitational radiation, equations of motion, post-Newtonian
  approximations, binary systems}

\bibitem{kww93}
Kidder, L.E., Will, C.M., and Wiseman, A.G., ``Spin effects in the inspiral of
  coalescing compact binaries'', {\em Phys. Rev. D}, {\bf 47}, R4183--R4187,
  (1993). Related online version (cited on 15 January 2001):
  \newline\url{http://arXiv.org/abs/gr-qc/9211025}.
  \epubtkKeywords{gravitational radiation, equations of motion, post-Newtonian
  approximations, binary systems}

\bibitem{KokkotasSchmidt99}
Kokkotas, K., and Schmidt, B., ``Quasi-Normal Modes of Stars and Black Holes'',
  {\em Living Rev. Relativity}, {\bf 2}, lrr-1999-2, (1999). URL (cited on 15
  July 2005): \newline\url{http://www.livingreviews.org/lrr-1999-2}.
  \epubtkKeywords{quasi-normal modes}

\bibitem{kopeikin02}
Kopeikin, S., and {Fomalont}, E.B., ``General relativistic model for
  experimental measurement of the speed of propagation of gravity by VLBI'', in
  Ros, E., Porcas, R.W., Lobanov, A.P., and Zensus, J.A., eds., {\em
  Proceedings of the 6th EVN Symposium}, 6th European VLBI Network Symposium on
  New Developments in VLBI Science and Technology, held in Bonn, June 25--28
  2002, ~49, (Max-Planck-Institut f\"ur Radioastronomie, Bonn, Germany, 2002).
  Related online version (cited on 15 July 2005):
  \newline\url{http://arXiv.org/abs/gr-qc/0206022}. \epubtkKeywords{Shapiro
  time delay, speed of gravity}

\bibitem{kopeikin05b}
{Kopeikin}, S.~M., ``{Comment on 'Model-dependence of Shapiro time delay and
  the "speed of gravity/speed of light" controversy'}'', {\em Class. Quantum
  Grav.}, {\bf 22}, 5181, (2005). URL (cited on 15 February 2006):
  \newline\url{http://arXiv.org/abs/gr-qc/0501048}. \epubtkKeywords{Shapiro
  time delay, speed of gravity}

\bibitem{kopeikin05a}
{Kopeikin}, S.~M., ``{Comments on the paper by S. Samuel "On the speed of
  gravity and the Jupiter/Quasar measurement"}'', (2005). URL (cited on 15
  February 2006): \newline\url{http://arXiv.org/abs/gr-qc/0501001}.
  \epubtkKeywords{Shapiro time delay, speed of gravity}

\bibitem{kopeikin05c}
{Kopeikin}, S.~M., ``{Note on the relationship between the speed of light and
  gravity in the bi-metric theory of gravity}'', (2005). URL (cited on 15
  February 2006): \newline\url{http://arXiv.org/abs/gr-qc/0512168}.
  \epubtkKeywords{Shapiro time delay, speed of gravity}

\bibitem{kopeikin01}
Kopeikin, S.M., ``Testing the relativistic effect of the propagation of gravity
  by very long baseline interferometry'', {\em Astrophys. J. Lett.}, {\bf 556},
  L1--L5, (2001). Related online version (cited on 15 July 2005):
  \newline\url{http://arXiv.org/abs/gr-qc/0105060}. \epubtkKeywords{Shapiro
  time delay, speed of gravity}

\bibitem{kopeikin03}
Kopeikin, S.M., ``The post-Newtonian treatment of the VLBI experiment on
  September 8, 2002'', {\em Phys. Lett. A}, {\bf 312}, 147--157, (2003).
  Related online version (cited on 15 July 2005):
  \newline\url{http://arXiv.org/abs/gr-qc/0212121}. \epubtkKeywords{Shapiro
  time delay, speed of gravity}

\bibitem{kopeikin04}
Kopeikin, S.M., ``The speed of gravity in general relativity and theoretical
  interpretation of the Jovian deflection experiment'', {\em Class. Quantum
  Grav.}, {\bf 21}, 3251--3286, (2004). Related online version (cited on 15
  July 2005): \newline\url{http://arXiv.org/abs/gr-qc/0310059}.
  \epubtkKeywords{Shapiro time delay, speed of gravity}

\bibitem{kosteleckylane99}
Kosteleck\'y, V.A., and Lane, C.~D., ``Constraints on Lorentz violation from
  clock-comparison experiments'', {\em Phys. Rev. D}, {\bf 60}, 116010, (1999).
  Related online version (cited on 15 July 2005):
  \newline\url{http://arXiv.org/abs/hep-ph/9908504}. \epubtkKeywords{tests of
  Lorentz invariance}

\bibitem{kosteleckymewes02}
Kosteleck\'y, V.A., and Mewes, M., ``Signals for Lorentz violation in
  electrodynamics'', {\em Phys. Rev. D}, {\bf 66}, 056005, (2002). Related
  online version (cited on 15 July 2005):
  \newline\url{http://arXiv.org/abs/hep-ph/0205211}. \epubtkKeywords{tests of
  Lorentz invariance}

\bibitem{kosteleckysamuel}
Kosteleck{\'y}, V.A., and Samuel, S., ``Gravitational phenomenology in
  higher-dimensional theories and strings'', {\em Phys. Rev. D}, {\bf 40},
  1886--1903, (1989). \epubtkKeywords{tests of Lorentz invariance}

\bibitem{kramer}
Kramer, M., ``Determination of the geometry of the PSR B1913+16 system by
  geodetic precession'', {\em Astrophys. J.}, {\bf 509}, 856--860, (1998).
  Related online version (cited on 15 January 2001):
  \newline\url{http://arXiv.org/abs/astro-ph/9808127}. \epubtkKeywords{binary
  pulsars}

\bibitem{krisher90a}
Krisher, T.P., Anderson, J.D., and Campbell, J.K., ``Test of the gravitational
  redshift effect at Saturn'', {\em Phys. Rev. Lett.}, {\bf 64}, 1322--1325,
  (1990). \epubtkKeywords{equivalence principle tests}

\bibitem{krisher90}
Krisher, T.P., Maleki, L., Lutes, G.F., Primas, L.E., Logan, R.T., Anderson,
  J.D., and Will, C.M., ``Test of the isotropy of the one-way speed of light
  using hydrogen-maser frequency standards'', {\em Phys. Rev. D}, {\bf 42},
  731--734, (1990). \epubtkKeywords{tests of Lorentz invariance}

\bibitem{krisher93}
Krisher, T.P., Morabito, D.D., and Anderson, J.D., ``The Galileo solar redshift
  experiment'', {\em Phys. Rev. Lett.}, {\bf 70}, 2213--2216, (1993).
  \epubtkKeywords{equivalence principle tests}

\bibitem{krolak95}
{Kr{\'o}lak}, A., {Kokkotas}, K.~D., and {Sch{\"a}fer}, G., ``{Estimation of
  the post-Newtonian parameters in the gravitational-wave emission of a
  coalescing binary}'', {\em Phys. Rev. D}, {\bf 52}, 2089--2111, (aug, 1995).
  Related online version (cited on 15 February 2006):
  \newline\url{http://arXiv:gr-qc/9503013}. \epubtkKeywords{gravitational
  radiation, parameter estimation}

\bibitem{lammer03}
L\"ammerzahl, C.L., ``The Einstein equivalence principle and the search for new
  physics'', in Giulini, D.J.W., Kiefer, C., and L\"ammerzahl, C.~L., eds.,
  {\em Quantum Gravity: From Theory to Experimental Search}, vol. 631 of
  Lecture Notes in Physics,  367--394, (Springer, Berlin, Germany, 2003).
  \epubtkKeywords{equivalence principle}

\bibitem{lamoreaux86}
Lamoreaux, S.K., Jacobs, J.P., Heckel, B.R., Raab, F.J., and Fortson, E.N.,
  ``New limits on spatial anisotropy from optically-pumped $^{201}$Hg and
  $^{199}$Hg'', {\em Phys. Rev. Lett.}, {\bf 57}, 3125--3128, (1986).
  \epubtkKeywords{tests of Lorentz invariance}

\bibitem{lebach}
Lebach, D.E., Corey, B.E., Shapiro, I.I., Ratner, M.I., Webber, J.C., Rogers,
  A.E.E., Davis, J.L., and Herring, T.A., ``Measurement of the solar
  gravitational deflection of radio waves using very-long-baseline
  interferometry'', {\em Phys. Rev. Lett.}, {\bf 75}, 1439--1442, (1995).
  \epubtkKeywords{Tests of relativistic gravity}

\bibitem{Lehner01}
Lehner, L., ``Numerical relativity: a review'', {\em Class. Quantum Grav.},
  {\bf 18}, 25, (2001). Related online version (cited on 15 July 2005):
  \newline\url{http://arXiv.org/abs/gr-qc/0106072}. \epubtkKeywords{numerical
  relativity}

\bibitem{lightmanlee}
Lightman, A.P., and Lee, D.L., ``Restricted proof that the weak equivalence
  principle implies the Einstein equivalence principle'', {\em Phys. Rev. D},
  {\bf 8}, 364--376, (1973). \epubtkKeywords{equivalence principle}

\bibitem{lineweaver96}
Lineweaver, C.H., Tenorio, L., Smoot, G.F., Keegstra, P., Banday, A.J., and
  Lubin, P., ``The dipole observed in the COBE DMR 4 year data'', {\em
  Astrophys. J.}, {\bf 470}, 38--42, (1996). \epubtkKeywords{cosmology, cosmic
  background}

\bibitem{lipa03}
Lipa, J.A., Nissen, J.A., Wang, S., Stricker, D.A., and Avaloff, D., ``New
  limit on signals of Lorentz violation in electrodynamics'', {\em Phys. Rev.
  Lett.}, {\bf 90}, 060403--1--4, (2003). Related online version (cited on 15
  July 2005): \newline\url{http://arXiv.org/abs/physics/0302093}.
  \epubtkKeywords{tests of Lorentz invariance}

\bibitem{lobo}
Lobo, J.A., ``Spherical GW detectors and geometry'', in Coccia, E., Veneziano,
  G., and Pizzella, G., eds., {\em Second Edoardo Amaldi Conference on
  Gravitational Waves}, Proceedings of the conference, held at CERN,
  Switzerland, 1--4 July, 1997, Edoardo Amaldi Foundation Series,  168--179,
  (World Scientific, Singapore, 1998). \epubtkKeywords{resonant gravitational
  wave detectors}

\bibitem{long03}
Long, J.C., Chan, H.W., Churnside, A.B., Gulbis, E.A., Varney, M.C.M., and
  Price, J.C., ``Upper limits to submillimetre-range forces from extra
  space-time dimensions'', {\em Nature}, {\bf 421}, 922--925, (February, 2003).
  \epubtkKeywords{inverse square law}

\bibitem{long99}
Long, J.C., Chan, H.W., and Price, J.C., ``Experimental status of
  gravitational-strength forces in the sub-centimeter regime'', {\em Nucl.
  Phys. B}, {\bf 539}, 23--34, (1999). Related online version (cited on 15
  January 2001): \newline\url{http://arXiv.org/abs/hep-ph/9805217}.
  \epubtkKeywords{inverse square law}

\bibitem{lopresto91}
LoPresto, J.C., Schrader, C., and Pierce, A.K., ``Solar gravitational redshift
  from the infrared oxygen triplet'', {\em Astrophys. J.}, {\bf 376}, 757--760,
  (1991). \epubtkKeywords{equivalence principle tests}

\bibitem{lorimer05}
Lorimer, D.~R., Stairs, I.~H., Freire, P. C.~C., Cordes, J.~M., Camilo, F.,
  Faulkner, A.~J., Lyne, A.~G., Nice, D.~J., Ransom, S.~M., Arzoumanian, Z.,
  Manchester, R.~N., Champion, D.~J., van Leeuwen, J., McLaughlin, M.~A.,
  Ramachandran, R., Hessels, J. W.~T., Vlemmings, W., Deshpande, A.~A., Bhat,
  N. D.~R., Chatterjee, S., Han, J.~L., Gaensler, B.~M., Kasian, L., Deneva,
  J.~S., Reid, B., Lazio, T. J.~W., Kaspi, V.~M., Crawford, F., Lommen, A.~N.,
  Backer, D.~C., Kramer, M., Stappers, B.~W., Hobbs, G.~B., Possenti, A.,
  D'Amico, N., and Burgay, M., ``The young, highly relativistic binary pulsar
  J1906+0746'', (2005). URL (cited on 15 February, 2006):
  \newline\url{http://arXiv.org/abs/astro-ph/0511523}. \epubtkKeywords{binary
  pulsars}

\bibitem{lorimer}
Lorimer, D.R., ``Binary and Millisecond Pulsars'', {\em Living Rev.
  Relativity}, {\bf 8}, lrr-2005-7, (1998). URL (cited on 03 February 2006):
  \newline\url{http://www.livingreviews.org/lrr-2005-7}.
  \epubtkKeywords{Pulsars}

\bibitem{lyne04}
Lyne, A.G., Burgay, M., Kramer, M., Possenti, A., Manchester, R.N., Camilo, F.,
  McLaughlin, M.A., Lorimer, D.R., D'Amico, N., Joshi, B.C., Reynolds, J., and
  Freire, P.C.C., ``A double-pulsar system: A rare laboratory for relativistic
  gravity and plasma physics'', {\em Science}, {\bf 303}, 1153--1157,
  (February, 2004). Related online version (cited on 15 July 2005):
  \newline\url{http://arXiv.org/abs/astro-ph/0401086}. \epubtkKeywords{binary
  pulsars}

\bibitem{maeda88}
Maeda, K.-I., ``On time variation of fundamental constants in superstring
  theories'', {\em Mod. Phys. Lett. A}, {\bf 3}, 243--249, (1988).
  \epubtkKeywords{string theory, equivalence principle}

\bibitem{maggiore}
Maggiore, M., and Nicolis, A., ``Detection strategies for scalar gravitational
  waves with interferometers and resonant spheres'', {\em Phys. Rev. D}, {\bf
  62}, 024004, (1999). Related online version (cited on 15 January 2001):
  \newline\url{http://arXiv.org/abs/gr-qc/9907055}.
  \epubtkKeywords{gravitational wave detectors, scalar-tensor gravity}

\bibitem{magueijo03}
Magueijo, J., ``New varying speed of light theories'', {\em Rep. Prog. Phys.},
  {\bf 66}, 2025--2068, (2003). \epubtkKeywords{theories of gravity}

\bibitem{malaney}
Malaney, R.A., and Mathews, G.J., ``Probing the early universe: A review of
  primordial nucleosynthesis beyond the standard big bang'', {\em Phys. Rep.},
  {\bf 229}, 147--219, (1993). \epubtkKeywords{cosmology, big bang
  nucleosynthesis}

\bibitem{maleki01}
Maleki, L., and Prestage, J.~D., ``SpaceTime Mission: Clock test of relativity
  at four solar radii'', in {\em Gyros, Clocks, and Interferometers: Testing
  Relativistic Gravity in Space}, vol. 562 of Lecture Notes in Physics,  369,
  (Springer, Berlin, Germany, 2001). \epubtkKeywords{equivalence principle
  tests}

\bibitem{salomon03}
Marion, H., Pereira Dos~Santos, F., Abgrall, M., Zhang, S., Sortais, Y., Bize,
  S., Maksimovic, I., Calonico, D., Grunert, J., Mandache, C., Lemonde, P.,
  Santarelli, G., Laurent, P., Clairon, A., and Salomon, C., ``Search for
  variations of fundamental constants using atomic fountain clocks'', {\em
  Phys. Rev. Lett.}, {\bf 90}, 150801--1--4, (2003). Related online version
  (cited on 15 July 2005): \newline\url{http://arXiv.org/abs/physics/0212112}.
  \epubtkKeywords{equivalence principle tests}

\bibitem{mattingly}
Mattingly, D., ``Modern Tests of Lorentz Invariance'', {\em Living Rev.
  Relativity}, {\bf 8}, lrr-2005-5, (2005). URL (cited on 15 July 2005):
  \newline\url{http://www.livingreviews.org/lrr-2005-5}. \epubtkKeywords{tests
  of Lorentz invariance}

\bibitem{mattingly02}
Mattingly, D., and Jacobson, T., ``Relativistic gravity with a dynamical
  preferred frame'', in Kosteleck\'y, V.A., ed., {\em CPT and Lorentz
  Symmetry}, Proceedings of the 2nd meeting held at Indiana University,
  Bloomington, August 15--18, 2001, (World Scientific, Singapore, 2002).
  Related online version (cited on 15 July 2005):
  \newline\url{http://arXiv.org/abs/gr-qc/0112012}. \epubtkKeywords{theories of
  gravity}

\bibitem{mecheri04}
Mecheri, R., Abdelatif, T., Irbah, A., Provost, J., and Berthomieu, G., ``New
  values of gravitational moments $J_{2}$ and $J_{4}$ deduced from
  helioseismology'', {\em Solar Phys.}, {\bf 222}, 191--197, (2004).
  \epubtkKeywords{helioseismology, solar interior}

\bibitem{narayan}
Menou, K., Quataert, E., and Narayan, R., ``Astrophysical evidence for
  black-hole event horizons'', in Piran, T., ed., {\em The Eighth Marcel
  Grossmann Meeting on Recent Developments in Theoretical and Experimental
  General Relativity, Gravitation and Relativistic Field Theories}, Proceedings
  of the meeting held at the Hebrew University of Jerusalem, June 22--27, 1997,
   204--224, (World Scientific, Singapore, 1999). Related online version (cited
  on 15 January 2001): \newline\url{http://arXiv.org/abs/astro-ph/9712015}.
  \epubtkKeywords{black hole accretion}

\bibitem{mm}
Michelson, A.~A., and Morley, E.~W., ``Relative motion of Earth and
  luminiferous ether'', {\em Am. J. Science}, {\bf 34}, 333, (1887).
  \epubtkKeywords{tests of Lorentz invariance}

\bibitem{milani02}
Milani, A., Vokrouhlick{\'y}, D., Villani, D., Bonanno, C., and Rossi, A.,
  ``Testing general relativity with the BepiColombo radio science experiment'',
  {\em Phys. Rev. D}, {\bf 66}, 082001, (2002). \epubtkKeywords{tests of
  relativistic gravity}

\bibitem{msstt97}
Mino, Y., Sasaki, M., Shibata, M., Tagoshi, H., and Tanaka, T., ``Black hole
  perturbation'', {\em Prog. Theor. Phys. Suppl.}, {\bf 128}, 1--121, (1997).
  Related online version (cited on 15 January 2001):
  \newline\url{http://arXiv.org/abs/gr-qc/9712057}. \epubtkKeywords{black hole
  perturbation theory}

\bibitem{MTW}
Misner, C.W., Thorne, K.S., and Wheeler, J.A., {\em Gravitation}, (W.H.
  Freeman, San Francisco, U.S.A., 1973). \epubtkKeywords{general relativity}

\bibitem{muller03}
M\"uller, H., Herrmann, S., Braxmaier, C., Schiller, S., and Peters, A.,
  ``Modern Michelson--Morley experiment using cryogenic optical resonators'',
  {\em Phys. Rev. Lett.}, {\bf 91}, 020401--1--4, (2003). Related online
  version (cited on 15 July 2005):
  \newline\url{http://arXiv.org/abs/physics/0305117}. \epubtkKeywords{tests of
  Lorentz invariance}

\bibitem{MullerPRD}
M{\"{u}}ller, J., Nordtvedt, K., and Vokrouhlick{\'{y}}, D., ``Improved
  constraint on the $\alpha_1$ PPN parameter from lunar motion'', {\em Phys.
  Rev. D}, {\bf 54}, R5927--R5930, (1996). \epubtkKeywords{tests of
  relativistic gravity, lunar laser ranging}

\bibitem{MullerMG}
M{\"{u}}ller, J., Schneider, M., Nordtvedt, K., and Vokrouhlick{\'{y}}, D.,
  ``What can LLR provide to relativity?'', in Piran, T., ed., {\em The Eighth
  Marcel Grossmann Meeting on Recent Developments in Theoretical and
  Experimental General Relativity, Gravitation and Relativistic Field
  Theories}, Proceedings of the meeting held at the Hebrew University of
  Jerusalem, June 22--27, 1997,  1151--1153, (World Scientific, Singapore,
  1999). \epubtkKeywords{tests of relativistic gravity, lunar laser ranging}

\bibitem{murphy01}
Murphy, M.T., Webb, J.K., Flambaum, V.V., Dzuba, V.~A., Churchill, C.~W.,
  Prochaska, J.~X., Barrow, J.D., and Wolfe, A.A., ``Possible evidence for a
  variable fine structure constant from QSO absorption lines: motivations,
  analysis and results'', {\em Mon. Not. R. Astron. Soc.}, {\bf 327},
  1208--1222, (2001). Related online version (cited on 15 July 2005):
  \newline\url{http://arXiv.org/abs/astro-ph/0012419}.
  \epubtkKeywords{equivalence principle tests}

\bibitem{Ni77}
Ni, W.-T., ``Equivalence principles and electromagnetism'', {\em Phys. Rev.
  Lett.}, {\bf 38}, 301--304, (1977). \epubtkKeywords{equivalence principle}

\bibitem{nordstrom13}
Nordstr\"om, G., ``Zur Theorie der Gravitation vom Standpunkt des
  Relativit\"atsprinzips'', {\em Annalen der Physik}, {\bf 42}, 533--554,
  (1913). \epubtkKeywords{Nordstrom's theory, theories of gravity}

\bibitem{nordtvedt1}
Nordtvedt, K., ``Equivalence principle for massive bodies. I. Phenomenology'',
  {\em Phys. Rev.}, {\bf 169}, 1014--1016, (1968). \epubtkKeywords{tests of
  relativistic gravity, equivalence principle}

\bibitem{nordtvedt2}
Nordtvedt, K., ``Equivalence principle for massive bodies. II. Theory'', {\em
  Phys. Rev.}, {\bf 169}, 1017--1025, (1968). \epubtkKeywords{theories of
  gravity, equivalence principle}

\bibitem{nordtvedt88b}
Nordtvedt, K., ``Existence of the gravitomagnetic interaction'', {\em Int. J.
  Theor. Phys.}, {\bf 27}, 1395--1404, (1988). \epubtkKeywords{tests of
  relativistic gravity, Lense-Thirring effect}

\bibitem{nordtvedt88a}
Nordtvedt, K., ``Gravitomagnetic interaction and laser ranging to Earth
  satellites'', {\em Phys. Rev. Lett.}, {\bf 61}, 2647--2649, (1988).
  \epubtkKeywords{tests of relativistic gravity, Lense-Thirring effect}

\bibitem{nordtvedt3}
Nordtvedt, K., ``$\dot G/G$ and a cosmological acceleration of gravitationally
  compact bodies'', {\em Phys. Rev. Lett.}, {\bf 65}, 953--956, (1990).
  \epubtkKeywords{tests of relativistic gravity, binary systems}

\bibitem{Nordtvedt95}
Nordtvedt, K., ``The relativistic orbit observables in lunar laser ranging'',
  {\em Icarus}, {\bf 114}, 51--62, (1995). \epubtkKeywords{tests of
  relativistic gravity, lunar laser ranging}

\bibitem{Nordtvedt00}
Nordtvedt, K., ``Improving gravity theory tests with solar system `grand
  fits''', {\em Phys. Rev. D}, {\bf 61}, 122001, (2000). \epubtkKeywords{tests
  of relativistic gravity}

\bibitem{nordtvedt01}
Nordtvedt, K., ``Testing Newton's third law using lunar laser ranging'', {\em
  Class. Quantum Grav.}, {\bf 18}, L133--L137, (2001). \epubtkKeywords{tests of
  relativistic gravity, lunar laser ranging}

\bibitem{Ohanian74}
Ohanian, H.~C., ``{Comment on the Schiff Conjecture}'', {\em Phys. Rev. D},
  {\bf 10}, 2041--2042, (1974). \epubtkKeywords{equivalence principle}

\bibitem{olive04}
Olive, K.A., Pospelov, M., Qian, Y.-Z., Manhes, G., Vangioni-Flam, E., Coc, A.,
  and Casse, M., ``Reexamination of the $^{187}$Re bound on the variation of
  fundamental couplings'', {\em Phys. Rev. D}, {\bf 69}, 027701--1--4, (2004).
  Related online version (cited on 15 July 2005):
  \newline\url{http://arXiv.org/abs/astro-ph/0309252}.
  \epubtkKeywords{equivalence principle}

\bibitem{rome}
Pallottino, G.V., ``The Resonant Mass Detectors of the Rome Group'', in Coccia,
  E., Veneziano, G., and Pizzella, G., eds., {\em Second Edoardo Amaldi
  Conference on Gravitational Waves}, Proceedings of the conference, held at
  CERN, Switzerland, 1--4 July, 1997, vol.~4 of Edoardo Amaldi Foundation
  Series,  105--114, (World Scientific, Singapore; River Edge, U.S.A., 1998).
  \epubtkKeywords{Resonant gravitational wave detectors}

\bibitem{paterno96}
Paterno, L., Sofia, S., and di~Mauro, M.P., ``The rotation of the Sun's core'',
  {\em Astron. Astrophys.}, {\bf 314}, 940--946, (1996).
  \epubtkKeywords{helioseismology, solar interior}

\bibitem{DIRE}
Pati, M.E., and Will, C.M., ``Post-Newtonian gravitational radiation and
  equations of motion via direct integration of the relaxed Einstein equations:
  Foundations'', {\em Phys. Rev. D}, {\bf 62}, 124015, (2000). Related online
  version (cited on 15 January 2001):
  \newline\url{http://arXiv.org/abs/gr-qc/0007087}.
  \epubtkKeywords{gravitational radiation, equations of motion, post-Newtonian
  approximations}

\bibitem{peik04}
Peik, E., Lipphardt, B., Schnatz, H., Schneider, T., Tamm, C., and Karshenboim,
  S.G., ``Limit on the present temporal variation of the fine structure
  constant'', {\em Phys. Rev. Lett.}, {\bf 93}, 170801--1--4, (2004). Related
  online version (cited on 15 July 2005):
  \newline\url{http://arXiv.org/abs/physics/0402132}.
  \epubtkKeywords{equivalence principle}

\bibitem{petrov05}
Petrov, Y.V., Nazarov, A.I., Onegin, M.S., Petrov, V.Y., and Sakhnovsky, E.G.,
  ``Natural nuclear reactor Oklo and variation of fundamental constants. Part
  1: Computation of neutronics of fresh core'', (2005). URL (cited on 15 July
  2005): \newline\url{http://arXiv.org/abs/hep-ph/0506186}.
  \epubtkKeywords{equivalence principle}

\bibitem{pijpers98}
Pijpers, F.P., ``Helioseismic determination of the solar gravitational
  quadrupole moment'', {\em Mon. Not. Roy. Astron. Soc.}, {\bf 297}, L76--L80,
  (1998). Related online version (cited on 15 July 2005):
  \newline\url{http://arXiv.org/abs/astro-ph/9804258}.
  \epubtkKeywords{helioseismology, solar interior}

\bibitem{pitjeva05}
Pitjeva, E.V., ``Relativistic effects and solar oblateness from radar
  observations of planets and spacecraft'', {\em Astron. Lett.}, {\bf 31},
  340--349, (2005). \epubtkKeywords{tests of relativistic gravity}

\bibitem{poissonBH}
Poisson, E., ``Measuring black-hole parameters and testing general relativity
  using gravitational-wave data from space-based interferometers'', {\em Phys.
  Rev. D}, {\bf 54}, 5939--5953, (1996). Related online version (cited on 15
  January 2001): \newline\url{http://arXiv.org/abs/gr-qc/9606024}.
  \epubtkKeywords{gravitational wave observations, black holes, parameter
  estimation}

\bibitem{poissonwill}
Poisson, E., and Will, C.M., ``Gravitational waves from inspiralling compact
  binaries: Parameter estimation using second-post-Newtonian waveforms'', {\em
  Phys. Rev. D}, {\bf 52}, 848--855, (1995). Related online version (cited on
  15 January 2001): \newline\url{http://arXiv.org/abs/gr-qc/9502040}.
  \epubtkKeywords{gravitational wave observations, parameter estimation,
  post-Newtonian approximations}

\bibitem{prestage85}
Prestage, J.D., Bollinger, J.J., Itano, W.M., and Wineland, D.J., ``Limits for
  spatial anisotropy by use of nuclear-spin-polarized $^9$Be$^+$ ions'', {\em
  Phys. Rev. Lett.}, {\bf 54}, 2387--2390, (1985). \epubtkKeywords{tests of
  Lorentz invariance}

\bibitem{prestage95}
Prestage, J.D., Tjoelker, R.L., and Maleki, L., ``Atomic clocks and variations
  of the fine structure constant'', {\em Phys. Rev. Lett.}, {\bf 74},
  3511--3514, (1995). \epubtkKeywords{equivalence principle tests}

\bibitem{auriga}
Prodi, G.A., Conti, L., Mezzena, R., Vitale, S., Taffarello, L., Zendri, J.P.,
  Baggio, L., Cerdonio, M., Colombo, A., Crivelli~Visconti, V., Macchietto, R.,
  Falferi, P., Bonaldi, M., Ortolan, A., Vedovato, G., Cavallini, E., and
  Fortini, P., ``Initial Operation of the Gravitational Wave Detector AURIGA'',
  in Coccia, E., Veneziano, G., and Pizzella, G., eds., {\em Second Edoardo
  Amaldi Conference on Gravitational Waves}, Proceedings of the conference,
  held at CERN, Switzerland, 1--4 July, 1997, vol.~4 of Edoardo Amaldi
  Foundation Series,  148--158, (World Scientific, Singapore; River Edge,
  U.S.A., 1998). \epubtkKeywords{Resonant gravitational wave detectors}

\bibitem{Psaltis04}
Psaltis, D., ``Measurements of black hole spins and tests of strong-field
  general relativity'', in Kaaret, P., Lamb, F.K., and Swank, J.H., eds., {\em
  X-ray Timing 2003: Rossi and Beyond}, Proceedings of the conference held 3--5
  November, 2003 in Cambridge, MA, vol. 714 of AIP Conference Proceedings,
  29--35, (American Institute of Physics, Melville, U.S.A., 2004). Related
  online version (cited on 15 July 2005):
  \newline\url{http://arXiv.org/abs/astro-ph/0402213}. \epubtkKeywords{black
  hole accretion, Lense-Thirring effect}

\bibitem{quast04}
Quast, R., Reimers, D., and Levshakov, S.A., ``Probing the variability of the
  fine-structure constant with the VLT/UVES'', {\em Astron. Astrophys. Lett.},
  {\bf 415}, L7--L11, (2004). Related online version (cited on 15 July 2005):
  \newline\url{http://arXiv.org/abs/astro-ph/0311280}.
  \epubtkKeywords{equivalence principle tests}

\bibitem{randall2}
Randall, L., and Sundrum, R., ``An Alternative to Compactification'', {\em
  Phys. Rev. Lett.}, {\bf 83}, 4690--4693, (1999). Related online version
  (cited on 15 July 2005): \newline\url{http://arXiv.org/abs/hep-ph/9906064}.
  \epubtkKeywords{inverse square law, extra dimensions}

\bibitem{randall1}
Randall, L., and Sundrum, R., ``Large Mass Hierarchy from a Small Extra
  Dimension'', {\em Phys. Rev. Lett.}, {\bf 83}, 3370--3373, (1999). Related
  online version (cited on 15 July 2005):
  \newline\url{http://arXiv.org/abs/hep-ph/9905021}. \epubtkKeywords{inverse
  square law, extra dimensions}

\bibitem{reasenberg}
Reasenberg, R.D., Shapiro, I.I., MacNeil, P.E., Goldstein, R.B., Breidenthal,
  J.C., Brenkle, J.P., Cain, D.L., Kaufman, T.M., Komarek, T.A., and
  Zygielbaum, A.I., ``Viking relativity experiment: Verification of signal
  retardation by solar gravity'', {\em Astrophys. J. Lett.}, {\bf 234},
  L219--L221, (1979). \epubtkKeywords{tests of relativistic gravity, Shapiro
  time dela}

\bibitem{reeves}
Reeves, H., ``On the origin of the light elements (Z $<$ 6)'', {\em Rev. Mod.
  Phys.}, {\bf 66}, 193--216, (1994). \epubtkKeywords{cosmology, big bang
  nucleosynthesis}

\bibitem{reynolds05}
{Reynolds}, C.~S., {Brenneman}, L.~W., and {Garofalo}, D., ``{Black Hole Spin
  in AGN and GBHCs}'', {\em Astrophys. Sp. Sci.}, {\bf 300}, 71--79, (nov,
  2005). Related online version (cited on 15 February 2006):
  \newline\url{http://arxiv.org/abs/astro-ph/0410116}. \epubtkKeywords{black
  hole accretion}

\bibitem{reynolds04}
{Reynolds}, C.~S., {Brenneman}, L.~W., {Wilms}, J., and {Kaiser}, M.~E.,
  ``{Iron line spectroscopy of NGC 4593 with XMM-Newton: where is the black
  hole accretion disc?}'', {\em Mon. Not. Roy. Astron. Soc.}, {\bf 352},
  205--210, (jul, 2004). Related online version (cited on 15 February 2006):
  \newline\url{http://arxiv.org/abs/astro-ph/0404187}. \epubtkKeywords{black
  hole accretion}

\bibitem{ries}
Ries, J.C., Eanes, R.J., Tapley, B.D., and Peterson, G.E., ``Prospects for an
  improved Lense-Thirring test with SLR and the GRACE gravity mission'', in
  Noomen, R., Klosko, S., Noll, C., and Pearlman, M., eds., {\em Proceedings of
  the 13th International Workshop on Laser Ranging: Science Session and Full
  Proceedings CD-ROM}, 13th International Workshop on Laser Ranging, NASA
  Conference Proceedings,  211--248. NASA, (2003).
  \epubtkKeywords{Lense-Thirring effect}

\bibitem{riis}
Riis, E., Anderson, L.-U.A., Bjerre, N., Poulson, O., Lee, S.A., and Hall,
  J.L., ``Test of the isotropy of the speed of light using fast-beam laser
  spectroscopy'', {\em Phys. Rev. Lett.}, {\bf 60}, 81--84, (1988).
  \epubtkKeywords{tests of Lorentz invariance}

\bibitem{dicke_2}
Roll, P.G., Krotkov, R., and Dicke, R.H., ``The equivalence of inertial and
  passive gravitational mass'', {\em Ann. Phys. (N.Y.)}, {\bf 26}, 442--517,
  (1964). \epubtkKeywords{equivalence principle tests}

\bibitem{rossi}
Rossi, B., and Hall, D.~B., ``Variation of the rate of decay of mesotrons with
  momentum'', {\em Phys. Rev.}, {\bf 59}, 223--228, (1941).
  \epubtkKeywords{tests of Lorentz invariance}

\bibitem{roxburgh01}
Roxburgh, I.W., ``Gravitational multipole moments of the Sun determined from
  helioseismic estimates of the internal structure and rotation'', {\em Astron.
  Astrophys.}, {\bf 377}, 688--690, (2001). \epubtkKeywords{helioseismology,
  solar interior}

\bibitem{ryan}
Ryan, F.D., ``Gravitational waves from the inspiral of a compact object into a
  massive, axisymmetric body with arbitrary multipole moments'', {\em Phys.
  Rev. D}, {\bf 52}, 5707--5718, (1995). \epubtkKeywords{Gravitational
  radiation, Binary systems}

\bibitem{samuel03}
Samuel, S., ``On the speed of gravity and the v/c corrections to the Shapiro
  time delay'', {\em Phys. Rev. Lett.}, {\bf 90}, 231101, (2003). Related
  online version (cited on 15 July 2005):
  \newline\url{http://arXiv.org/abs/astro-ph/0304006}. \epubtkKeywords{speed of
  gravity, Shapiro time delay}

\bibitem{samuel04}
Samuel, S., ``On the Speed of Gravity and the Jupiter/quasar Measurement'',
  {\em Int. J. Mod. Phys. D}, {\bf 13}, 1753--1770, (2004). Related online
  version (cited on 15 July 2005):
  \newline\url{http://arXiv.org/abs/astro-ph/0412401}. \epubtkKeywords{speed of
  gravity, Shapiro time delay}

\bibitem{Santiago97}
Santiago, D.I., Kalligas, D., and Wagoner, R.V., ``Nucleosynthesis constraints
  on scalar-tensor theories of gravity'', {\em Phys. Rev. D}, {\bf 56},
  7627--7637, (1997). \epubtkKeywords{big bang nucleosynthesis, scalar-tensor
  gravity}

\bibitem{SasakiTagoshi03}
Sasaki, M., and Tagoshi, H., ``Analytic black hole perturbation approach to
  gravitational radiation'', {\em Living Rev. Relativity}, {\bf 6}, lrr-2003-6,
  (2003). URL (cited on 15 July 2005):
  \newline\url{http://www.livingreviews.org/lrr-2003-6}. \epubtkKeywords{black
  hole perturbations}

\bibitem{scharrewill}
Scharre, P.D., and Will, C.M., ``Testing scalar-tensor gravity using space
  gravitational-wave interferometers'', {\em Phys. Rev. D}, {\bf 65}, 042002,
  (2002). Related online version (cited on 15 July 2005):
  \newline\url{http://arXiv.org/abs/gr-qc/0109044}.
  \epubtkKeywords{scalar-tensor gravity, parameter estimation}

\bibitem{shankland}
Shankland, R.S., McCuskey, S.W., Leone, F.C., and Kuerti, G., ``New analysis of
  the interferometer observations of Dayton C. Miller'', {\em Rev. Mod. Phys.},
  {\bf 27}, 167--178, (1955). \epubtkKeywords{tests of Lorentz invariance}

\bibitem{shapiroGR12}
Shapiro, I.I., ``Solar system tests of general relativity: Recent results and
  present plans'', in Ashby, N., Bartlett, D.F., and Wyss, W., eds., {\em
  General Relativity and Gravitation}, Proceedings of the 12th International
  Conference on General Relativity and Gravitation, University of Colorado at
  Boulder, July 2--8, 1989,  313--330, (Cambridge University Press, Cambridge,
  U.K., New York, U.S.A., 1990). \epubtkKeywords{Tests of relativistic gravity}

\bibitem{shapiro}
Shapiro, I.I., ``A century of relativity'', {\em Rev. Mod. Phys.}, {\bf 71},
  S41--S53, (1999). \epubtkKeywords{Tests of relativistic gravity}

\bibitem{sshapiro04}
Shapiro, S.S., Davis, J.L., Lebach, D.E., and Gregory, J.S., ``Measurement of
  the solar gravitational deflection of radio waves using geodetic
  very-long-baseline interferometry data, 1979--1999'', {\em Phys. Rev. Lett.},
  {\bf 92}, 121101, (2004). \epubtkKeywords{Tests of relativistic gravity}

\bibitem{shlyakter}
Shlyakter, A.I., ``Direct test of the constancy of fundamental nuclear
  constants'', {\em Nature}, {\bf 264}, 340, (1976).
  \epubtkKeywords{equivalence principle tests}

\bibitem{petitjean1}
Srianand, R., Chand, H., Petitjean, P., and Aracil, B., ``Limits on the time
  variation of the electromagnetic fine-structure constant in the low energy
  limit from absorption lines in the spectra of distant quasars'', {\em Phys.
  Rev. Lett.}, {\bf 92}, 121302--1--4, (2004). Related online version (cited on
  15 July 2005): \newline\url{http://arXiv.org/abs/astro-ph/0402177}.
  \epubtkKeywords{equivalence principle tests}

\bibitem{stairs05}
Stairs, I.~H., Faulkner, A.~J., Lyne, A.~G., Kramer, M., Lorimer, D.~R.,
  McLaughlin, M.~A., Manchester, R.~N., Hobbs, G.~B., Camilo, F., Possenti, A.,
  Burgay, M., D'Amico, N., Freire, P.~C., and Gregory, P.~C., ``Discovery of
  three wide-orbit binary pulsars: Implications for binary evolution and
  equivalence principles'', {\em Astrophys. J.}, {\bf 632}, 1060--1068, (2005).
  Related online version (cited on 15 February 2006):
  \newline\url{http://arXiv.org/abs/astro-ph/0506188}. \epubtkKeywords{binary
  pulsars, equivalence principle tests}

\bibitem{StairsLRR}
Stairs, I.H., ``Testing General Relativity with Pulsar Timing'', {\em Living
  Rev. Relativity}, {\bf 6}, lrr-2003-5, (2003). URL (cited on 15 July 2005):
  \newline\url{http://www.livingreviews.org/lrr-2003-5}. \epubtkKeywords{binary
  pulsars}

\bibitem{stairs2}
Stairs, I.H., Nice, D.J., Thorsett, S.E., and Taylor, J.H., ``Recent Arecibo
  timing of the relativistic binary PSR B1534+12'', in {\em Gravitational Waves
  and Experimental Gravity}, Proceedings of the XXXIV Rencontres de Moriond,
  Les Arcs, Savoie, France, March 13--20, 1999, Les Arcs, Savoie, France,
  (World Publishers, Hanoi, 2000). Related online version (cited on 15 January
  2001): \newline\url{http://arXiv.org/abs/astro-ph/9903289}.
  \epubtkKeywords{binary pulsars}

\bibitem{gpbwebsite}
Stanford University, ``Gravity Probe B: Testing Einstein's Universe'', project
  homepage. URL (cited on 15 July 2005):
  \newline\url{http://einstein.stanford.edu/}. \epubtkKeywords{tests of
  relativistic gravity, Lense-Thirring effect}

\bibitem{STEP}
Stanford University, ``STEP: Satellite Test of the Equivalence Principle'',
  project homepage, (2005). URL (cited on 15 July 2005):
  \newline\url{http://einstein.stanford.edu/STEP/}. \epubtkKeywords{equivalence
  principle tests}

\bibitem{stanwix05}
{Stanwix}, P.~L., {Tobar}, M.~E., {Wolf}, P., {Susli}, M., {Locke}, C.~R.,
  {Ivanov}, E.~N., {Winterflood}, J., and {van Kann}, F., ``{Test of Lorentz
  Invariance in Electrodynamics Using Rotating Cryogenic Sapphire Microwave
  Oscillators}'', {\em Phys. Rev. Lett.}, {\bf 95}(4), 040404, (jul, 2005).
  Related online version (cited on 15 February 2006):
  \newline\url{http://arXiv.org/abs/hep-ph/0506074}. \epubtkKeywords{tests of
  Lorentz invariance}

\bibitem{Su94}
Su, Y., Heckel, B.R., Adelberger, E.G., Gundlach, J.H., Harris, M., Smith,
  G.L., and Swanson, H.E., ``New tests of the universality of free fall'', {\em
  Phys. Rev. D}, {\bf 50}, 3614--3636, (1994). \epubtkKeywords{equivalence
  principle tests}

\bibitem{talmadge}
Talmadge, C.L., Berthias, J.-P., Hellings, R.W., and Standish, E.M.,
  ``Model-Independent Constraints on Possible Modifications of Newtonian
  Gravity'', {\em Phys. Rev. Lett.}, {\bf 61}, 1159--1162, (1988).
  \epubtkKeywords{tests of relativistic gravity, graviton}

\bibitem{Taylor87}
Taylor, J.~H., ``Astronomical and Space Experiments to Test Relativity'', in
  MacCallum, M. A.~H., ed., {\em General Relativity and Gravitation},  209,
  (Cambridge University Press, Cambridge, U.K.; New York, U.S.A., 1987).
  \epubtkKeywords{binary pulsars}

\bibitem{Taylor94}
Taylor, J.H., ``Nobel Lecture: Binary pulsars and relativistic gravity'', {\em
  Rev. Mod. Phys.}, {\bf 66}, 711--719, (1994). \epubtkKeywords{binary pulsars}

\bibitem{TWDW92}
Taylor, J.H., Wolszczan, A., Damour, T., and Weisberg, J.M., ``Experimental
  constraints on strong-field relativistic gravity'', {\em Nature}, {\bf 355},
  132--136, (1992). \epubtkKeywords{binary pulsars}

\bibitem{TaylorVeneziano}
Taylor, T.R., and Veneziano, G., ``Dilaton coupling at large distance'', {\em
  Phys. Lett. B}, {\bf 213}, 450--454, (1988). \epubtkKeywords{string theory,
  scalar-tensor gravity}

\bibitem{Thorne300}
Thorne, K.S., ``Gravitational radiation'', in Hawking, S.W., and Israel, W.,
  eds., {\em Three Hundred Years of Gravitation},  330--458, (Cambridge
  University Press, Cambridge, U.K.; New York, U.S.A., 1987).
  \epubtkKeywords{Gravitational radiation, Gravitational wave detectors}

\bibitem{snowmass}
Thorne, K.S., ``Gravitational waves'', in Kolb, E.W., and Peccei, R., eds.,
  {\em Particle and Nuclear Astrophysics and Cosmology in the Next Millennium},
  Proceedings of the 1994 Snowmass Summer Study, Snowmass, Colorado, June 29 --
  July 14, 1994,  160--184, (World Scientific, Singapore; River Edge, U.S.A.,
  1995). Related online version (cited on 15 January 2001):
  \newline\url{http://arXiv.org/abs/gr-qc/9506086}.
  \epubtkKeywords{Gravitational radiation, Gravitational wave detectors}

\bibitem{treuhaft}
Treuhaft, R.N., and Lowe, S.T., ``A measurement of planetary relativistic
  deflection'', {\em Astron. J.}, {\bf 102}, 1879--1888, (1991).
  \epubtkKeywords{Tests of relativistic gravity}

\bibitem{turneaure}
Turneaure, J.P., Will, C.M., Farrell, B.F., Mattison, E.M., and Vessot, R.F.C.,
  ``Test of the principle of equivalence by a null gravitational redshift
  experiment'', {\em Phys. Rev. D}, {\bf 27}, 1705--1714, (1983).
  \epubtkKeywords{equivalence principle tests}

\bibitem{turyshev04a}
Turyshev, S.G., Shao, M., and Nordtvedt, K., ``Experimental design for the
  LATOR mission'', {\em Int. J. Mod. Phys. D}, {\bf 13}, 2035--2063, (2004).
  Related online version (cited on 15 July 2005):
  \newline\url{http://arXiv.org/abs/gr-qc/0410044}. \epubtkKeywords{tests of
  relativistic gravity, Shapiro time delay}

\bibitem{turyshev04b}
Turyshev, S.G., Shao, M., and Nordtvedt, K., ``The laser astrometric test of
  relativity mission'', {\em Class. Quantum Grav.}, {\bf 21}, 2773--2799,
  (2004). Related online version (cited on 15 July 2005):
  \newline\url{http://arXiv.org/abs/gr-qc/0311020}. \epubtkKeywords{tests of
  relativistic gravity, Shapiro time delay}

\bibitem{uzan03}
Uzan, J.-P., ``The fundamental constants and their variation: observational and
  theoretical status'', {\em Rev. Mod. Phys.}, {\bf 75}, 403, (2003). Related
  online version (cited on 15 July 2005):
  \newline\url{http://arXiv.org/abs/hep-ph/0205340}.
  \epubtkKeywords{equivalence principle, equivalence principle tests}

\bibitem{vdv70}
{van Dam}, H., and {Veltman}, M., ``{Massive and mass-less Yang-Mills and
  gravitational fields}'', {\em Nuclear Physics B}, {\bf 22}, 397--411, (sep,
  1970). \epubtkKeywords{graviton}

\bibitem{vessot}
Vessot, R.F.C., Levine, M.W., Mattison, E.M., Blomberg, E.L., Hoffman, T.E.,
  Nystrom, G.U., Farrell, B.F., Decher, R., Eby, P.B., Baugher, C.R., Watts,
  J.W., Teuber, D.L., and Wills, F.O., ``Test of relativistic gravitation with
  a space-borne hydrogen maser'', {\em Phys. Rev. Lett.}, {\bf 45}, 2081--2084,
  (1980). \epubtkKeywords{equivalence principle tests}

\bibitem{visser}
Visser, M., ``Mass for the graviton'', {\em Gen. Relativ. Gravit.}, {\bf 30},
  1717--1728, (1998). Related online version (cited on 15 January 2001):
  \newline\url{http://arXiv.org/abs/gr-qc/9705051}. \epubtkKeywords{theories of
  gravity, graviton}

\bibitem{wagoner}
Wagoner, R.V., ``Resonant-mass detection of tensor and scalar waves'', in
  Marck, J.A., and Lasota, J.P., eds., {\em Relativistic Gravitation and
  Gravitational Radiation}, Proceedings of the Les Houches School of Physics,
  held in Les Houches, Haute Savoie, 26 September -- 6 October, 1995,
  419--432, (Cambridge University Press, Cambridge, U.K., 1997).
  \epubtkKeywords{Resonant gravitational wave detectors, scalar-tensor gravity}

\bibitem{wagoner2}
Wagoner, R.V., and Kalligas, D., ``Scalar-tensor theories and gravitational
  radiation'', in Marck, J.A., and Lasota, J.P., eds., {\em Relativistic
  Gravitation and Gravitational Radiation}, Proceedings of the Les Houches
  School of Physics, held in Les Houches, Haute Savoie, 26 September -- 6
  October, 1995,  433--446, (Cambridge University Press, Cambridge, U.K.,
  1997). \epubtkKeywords{Resonant gravitational wave detectors, scalar-tensor
  gravity}

\bibitem{wagwill}
Wagoner, R.V., and Will, C.M., ``Post-Newtonian gravitational radiation from
  orbiting point masses'', {\em Astrophys. J.}, {\bf 210}, 764--775, (1976).
  \epubtkKeywords{gravitational radiation, post-Newtonian approximations,
  binary systems}

\bibitem{webb99}
Webb, J.~K., Flambaum, V.~V., Churchill, C.~W., Drinkwater, M.~J., and Barrow,
  J.~D., ``Search for time variation of the fine structure constant'', {\em
  Phys. Rev. Lett.}, {\bf 82}, 884--887, (1999). Related online version (cited
  on 15 July 2005): \newline\url{http://arXiv.org/abs/astro-ph/9803165}.
  \epubtkKeywords{equivalence principle tests}

\bibitem{Weinberg}
Weinberg, S., {\em Gravitation and Cosmology: Principles and Applications of
  the General Theory of Relativity}, (Wiley, New York, U.S.A., 1972).
  \epubtkKeywords{general relativity}

\bibitem{WeisbergTaylor02}
Weisberg, J.M., and Taylor, J.H., ``General relativistic geodetic spin
  precession in binary pulsar B1913+16: Mapping the emission beam in two
  dimensions'', {\em Astrophys. J.}, {\bf 576}, 942--949, (2002). Related
  online version (cited on 15 July 2005):
  \newline\url{http://arXiv.org/abs/astro-ph/0205280}. \epubtkKeywords{binary
  pulsars}

\bibitem{WeisbergTaylor05}
Weisberg, J.M., and Taylor, J.H., ``The relativistic binary pulsar B1913+16:
  Thirty years of observations and analysis'', in Rasio, F.A., and Stairs,
  I.H., eds., {\em Binary Radio Pulsars}, Proceeding of the Aspen Conference,
  vol. 328 of ASP Conference Series,  25--32, (Astronomical Society of the
  Pacific, San Francisco, U.S.A., 2005). Related online version (cited on 15
  July 2005): \newline\url{http://arXiv.org/abs/astro-ph/0407149}.
  \epubtkKeywords{binary pulsars}

\bibitem{WenSchutz05}
Wen, L., and Schutz, B.F., ``Coherent network detection of gravitational waves:
  the redundancy veto'', {\em Class.Quantum Grav.}, {\bf 22}, S1321--S1336,
  (2005). Related online version (cited on 15 July 2005):
  \newline\url{http://arXiv.org/abs/gr-qc/0508042}. Proceedings of the 9th
  Gravitational Wave Data Analysis Workshop, Annecy, France, 15--18 December
  2004. \epubtkKeywords{gravitational wave detectors}

\bibitem{Will71a}
Will, C.M., ``Theoretical frameworks for testing relativistic gravity. II.
  Parametrized post-Newtonian hydrodynamics and the Nordtvedt effect'', {\em
  Astrophys. J.}, {\bf 163}, 611--628, (1971). \epubtkKeywords{tests of
  relativistic gravity}

\bibitem{Will76}
Will, C.M., ``Active mass in relativistic gravity: Theoretical interpretation
  of the Kreuzer experiment'', {\em Astrophys. J.}, {\bf 204}, 224--234,
  (1976). \epubtkKeywords{tests of relativistic gravity}

\bibitem{Will300}
Will, C.M., ``Experimental gravitation from Newton's Principia to Einstein's
  general relativity'', in Hawking, S.W., and Israel, W., eds., {\em Three
  Hundred Years of Gravitation},  80--127, (Cambridge University Press,
  Cambridge, U.K.; New York, U.S.A., 1987). \epubtkKeywords{tests of
  relativistic gravity}

\bibitem{willcavendish}
Will, C.M., ``Henry Cavendish, Johann von Soldner, and the deflection of
  light'', {\em Am. J. Phys.}, {\bf 56}, 413--415, (1988).
  \epubtkKeywords{tests of relativistic gravity}

\bibitem{WillSky}
Will, C.M., ``Twilight time for the fifth force?'', {\em Sky and Telescope},
  {\bf 80}, 472--479, (1990). \epubtkKeywords{tests of relativistic gravity}

\bibitem{Will92b}
Will, C.M., ``Clock synchronization and isotropy of the one-way speed of
  light'', {\em Phys. Rev. D}, {\bf 45}, 403--411, (1992).
  \epubtkKeywords{tests of Lorentz invariance}

\bibitem{Will92c}
Will, C.M., ``Is momentum conserved? A test in the binary system PSR 1913+16'',
  {\em Astrophys. J. Lett.}, {\bf 393}, L59--L61, (1992). \epubtkKeywords{tests
  of relativistic gravity, binary pulsars}

\bibitem{tegp}
Will, C.M., {\em Theory and experiment in gravitational physics}, (Cambridge
  University Press, Cambridge, U.K.; New York, U.S.A., 1993), 2nd edition.
  \epubtkKeywords{tests of relativistic gravity}

\bibitem{WER}
Will, C.M., {\em Was Einstein Right?: Putting General Relativity to the Test},
  (Basic Books, New York, U.S.A., 1993), 2nd edition. \epubtkKeywords{tests of
  relativistic gravity}

\bibitem{willbd}
Will, C.M., ``Testing scalar-tensor gravity with gravitational-wave
  observations of inspiralling compact binaries'', {\em Phys. Rev. D}, {\bf
  50}, 6058--6067, (1994). Related online version (cited on 15 January 2001):
  \newline\url{http://arXiv.org/abs/gr-qc/9406022}.
  \epubtkKeywords{gravitational radiation, scalar-tensor gravity, binary
  systems}

\bibitem{sussp}
Will, C.M., ``The confrontation between general relativity and experiment: A
  1995 update'', in Hall, G.S., and Pulham, J.R., eds., {\em General
  Relativity}, Proceedings of the Forty Sixth Scottish Universities Summer
  School in Physics, Aberdeen, July 1995, vol.~46 of Scottish Graduate Series,
  239--282, (Institute of Physics Publishing, Bristol, U.K., 1996).
  \epubtkKeywords{tests of relativistic gravity}

\bibitem{graviton}
Will, C.M., ``Bounding the mass of the graviton using gravitional-wave
  observations of inspiralling compact binaries'', {\em Phys. Rev. D}, {\bf
  57}, 2061--2068, (1998). Related online version (cited on 15 January 2001):
  \newline\url{http://arXiv.org/abs/gr-qc/9709011}.
  \epubtkKeywords{gravitational radiation, theories of gravity, binary systems}

\bibitem{slac}
Will, C.M., ``The confrontation between general relativity and experiment: A
  1998 update'', in Dixon, L.J., ed., {\em Gravity: From the Hubble Length to
  the Planck Length}, Proceedings of the 26th SLAC Summer Institute on Particle
  Physics (SSI 98), Stanford, USA, 3--14 August 1998, vol. 538, (SLAC,
  Springfield, U.S.A., 1998). URL (cited on 15 January 2001):
  \newline\url{http://www.slac.stanford.edu/pubs/confproc/ssi98/ssi98-002.html%
}. Related online version (cited on 15 January 2001):
  \newline\url{http://arXiv.org/abs/gr-qc/9811036}. \epubtkKeywords{tests of
  relativistic gravity}

\bibitem{physicscentral}
Will, C.M., ``Einstein's relativity and everyday life'', electronic, Americal
  Physical Society, (2000). URL (cited on 15 January 2001):
  \newline\url{http://www.physicscentral.com/writers/writers-00-2.html}.
  \epubtkKeywords{global positioning system}

\bibitem{willspeed03}
Will, C.M., ``Propagation speed of gravity and the relativistic time delay'',
  {\em Astrophys. J.}, {\bf 590}, 683--690, (2003). Related online version
  (cited on 15 July 2005): \newline\url{http://arXiv.org/abs/astro-ph/0301145}.
  \epubtkKeywords{speed of gravity, Shapiro time delay}

\bibitem{DIRE3}
{Will}, C.M., ``Post-Newtonian gravitational radiation and equations of motion
  via direct integration of the relaxed Einstein equations. III. Radiation
  reaction for binary systems with spinning bodies'', {\em Phys. Rev. D}, {\bf
  71}, 084027, (2005). Related online version (cited on 15 July 2005):
  \newline\url{http://arXiv.org/abs/gr-qc/0502039}.
  \epubtkKeywords{gravitational radiation, equations of motion, post-Newtonian
  approximations}

\bibitem{willnordtvedt72}
Will, C.M., and Nordtvedt, K., ``Conservation laws and preferred frames in
  relativistic gravity. I. Preferred-frame theories and an extended PPN
  formalism'', {\em Astrophys. J.}, {\bf 177}, 757--774, (1972).
  \epubtkKeywords{theories of gravity, post-Newtonian approximations}

\bibitem{opus}
Will, C.M., and Wiseman, A.G., ``Gravitational radiation from compact binary
  systems: Gravitational waveforms and energy loss to second post-Newtonian
  order'', {\em Phys. Rev. D}, {\bf 54}, 4813--4848, (1996). Related online
  version (cited on 15 January 2001):
  \newline\url{http://arXiv.org/abs/gr-qc/9608012}.
  \epubtkKeywords{gravitational radiation, post-Newtonian approximations,
  binary systems}

\bibitem{willyunes}
Will, C.M., and Yunes, N., ``Testing alternative theories of gravity using
  LISA'', {\em Class. Quantum Grav.}, {\bf 21}, 4367--4381, (2004). Related
  online version (cited on 15 July 2005):
  \newline\url{http://arXiv.org/abs/gr-qc/0403100}. \epubtkKeywords{parameter
  estimation, gravitational radiation, theories of gravity}

\bibitem{zaglauer}
Will, C.M., and Zaglauer, H.W., ``Gravitational radiation, close binary systems
  and the Brans-Dicke theory of gravity'', {\em Astrophys. J.}, {\bf 346},
  366--377, (1989). \epubtkKeywords{gravitational radiation, scalar-tensor
  gravity, binary systems}

\bibitem{Williams}
Williams, J.G., Newhall, X.X., and Dickey, J.O., ``Relativity parameters
  determined from lunar laser ranging'', {\em Phys. Rev. D}, {\bf 53},
  6730--6739, (1996). \epubtkKeywords{tests of relativistic gravity, lunar
  laser ranging}

\bibitem{williams04}
Williams, J.G., Turyshev, S.G., and Boggs, D.H., ``Progress in lunar laser
  ranging tests of relativistic gravity'', {\em Phys. Rev. Lett.}, {\bf 93},
  261101--1--4, (2004). Related online version (cited on 15 July 2005):
  \newline\url{http://arXiv.org/abs/gr-qc/0411113}. \epubtkKeywords{tests of
  relativistic gravity, lunar laser ranging}

\bibitem{williams04ijmp}
Williams, J.G., Turyshev, S.G., and Murphy, T.W., ``Improving LLR tests of
  gravitational theory'', {\em Int. J. Mod. Phys. D}, {\bf 13}, 567--582,
  (2004). Related online version (cited on 15 July 2005):
  \newline\url{http://arXiv.org/abs/gr-qc/0311021}. \epubtkKeywords{tests of
  relativistic gravity, lunar laser ranging}

\bibitem{wolf03}
Wolf, P., Bize, S., Clairon, A., Luiten, A.N., Santarelli, G., and Tobar, M.E.,
  ``Tests of Lorentz invariance using a microwave resonator'', {\em Phys. Rev.
  Lett.}, {\bf 90}, 060402--1--4, (2003). Related online version (cited on 15
  July 2005): \newline\url{http://arXiv.org/abs/gr-qc/0210049}.
  \epubtkKeywords{tests of Lorentz invariance}

\bibitem{wolfe76}
Wolfe, A.M., Brown, R.L., and Roberts, M.S., ``Limits on the variation of
  fundamental atomic quantities over cosmic time scales'', {\em Phys. Rev.
  Lett.}, {\bf 37}, 179--181, (1976). \epubtkKeywords{equivalence principle
  tests}

\bibitem{zakharov70}
{Zakharov}, V.~I., ``{Linearized gravitation theory and the graviton mass}'',
  {\em Journal of Experimental and Theoretical Physics Letters}, {\bf 12},
  312+, (1970). \epubtkKeywords{graviton}

\end{thebibliography}

\end{document}